%% file: main.tex
\documentclass[a4paper,11pt]{article}
\pdfoutput=1

\usepackage{jcappub}
\usepackage[normalem]{ulem}
\usepackage{graphicx}
\usepackage{aas_macros}
\usepackage{amsmath, bm}	% Advanced maths commands
\usepackage{amssymb}	% Extra maths symbols
\usepackage[dvipsnames]{xcolor}     % colorful comments 
\usepackage{float} %required for the placement specifier H
\usepackage{ulem}
\usepackage{xspace}
\usepackage{fontawesome}
\usepackage{hyperref}
\usepackage{orcidlink}
\usepackage{hanging}

\newcommand{\hMpc}{\,h\, \mathrm{Mpc}^{-1}}
\newcommand{\hMpccube}{\,h^{3} \mathrm{Mpc}^{-3}}
\newcommand{\Mpchcube}{\,h^{-3} \mathrm{Mpc}^{3}}

\newcommand{\Mpch}{\,h^{-1} \mathrm{Mpc}}
\newcommand{\Gpch}{\,h^{-1} \mathrm{Gpc}}

\newcommand{\Sigmasm}{\Sigma_{\text{sm}}}
\newcommand{\beq}{\begin{eqnarray}}
\newcommand{\eeq}{\end{eqnarray}}
\newcommand{\qper}{\alpha_{\perp}}
\newcommand{\qpar}{\alpha_{\parallel}}
\newcommand{\qiso}{\alpha_{\text{iso}}}
\newcommand{\qap}{\alpha_{\text{AP}}}
\newcommand{\sigmaper}{\Sigma_{\perp}}
\newcommand{\sigmapar}{\Sigma_{\parallel}}
\newcommand{\sigmas}{\Sigma_{\text{s}}}
\newcommand{\rhat}{{\bf \hat{r}}}
\newcommand{\gauss}{\mathcal{N}}    
\newcommand{\unif}{\mathcal{U}}

\newcommand{\recsym}{{\tt RecSym}}
\newcommand{\reciso}{{\tt RecIso}}
\newcommand{\pyrecon}{{\tt pyrecon}}
\newcommand{\ifft}{{\tt IFFT}}
\newcommand{\multigrid}{{\tt multigrid}}
\newcommand{\abacussummit}{{\tt AbacusSummit}}
\newcommand{\ezmocks}{{\tt EZmocks}}
\newcommand{\bgs}{{\tt BGS}}
\newcommand{\elgo}{{\tt ELG1}}
\newcommand{\elgt}{{\tt ELG2}}
\newcommand{\elgs}{{\tt ELGs}}

\newcommand{\lrgo}{{\tt LRG1}}
\newcommand{\lrgt}{{\tt LRG2}}
\newcommand{\lrgth}{{\tt LRG3}}
\newcommand{\lrgs}{{\tt LRGs}}
\newcommand{\qso}{{\tt QSO}}

\usepackage[nameinlink,noabbrev]{cleveref}
\crefname{equation}{Eq.}{Eqs.}
\crefname{section}{Section}{Sections}
\crefname{figure}{Fig.}{Figs.}
\crefname{table}{Table}{Tables}
\crefname{appendix}{Appendix}{Appendices}
\Crefname{figure}{Figure}{Figures}
\Crefname{equation}{Equation}{Equations}
\Crefname{section}{Section}{Sections}
\Crefname{table}{Table}{Tables}

\def\pvmhid#1{}

\title{\boldmath Optimal Reconstruction of Baryon Acoustic Oscillations for DESI 2024}

\input{authors}

\abstract{Baryon acoustic oscillations (BAO) provide a robust standard ruler to measure the expansion history of the Universe through galaxy clustering. Density-field reconstruction is now a widely adopted procedure for increasing the precision and accuracy of the BAO detection. With the goal of finding the optimal reconstruction settings to be used in the DESI 2024 galaxy BAO analysis, we assess the sensitivity of the post-reconstruction BAO constraints to different choices in our analysis configuration, performing tests on blinded data from the first year of DESI observations (DR1), as well as on mocks that mimic the expected clustering and selection properties of the DESI DR1 target samples. Overall, we find that BAO constraints remain robust against multiple aspects in the reconstruction process, including the choice of smoothing scale, treatment of redshift-space distortions, fiber assignment incompleteness, and parameterizations of the BAO model. We also present a series of tests that DESI followed in order to assess the maturity of the end-to-end galaxy BAO pipeline before the unblinding of the large-scale structure catalogs.}

\begin{document}
\maketitle
\flushbottom

\section{Introduction}

Galaxy redshift surveys have shaped our understanding of the origin and evolution of the Cosmos. Measurements of baryon acoustic oscillations \citep[BAO;][]{Peebles1970, Sunyaev1970, Bond1987} in the galaxy distribution have allowed us to measure the expansion history of the Universe with high precision, establishing $\Lambda$CDM as our standard cosmological paradigm \citep{Eisenstein2005, Cole2005,Alam2017, eboss2020, Moon2023:2304.08427}. These oscillations are a characteristic pattern of clustering in the large-scale structure (LSS), corresponding to remnants of acoustic waves that traveled through the photon-baryon plasma in the early Universe. These sound waves imprinted a characteristic scale in the distribution of matter, given by the distance the perturbations traveled before they froze in place once hydrogen recombined and the baryons stopped being dragged by the photons (the \textit{baryon drag} epoch). Over time, these cosmic ripples provided sites where galaxies preferentially formed, and, as a consequence, this \textit{sound horizon} scale is also found in the spatial distribution of galaxies, either as a localized bump in the two-point correlation function or as wiggles in the power spectrum \citep{Eisenstein1998, Matsubara2004}.

Non-linear gravitational evolution broadens the BAO peak in the correlation function or damps high-k oscillations in the power spectrum. In addition, the location of the peak can also be shifted due to non-linear evolution and galaxy biasing \citep{Eisenstein2007:astro-ph/0604361,KP4s2-Chen}. These effects degrade the precision of BAO measurements but can be partially corrected. A technique known as \textit{reconstruction} has been designed to reverse the non-linear effects by shifting overdensities back to their initial positions, so that the sound horizon scale can be recovered with greater statistical significance \citep{Eisenstein2007:astro-ph/0604362}. In addition, reconstruction has been shown to be effective in removing the shifts in the acoustic peak caused by non-linearities and galaxy biasing, reducing the systematic errors in BAO measurements \citep[e.g.][]{Seo2008, Seo2010, Schmittfull2015, Ding2018}.

Clustering analyses from the last decade, based on the Sloan Digital Sky Survey \citep[SDSS;][]{York2000}, the Baryon Oscillation Spectroscopic Survey \citep[BOSS;][]{Dawson2013}, the extended Baryon Oscillation Spectroscopic Survey \citep[eBOSS;][]{Dawson2016}, and WiggleZ \cite{Blake2011}, have routinely adopted reconstruction as a standard to extract cosmological information from BAO \citep{Padmanabhan2012, Anderson2014:1303.4666, Tojeiro2014:1401.1768, Kazin2014:1401.0358, Ross2015:1409.3242, Alam2017, eboss2020}. Recently, \cite{Moon2023:2304.08427} reported the first detection of BAO from the early data of the Dark Energy Spectroscopic Instrument (DESI) survey \citep{DESI2023a.KP1.SV, OnePercent_b}, showing that reconstruction effectively increases the statistical significance of BAO detection for various DESI target samples.

Running reconstruction on galaxy maps requires certain modeling and algorithmic choices that can potentially affect the performance of the reconstruction process. Many different reconstruction algorithms have been proposed in the literature \citep[e.g.][]{Padmanabhan2012, White2015:1504.03677, Burden2015:1504.02591}, and for a given model, several hyperparameters must be calibrated to optimally recover the linear density field. With the standard reconstruction algorithms that have been applied to galaxy surveys, one needs to first assume a fiducial cosmology to convert galaxy redshifts to distances, as well as a fiducial value for the growth rate of cosmic structure and the linear bias of the galaxy sample, both of which are necessary to calculate the Zel'dovich displacement field that is at the core of the reconstruction formalism \citep{Zeldovich1970, Eisenstein2007:astro-ph/0604362}. One also needs to make choices related to the numerical implementation of the reconstruction recipes, such as how exactly the discrete galaxy positions are mapped to a smoothed overdensity field on a grid, which can also play a role in determining the displacement field \citep{White2010:1004.0250, Burden2014:1408.1348, Anderson2014:1303.4666, Vargas-Magaña:1509.06384}.

Ultimately, the optimal reconstruction settings depend on the specific galaxy sample that is being analyzed, with considerations for the sample number density, redshift range, and survey geometry. The goal of this article is to present a thorough study of BAO reconstruction as applied to the DESI 2024 results, which use the first data release of the DESI cosmological survey (DESI DR1; \cite{DESI2024.I.DR1}). For this purpose, the first half of our paper analyzes mocks that match the DESI DR1 target samples with different reconstruction schemes and settings, with the aim of optimizing the pipeline that DESI will use to perform the main BAO analysis, presented in \cite{DESI2024.III.KP4}, and the corresponding cosmological interpretation of the results, presented in \cite{DESI2024.VI.KP7A}.

To minimize potential confirmation biases during parameter inference, the LSS catalogs analyzed by the DESI Collaboration are blinded to cosmology, as detailed in \citep{KP3s9-Andrade}. In this context, blinding involves the deliberate concealment or modification of the catalogs, ensuring that the choices and interpretations in our analysis remain unbiased. The second half of our paper concerns a series of tests performed on the blinded DESI DR1 data and mocks that assess the robustness of the BAO constraints against different measurement and modeling choices in our pipeline, which informed the decision of unblinding the catalogs to perform the final cosmological analysis from BAO.

The paper is organized as follows. In \cref{sec:observations} we describe the observables used in this work, including the clustering and the reconstruction of the blinded DESI DR1 catalogs. In \cref{sec:modeling} we present the theoretical aspect, including the description of the BAO model and parameter inference, as well as a description of the mock galaxy catalogs that mimic the DESI DR1 selection and clustering properties. The results are presented in \cref{sec:results}, where we present parameter recovery tests on the mock galaxy catalogs and unblinding tests using the DESI DR1 data.

Throughout this paper, we adopt a fiducial cosmology that is characterized by $\omega_{\rm cdm} = 0.1200$, $\omega_{\rm b} = 0.02237$, $h = 0.6736$, $N_{\rm ur} = 2.0328$, and one massive neutrino with $\omega_{\nu} = 0.00064420$, which matches the Planck 2018 base-$\Lambda$CDM parameters \citep{Planck2020}, as well as the baseline cosmology from the \abacussummit\ simulations \citep{Maksimova2021}, described in \cref{subsec:mock_catalogues}. A careful study of the impact of this choice of fiducial cosmology on the BAO analysis is presented in \cite{KP4s9-Perez-Fernandez}.

\section{Observations}
\label{sec:observations}

\subsection{DESI DR1 samples}

The Dark Energy Spectroscopic Instrument (DESI; \cite{DESI2016a.Science}) is a multi-fiber spectrograph installed on the Nicholas U. Mayall 4-meter telescope at Kitt Peak National Observatory in Arizona, which can simultaneously take spectra of 5,000 objects using fiber-fed robotic positioners \citep{DESI2022.KP1.Instr}. The DESI data release 1 \cite[DR1;][]{DESI2024.I.DR1} comprises observations from May 14, 2021, following a period of successful survey validation \citep{DESI2023a.KP1.SV}, which resulted in DESI's first Early Data Release \citep{DESI2023b.KP1.EDR}, through June 14, 2022. Depending on the observerving conditions, DESI dynamically allocates its observing time into a `bright-time' program, and a `dark-time' program. The bright-time program comprises the Bright Galaxy Survey \cite{Hahn2022:2208.08512}, while the dark-time program observes luminous red galaxies, emission-line galaxies, and quasars, as well as a sample of Milky Way stars \citep[MWS,][]{Cooper2023:2208.08514}. In addition, DESI targets Lyman-$\alpha$ forest quasars to trace the distribution of neutral hydrogen \cite{2023MNRAS.tmp.3626R}.

For this study, we use the large-scale structure (LSS) catalogs presented in \cite{DESI2024.II.KP3}, focusing on the BAO measurements of the following DR1 target samples:
\begin{itemize}
    \item {\bf The Bright Galaxy Sample (\bgs)}: This sample consists of 300,017 good redshifts in the redshift range $0.1 < z < 0.4$ over an area of 7,473\,$\deg^2$. Although the nominal Bright Galaxy Sample is flux-limited and has a number density that has a strong redshift dependence, the sample used for the DR1 analysis was engineered to have a roughly constant number density of at $0.1 < z < 0.4$, which was achieved by using $r$ band absolute magnitude cuts with a $k$+E correction \citep{DESI2024.II.KP3}. The resulting number density for this \bgs\ sample is roughly constant at $3\times10^{-4} \hMpccube$ when averaged over the DR1 footprint, having a clustering amplitude similar to that of \lrgs\ (see below), making them similarly biased with respect to the dark matter density field.

    \item {\bf The Luminous Red Galaxy Sample (\lrgs)}: This sample includes 2,138,600 good redshifts over 5,840$\,\deg^2$ in the redshift range $0.4 < z < 1.1$. These are red, passive galaxies that are highly biased with respect to the underlying matter distribution, making them specially suitable for BAO analyses, similar to the previous LRG samples from SDSS \cite{Dawson2013, Dawson2016}. It has a roughly constant number density of $3.5\times10^{-4} \hMpccube$ up to $z = 0.8$, with a decrease in density at $0.8 < z < 1.1$ (see Figure~1 from \cite{DESI2024.III.KP4}). We further split this sample into three redshift bins when calculating the two-point clustering: $0.4 < z < 0.6$ (\lrgo), $0.6 < z < 0.8$ (\lrgt), and $0.8 < z < 1.1$ (\lrgth).

    \item {\bf The Emissison-line Galaxy Sample (\elgs)}: This sample comprises 2,432,022 good redshifts in the range $0.8 < z < 1.6$ over an area of 5,914\,$\deg^{2}$. These are [OII] emission-line galaxies that are generally active star-forming galaxies. They are less clustered with respect to the matter density field compared to passive galaxies such as the \lrgs. We subdivide them into two disjoint redshift bins for the clustering measurements: $0.8 < z < 1.1$ (\elgo) and $1.1 < z < 1.6$ (\elgt).

    \item {\bf The Quasar Sample (\qso)}: This sample consists of $856,652$ good redshifts at $0.8 < z < 2.1$, covering an area of 7,249\,$\deg^2$. This is the sample that has the largest bias with respect to the underlying dark matter distribution, since it has a clustering amplitude that is somewhere in between the \lrgs\ and \elgs, but at a much higher effective redshift.
\end{itemize}

To avoid confirmation bias in the cosmological analysis, the LSS catalogs of each target sample were blinded to cosmology. In practice, the redshifts of the tracers were shifted with a certain prescription, which changes the position of the BAO feature and the redshift-space distortions (RSD) signal. The catalogs were only unblinded once 1) the full end-to-end analysis pipeline was fixed and 2) the data and results run through this pipeline had passed a series of predefined tests, ensuring that the choices and interpretations in our analysis remain unbiased. A detailed description of the blinding scheme is presented in \cite{KP3s9-Andrade}, while unblinded measurements are presented in \cite{DESI2024.III.KP4}. Beyond the above samples, the DESI 2024 results will also include BAO in the Lyman-$\alpha$ forest \citep{DESI2024.IV.KP6}, and both sets will be combined to infer cosmology \citep{DESI2024.VI.KP7A}. Analyses using clustering information beyond just the BAO scale will be presented in \cite{DESI2024.V.KP5}, and will provide additional cosmological constraints, including on primordial non-Gaussianities \citep{DESI2024.VII.KP7B,DESI2024.VIII.KP7C}.

In addition to the tracer LSS catalogs, we also use a series of random catalogs that follow the footprint and radial selection of the DESI samples but with no intrinsic clustering \cite{KP3s15-Ross}, which we use to estimate the overdensity field for reconstruction and the two-point clustering.

\subsection{Density-field reconstruction} \label{subsec:reconstruction}

In a Lagrangian framework, we can relate the Eulerian position of a galaxy at a time $t$, denoted by ${\bf x}({\bf q}, t)$, to its initial position ${\bf q}$ in Lagrangian space by
\begin{align}
{\bf x}({\bf q}, t) = {\bf q} + \bm{\Psi}({\bf q}, t),
\end{align}
where $\bm{\Psi}({\bf q}, t)$ is the Lagrangian displacement vector in real space. The galaxy velocity flows that are sourced by structure formation, which are captured by the displacement vector, smear and shift the BAO signal, degrading its detection significance and biasing the distance-redshift measurements. The goal of BAO reconstruction \citep{Eisenstein2007:astro-ph/0604361} is to reverse the non-linear motions in the density field, increasing the accuracy and precision of the BAO method. Reconstruction achieves this by calculating the displacement field based on an estimate of the galaxy velocity field, which is in turn estimated from the observed galaxy density field.

The first-order term in a perturbative expansion of $\bm{\Psi}({\bf q})$ using Lagrangian perturbation theory, also known as the Zel'dovich approximation \citep{Zeldovich1970}, reads
\begin{equation}
    \bm{\Psi}({\bf q}) = \int \frac{d^3k}{(2\pi)^3}\frac{i{\bf k}}{k^2}\delta({\bf k})e^{i{\bf k} \cdot {\bf q}} \,.
\end{equation}
In addition, the displacement field can be related to the observed redshift-space galaxy field $\delta_{\text{g}}$ via the linear continuity equation \citep{Nusser1994, Padmanabhan2012},
\begin{align} \label{eq:continuity_psi}
\nabla \cdot \bm{\Psi} + \frac{f}{b} \nabla \cdot \left[(\bm{\Psi} \cdot \rhat) \rhat\right] = -\frac{\delta_{\text{g}}}{b},
\end{align}
where $f$ is the linear structure growth rate, $\rhat$ is the line-of-sight direction, and $b$ is the linear galaxy bias. On linear scales, we can assume $\bm{\Psi}$ is an irrotational field. Then, we can write the displacement as the gradient of a potential, $\bm{\Psi} = - \nabla \phi$, and rewrite \cref{eq:continuity_psi} as
\begin{equation} \label{eq:continuity_phi}
    \nabla^2 \phi + \frac{f}{b} \nabla \cdot (\nabla \phi_r) \rhat = \frac{\delta_{\rm g}}{b}
\end{equation}

The solution to \cref{eq:continuity_phi} can be solved in configuration space using finite-difference approximations \cite{Padmanabhan2012} or \textit{multigrid} relaxation techniques \cite{White2015:1504.03677}. Alternatively, \cref{eq:continuity_psi} can be solved in Fourier space, taking advantage of the computational efficiency of Fast Fourier Transforms (FFTs) \cite{Burden2014:1408.1348, Burden2015:1504.02591}. 

\begin{table}
    \centering
    \begin{tabular}{|l|c|c|c|c|}
    \hline
    Tracer & Redshift range & Linear bias $b$ & Growth rate $f$ & Smoothing scale $\Sigmasm$\\
    \hline
    \bgs & 0.1--0.4 & 1.5 & 0.682 & $15 \Mpch$\\
    \lrgs & 0.4--1.1 & 2.0 & 0.834 & $15 \Mpch$ \\
    \elgs & 0.8--1.6 & 1.2 & 0.900 & $15 \Mpch$ \\
    \qso & 0.8--2.1 & 2.1 & 0.928 & $30 \Mpch$ \\
    \hline
    \end{tabular}
    \caption{Fiducial redshift range, tracer linear bias, growth rate of structure, and smoothing scale assumed when reconstructing each DESI target sample. The smoothing scale was chosen after testing different values for each tracer. The growth rate of structure is determined from our fiducial cosmology and the effective redshift of each sample.}
    \label{tab:reconstruction_params}
\end{table}

Although the methods mentioned above use different numerical techniques to solve \cref{eq:continuity_psi,eq:continuity_phi}, their base underlying algorithm can be summarized in the following steps \citep{White2015:1504.03677}:
\begin{enumerate}
    \item Smooth the galaxy overdensity field using a kernel $\mathcal{S}$ to filter out high-k modes that are not well described by linear theory.
    \item Assuming a value for the structure growth rate $f$ and the linear galaxy bias $b$, estimate the displacement field $\bm{\Psi}$ by solving \cref{eq:continuity_psi}. The line-of-sight component of the displacement is multiplied by $1 + f$ to account for RSD.
    \item Move galaxy positions by $-\bm{\Psi}$ to obtain the displaced density field $\delta_d$.
    \item Move an initially unclustered catalogue of random particles by $-\bm{\Psi}$ to obtain the `shifted' field $\delta_s$. In this step, one can choose to include or ignore the $1 + f$ factor in the displacement along the line of sight, which results in two different reconstruction conventions. The \recsym\ convention applies the $1 + f$ factor to both galaxies and randoms, preserving linear RSD in the reconstructed field. The \reciso\ convention does not include the $1 + f$ factor when displacing randoms, and therefore results in a more isotropically-clustered sample after reconstruction.
\end{enumerate}

The reconstructed density field can then be estimated as $\delta_r \equiv \delta_d - \delta_s$, with a power spectrum $P_r \propto \langle | \delta^2_r | \rangle$. In configuration space, the post-reconstruction correlation function can be calculated using a modified version of the Landy-Szalay estimator \cite{Landy1993}, replacing the DR and RR terms in the numerator by DS and SS, where D and S denote the displaced data catalog and the shifted random catalogue, respectively.

It should be noted that other improved reconstruction algorithms have been proposed in the literature \cite[e.g., ][]{Seo2010:0910.5005, Schmittfull2017:1704.06634, Hada2018:1804.04738}, which follow different recipes. Although these methods hold significant promise for reconstructing the linear small-scale density field in the very low shot noise regime, the expected improvements in the BAO distance measurements are marginal at the galaxy number densities of DESI DR1.

For this work and the main DESI BAO analysis presented in \cite{DESI2024.III.KP4}, we adopt the iterative reconstruction in Fourier space proposed in \cite{Burden2014:1408.1348} as our default reconstruction algorithm. In \citep{KP4s3-Chen}, we show that this implementation gives consistent results with the \multigrid\ algorithm, but with a reduced computational cost. Below, we present a brief description of the algorithm, and refer the reader to \cite{KP4s3-Chen} for a detailed comparison with reconstruction algorithms in the context of BAO analyses.

\subsubsection{Reconstruction based on iterative Fast Fourier Transforms}

We start by decomposing $(\bm{\Psi} \cdot \rhat) \rhat$ into the gradient of a scalar potential field $A$ and the curl of a vector field ${\bf B}$ using Helmholtz's Theorem:
\begin{equation} \label{eq:helmholtz_theorem}
    (\bm{\Psi} \cdot \rhat) \rhat = \nabla A + \nabla \times {\bf B}\,.
\end{equation}
Substituting this in \cref{eq:continuity_psi}, we obtain
\begin{equation}
    \nabla(\phi + fA) = -\nabla \nabla^{-2} \frac{\delta_{\rm g}}{b}\,,
\end{equation}
where we have assumed that $\bm{\Psi}$ can be written as the gradient of a potential, as in \cref{eq:continuity_phi}. The formula above is exact, but cannot be solved directly using FFTs. The difficulty arises from the second term on the right-hand side of \cref{eq:helmholtz_theorem}. If we ignore this term by assuming $(\bm{\Psi} \cdot \rhat) \rhat$ is irrotational,
\begin{equation}
    (\bm{\Psi} \cdot \rhat) \rhat \approx \nabla A \,,
\end{equation}
then we can write
\begin{equation}
    \bm{\Psi} + \frac{f}{b}(\bm{\Psi} \cdot \rhat) \rhat = -\nabla \nabla^{-2} \frac{\delta_{\rm g}}{b} \,.
\end{equation}
The right-hand side of this equation can be computed using FFTs, with a solution given by
\begin{equation} \label{eq:ifft_start}
    \bm{\Psi} = \ifft \left[ - \frac{i {\bf k} \delta ({\bf k})}{k^2b} \right] - \frac{\beta}{1 + \beta} \left\{\ifft \left[ \frac{i {\bf k} \delta ({\bf k})}{k^2b} \right] \cdot \rhat  \right\} \rhat \,
\end{equation}
where \ifft\ denotes the inverse Fourier Transform, and $\beta = f/b$. 

In \cite{Burden2015:1504.02591}, it has been shown that assuming $(\bm{\Psi} \cdot \rhat) \rhat$ is irrotational overcorrects for RSD in the reconstructed field, which is not desirable. However, they propose to start from this approximation and then use an iterative scheme to recover the real-space displacement field, provisional on knowing the value of the growth rate of structure $f$.

If we start from \cref{eq:ifft_start} and completely ignore the RSD component, we can write the displacement as
\begin{equation}
    \bm{\Psi} = \ifft \left[ - \frac{i {\bf k} \delta_{\rm g, rsd} ({\bf k})}{k^2b} \right]\,
\end{equation}
or, in terms of the potential, as
\begin{equation}
    \phi_{\rm est, 1} = \ifft \left[ - \frac{\delta_{\rm g, rsd} ({\bf k})}{k^2b} \right]\,.
\end{equation}
Here, $\delta_{\rm g, rsd}$ is the observed redshift-space galaxy overdensity field. We can use this first estimate of the potential to estimate the real-space overdensity from the redshift-space one,
\begin{equation} \label{eq:deltareal_from_deltarsd}
    \frac{\delta_{\rm g, real, 1}}{b} = \frac{\delta_{\rm g, rsd}}{b} + \frac{f}{b} \nabla \cdot (\nabla \phi_{\rm est, 1} \cdot \rhat )\rhat
\end{equation}
In the next iteration, we build upon this estimate of the real-space overdensity to re-calculate the potential,
\begin{equation}
    \phi_{\rm est, 2} = \ifft \left[ - \frac{\delta_{\rm real, 1} ({\bf k})}{k^2b} \right]\,.
\end{equation}
and the operation continues. Once the algorithm has converged, we can estimate the displacement as
\begin{equation}
    \Psi_{{\rm FFT}, n} = \nabla \nabla^{-2}\left(\frac{\delta_{{\rm g, real}, n}}{b} \right)\,,
\end{equation}
where $\delta_{{\rm g, real}, n}$ is the estimation of the real-space galaxy overdensity field on the $n$-th iteration.

The description above constitutes the \ifft\ algorithm in its basic form. However, to speed up convergence, different starting points can be considered for the first iteration. In \pyrecon, during the first iteration, the second term on the right-hand side of \cref{eq:deltareal_from_deltarsd} takes an additional factor $1/(1 + \beta)$, motivated by \cref{eq:ifft_start}.

We have previously run convergence tests to assess the number of iterations required to obtain an optimal reconstruction of the DESI target samples. Based on results that are presented in \cite{KP4s3-Chen}, we choose $n = 3$ as the default number of iterations for the \ifft\ algorithm throughout the rest of this work.

It should be noted that this method differs from the one adopted in the eBOSS clustering analysis. The eBOSS method \citep{Bautista2021, eboss2020} iteratively updates the galaxy and random particle positions, while the {\tt IFFT} algorithm we adopt here updates the overdensity on the rectangular meshes, matching the algorithm proposed in \cite{Burden2014:1408.1348, Burden2015:1504.02591}. This is the first time that this algorithm has been implemented in a BAO data analysis.

\subsubsection{The \pyrecon\ reconstruction toolkit}
\label{subsubsec:pyrecon}

We reconstruct the DESI DR1 samples using \pyrecon\footnote{\url{https://github.com/cosmodesi/pyrecon}}, a comprehensive reconstruction toolkit developed by the DESI collaboration, which implements multiple reconstruction algorithms, accommodates various conventions, and provides the flexibility to process periodic-box simulations or survey data with non-uniform geometries.
\pyrecon\ begins by drawing two identical rectangular grids with a given resolution (which we set to $4 \Mpch$ for this work\footnote{We have performed convergence tests to choose an optimal value for the grid resolution, given our choice of smoothing scales and the computational restrictions due to the increase in memory allocation as the grid resolution increases. We also add some padding to the grid borders so that the size of the grid is 20\% larger on each side than what would be strictly necessary to perfectly contain the catalogs, which avoids introducing spurious correlations on the scales of interest when performing FFTs, which inherently assume periodicity for the grid operations.}) that encompasses the survey volume. The galaxy and random particles are then painted on each grid using triangular-shaped cloud interpolation \cite{Sefusatti2016:1512.07295}. The galaxy and random fields are smoothed in Fourier space using a Gaussian kernel of width $\Sigmasm$,
\begin{equation}
   \mathcal{S} = e^{-(k \Sigmasm)^2/2} \,,
\end{equation}
after which they are converted back to configuration space (we test different $\Sigmasm$ values for each tracer, and determine the optimal smoothing scale in \cref{subsec:smoothing_scale}). The overdensity field is then estimated by taking the ratio of the galaxy and random grids. Grid cells that contain fewer random particles than a given threshold are assigned a density contrast of zero. We set this threshold to 75\% of the mean number of random particles per cell. This recipe is common for all reconstruction algorithms implemented in \pyrecon, after which the displacement field depends on the specifics of each algorithm (in our case, the \ifft\ algorithm).

Throughout this work, we adopt some fiducial values for the linear galaxy bias and the growth rate of structure when running reconstruction, which are listed in \cref{tab:reconstruction_params}. The linear galaxy bias parameters have been determined by \cite{KP4s10-Mena-Fernandez,KP4s11-Garcia-Quintero}, and correspond to 1.5 for \bgs, 2.0 for \lrgs, 1.2 for \elgs, and 2.1 for \qso. We calculate the linear growth rate of structure from {\tt cosmoprimo}\footnote{\url{https://github.com/cosmodesi/cosmoprimo}} as $f = \sigma_8^{vv}(z_{\rm eff}) / \sigma_8^{dd}(z_{\rm eff})$, where $\sigma_8^{vv}$ and $\sigma_8^{vv}$ are the root mean square of velocity and density perturbations in spheres of $8 \Mpch$, respectively \citep{Song2009:0807.0810}. This is calculated using our fiducial cosmology model, at the effective redshift of each target sample, which is 0.296 for \bgs, 0.780 for \lrgs, 1.194 for \elgs, and 1.495 for \qso\ \citep{DESI2024.II.KP3}. The impact on reconstruction due to changes in fiducial cosmology has been studied in \cite{Sherwin2019:1808.04384, Carter2020:1906.03035}, and a general study of fiducial cosmology effects on the DESI BAO analysis is presented in detail in \cite{KP4s9-Perez-Fernandez}. In addition, \cite{KP4s2-Chen} present theoretical calculations showing the expected (small) impact of assuming an incorrect galaxy bias or growth rate on the BAO constraints.

\subsection{Clustering measurements}

We study the clustering properties of the DESI DR1 samples by means of anisotropic two-point statistics in configuration and Fourier space. In particular, we focus on the multipole moments of the anisotropic galaxy correlation function and the power spectrum of each sample. We calculate the correlation function multipoles with {\tt pycorr}\footnote{\url{https://github.com/cosmodesi/pycorr}}, which is a wrapper around a modified version of the {\tt corrfunc} pair-counting code \cite{corrfunc}. Before its decomposition into multipoles, the correlation function is binned in $s$ and $\mu$, where $s$ is the redshift-space scalar separation between the pair and $\mu$ is the cosine of the angle between the vector connecting the pair and the observer's line of sight. We use linear bins of width $4 \Mpch$ in $s$ from $50 \Mpch$ to $150 \Mpch$. Similarly, the power spectrum multipoles are calculated with {\tt pypower}\footnote{\url{https://github.com/cosmodesi/pypower.}}, where the power spectrum estimator takes reference from \citep{Hand2017}. We use $k$ bins of width $0.005 \hMpc$ between $0.02 \hMpc$ and $0.3 \hMpc$. These scales were identified as optimal for BAO fitting by \cite{KP4s2-Chen}. More details of the calculation of these clustering statistics, including how galaxies are weighed to optimally measure the clustering signal, are presented in \cite{DESI2024.II.KP3}.

To estimate the errors associated with the correlation functions, we use semi-analytical semi-empirical covariance matrices generated with the {\tt RascalC} code \citep{Philcox2020:1904.11070, 2023MNRAS.524.3894R}, while the errors associated with the power spectra are based on {\tt CovaPT} \citep{Wadekar2020:1910.02914, KP4s8-Alves}. Refs. \cite{KP4s6-Forero-Sanchez, KP4s7-Rashkovetskyi, KP4s8-Alves} present a comprehensive description of the construction and validation of these covariance matrices for DESI DR1. 

\begin{table}
    \centering
        \begin{tabular}{|l|c|c|c|}
            \hline
           Parameter & $P(k)$ prior & $\xi(r)$ prior & Interpretation \\ \hline
           $\qiso$ & $\unif(0.8, 1.2)$  & $\unif(0.8, 1.2)$  & Isotropic BAO dilation \\
          $\qap^*$ & $\unif(0.8, 1.2)$ & $\unif(0.8, 1.2)$ &  Anisotropic (AP) BAO dilation \\
           $\Sigma{\perp}$ & $\mathcal{N}(\Sigma^{\mathrm{fid}}_{\perp},1.0)$ &$\mathcal{N}(\Sigma^{\mathrm{fid}}_{\perp}, 1.0)$ & Transverse BAO damping [$\Mpch$]\\
            $\Sigma{\parallel}$ & $\mathcal{N}(\Sigma^{\mathrm{fid}}_{\parallel},2.0)$ &$\mathcal{N}(\Sigma^{\mathrm{fid}}_{\parallel}, 2.0)$ & Line-of-sight BAO damping [$\Mpch$]\\
            $\Sigma_s$ & $\mathcal{N}(2.0,2.0)$ & $\mathcal{N}(2.0,2.0)$ & Fingers-of-God damping [$\Mpch$] \\
            $b_1$ & $\unif(0.2, 4)$ & $\unif(0.2, 4)$ & Linear galaxy bias \\
            $d\beta^*$ & $\unif(0.7, 1.3)$ & $\unif(0.7, 1.3)$ & Linear RSD parameter\\
            $a_{0, n}$ & $\mathcal{N}(0, 10^4)$ & N/A  & Spline parameters for the monopole\\        
            $a_{2,n}^*$ & $\mathcal{N}(0, 10^4)$ & $\mathcal{N}(0, 10^4)$ & Spline parameters for the quadrupole\\
            $b_{0,n}$ & N/A & $\unif(-\infty, \infty)$  & Unknown large scale systematics\\
            $b_{2,n}^*$ & N/A & $\unif(-\infty, \infty)$  & Unknown large scale systematics\\
        \hline
        \end{tabular}
    \caption{Priors on the parameters that are varied during fits to the power spectrum (second column) and the correlation function (third column). $\unif$ and $\gauss$ denote uniform and Gaussian distributions, respectively. The parameters marked with a star are fixed to their default values for isotropic BAO fits (\bgs, \elgo\, and \qso), where the default values are $d\beta = 1$, $a_{2,n} = 0$, $b_{2,n} = 0$. The mean values of the priors on the damping parameters for each tracer are shown in \cref{tab:damping_parameters}.}
    \label{tab:priors}
\end{table}

\section{Modeling}
\label{sec:modeling}

\subsection{DESI mocks}
\label{subsec:mock_catalogues}

To test the effects of reconstruction in a controlled setting, we use two collections of mock galaxy catalogs that were calibrated to match the clustering and selection properties of the DESI DR1 target samples, as described in \cite{KP3s8-Zhao}: full N-body mocks based on the \textsc{AbacusSummit} suite of simulations \citep{Maksimova2021}, and approximate mocks based on the EZmock algorithm \cite{Chuang2015:1409.1124}. All mocks were constructed assuming a Planck 2018 base-$\Lambda$CDM cosmology, more specifically using the mean of the marginalized parameter distributions from the Planck TT,TE,EE+lowE+lensing likelihood \citep{Planck2020}: $\omega_{\rm cdm} = 0.1200$, $\omega_{\rm b} = 0.02237$, $h = 0.6736$, and $N_{\rm ur} = 2.0328$.

The N-body mocks start from halo catalogs from the $2 \Gpch$ {\tt AbacusSummit} simulation boxes \citep{Maksimova2021}, at snapshots within the redshift range of each DESI sample. Dark matter haloes are populated with galaxies using the Halo Occupation Distribution (HOD) model, with parameters that are calibrated to match the number density and clustering of each galaxy sample (an assessment of the impact of the HOD modelling on the DESI error budget is presented in \cite{KP4s10-Mena-Fernandez, KP4s11-Garcia-Quintero}). Mocks with cutsky geometry are then constructed by replicating and patching the $2 \Gpch$ boxes from different snapshots, converting them to sky coordinates and trimming them to match the DESI footprint and radial selection of each sample. For \lrgs, we use the $z = 0.5$ snapshot to construct the $z < 0.6$ portion of the cutsky, and the $z = 0.8$ snapshot to cover the $z > 0.6$ portion. For \elgs, snapshots at $z = 0.950$ and $z = 1.325$ are used to cover  $z < 1.1$ and $z > 1.1$, respetively. The \qso\ cutsky reads data from the $z = 1.400$ snapshot for all redshifts. We note that these cutsky mocks differ from the full lightcones described in \cite{Hadzhiyska2022:2110.11413}.

The approximate mocks consist of a series of 1000 galaxy catalogs generated with the {\tt EZmock} algorithm \cite{Chuang2015:1409.1124} with the same volume, number density, and HOD as the \textsc{AbacusSummit} simulations. The \ezmocks\ are based on the Zel'dovich approximation, including prescriptions for the missing physical ingredients, such as stochastic scale-dependent, non-local and non-linear biasing contributions.

Since DESI measures thousands of galaxy redshifts simultaneously using fiber-fed robotic positioners, the galaxy clustering can be artificially decreased at small scales due to the fiber assignment scheme, which is sometimes unable to simultaneously target galaxy pairs that are positioned too close on the sky. To ensure that both the \abacussummit\ and \ezmocks\ reproduce this systematic error on the clustering measurements, they are passed through a pipeline that mimics the fiber assignment scheme of the DESI DR1 samples \citep{DESI2024.II.KP3}.

\subsection{Modelling the BAO signal}
\label{sec:bao_model}

An extensive discussion on the modeling of the pre- and post-reconstruction BAO signal is presented in \cite{KP4s2-Chen}. Here, we present a brief description of the main constituents of the model in configuration and Fourier space.

\subsubsection{Fourier space}

Our model for the observed galaxy power spectrum is based on the work of \cite{KP4s2-Chen} and can be written generically as
\begin{equation} \label{eq:generic_model}
    P(k, \mu) = \mathcal{B}(k, \mu) P_{\rm nw}(k) + \mathcal{C}(k, \mu)P_{\rm w}(k) + \mathcal{D}(k)\,,
\end{equation}
where $P_{\rm nw}(k)$ and $P_{\rm w}(k)$ denote the smooth (no-wiggle) and BAO (wiggle) components of the linear power spectrum, respectively, which are obtained following the \textit{peak average} method from \cite{Brieden2022:2204.11868}. The linear matter power spectrum template is predicted from {\tt CLASS}\footnote{\url{https://github.com/lesgourg/class_public}}\cite{Blas2011:1104.2933} using our fiducial cosmology.

Following \cite{Seo2016, Beutler2017}, we adopt the following parametric form for $\mathcal{B}(k, \mu)$: 
\begin{equation}
    \mathcal{B}(k,\mu) = \left(b+f\mu^{2}(1 - s(k)\right)^{2} F_{\rm fog}\,,
\end{equation}
where $F_{\rm fog} = \left(1 + \frac{1}{2} k^2\mu^{2} \Sigma_s^2\right)^{-2}$ accounts for the `Fingers of God' effect due to halo virialization \citep{Park1994}. For pre-reconstruction and the \recsym\ convention, $s(k) = 1$, while for \reciso, $s(k) = \exp\left[-(k \Sigma_{\rm sm})^2/2 \right]$, where $\Sigma_{\rm sm}$ is the reconstruction smoothing scale.

$\mathcal{C}(k, \mu)$ captures the anisotropic non-linear damping on the BAO feature,
\begin{equation}
    \mathcal{C}(k,\mu) = \left(b+f\mu^{2}\right)^{2}\exp\left[-\frac{1}{2}k^2\biggl(\mu^{2}\Sigma_{||}^2 + (1-\mu^{2})\Sigma^{2}_{\perp}\biggl)\right]
\end{equation}
where $\Sigma_{||}$ and $\Sigma_{||}$ model the damping for modes along and perpendicular to the line of sight.

The $\mathcal{D}(k)$ factor captures any deviation from linear theory in the broadband shape of the power spectrum multipoles. Motivated by \cite{KP4s2-Chen}, we parameterize it separately for each clustering multipole using a spline basis with bases separated by $\Delta$,
\begin{equation} \label{eq:spline_basis}
    \mathcal{D}_\ell(k > k_{\rm min}) = \sum_{n=-1}^7 a_n W_3\left(\frac{k}{\Delta} - n\right)\,,
\end{equation}
where $W_3$ is a piecewise cubic spline kernel \citep{Chaniotis2004, Sefusatti2016:1512.07295}. We set $\Delta = 0.06 \hMpc$, which is chosen such that the spline basis is able to match the broadband shape of the power spectrum, without reproducing the BAO wiggles themselves.

If the fiducial cosmology used to convert redshifts to distances differs from the true cosmology of the Universe, it can induce geometric distortions in the galaxy distribution. Moreover, the sound horizon in the template cosmology could also differ from the true sound horizon. To account for both effects, we parameterize the position of the BAO wiggles via the dilation parameters $\alpha_\perp$ and $\alpha_\parallel$, which scale the separations in the directions across and along the line of sight:
\begin{align}
    k^{\rm fid} &= \frac{k}{\alpha_\perp}\left[1 + \mu^2 \left(\frac{1}{F^2} - 1 \right) \right]^{1/2} \\
    \mu^{\rm fid} &= \frac{\mu}{F} \left[1 + \mu^2 \left(\frac{1}{F^2} - 1 \right) \right]^{-1/2}\,,
\end{align}
where $F = \alpha_{\parallel}/\alpha_{\perp}$. Here, the $^{\rm fid}$ superscript denote quantities in the fiducial cosmology. The dilation parameters can be interepreted in terms of cosmology as 
\begin{align}
    \alpha_\parallel &= \frac{H^{\rm fid}(z)}{H(z)}\frac{r_s^{\rm fid}}{r_s(z_d)} \\
    \alpha_\perp &= \frac{D_{\rm M}(z)}{D_{\rm M}^{\rm fid}(z)}\frac{r_s^{\rm fid}(z_d)}{r_s(z_d)} \,,
\end{align}
where $D_{\rm M}(z)$ is the transverse comoving distance to redshift $z$, $H(z)$ is the Hubble parameter, and $r_s(z_d)$ is the sound horizon at the baryon drag epoch $z_d$. From here, we can define the following quantities:
\begin{align}
    \alpha_{\text{iso}} &\equiv (\alpha_\perp^2 \alpha_\parallel)^{\frac{1}{3}},\label{eq:qiso} \\
    \alpha_{\text{AP}} &\equiv \frac{\alpha_{\parallel}}{\alpha_{\perp}}.\label{eq:qap}
\end{align}
which encode the isotropic and anisotropic BAO dilation, respectively.

\subsubsection{Configuration space}

Our starting point is the anisotropic power spectrum obtained from \cref{eq:generic_model} without the $\mathcal{D}(k)$ term, which are Hankel-transformed to configuration space as
\begin{equation}
    \xi_l(s) = \frac{i^\ell}{2\pi^2} \int_0^\infty k^2 j_\ell (ks) P_\ell (k) \mathrm{d}k \,,
\end{equation}
where $j_\ell$ are the spherical Bessel functions.

To parameterize the broadband part of the correlation function, we take the Hankel transform of the $n=[0,1]$ spline bases from \cref{eq:spline_basis} for the quadrupole, and we additionally include even-power polynomial terms for both monopole and quadrupole:
\begin{equation}
    \mathcal{\tilde{D}}_\ell(s > s_{\rm min}) = \sum_{n = 0}^{1} b_n \left( \frac{r}{\Delta_s}\right)^{2n}\,,
\end{equation}
where $\Delta_s$ = $k_{\rm min} / 2\pi$.

\begin{table*}
    \renewcommand{\arraystretch}{1.2}
    \centering
    \begin{tabular}{| c | c | c| c | c| c | c |}
        \hline
        Parameter & Recon& \bgs & \lrgs & \elgs  & \qso\\
        \hline
        $\sigmaper^{\rm fid}\, [\Mpch]$  & Pre & $6.5$ & $4.5$ & $4.5$ & $3.5$\\
        $\sigmapar^{\rm fid}\, [\Mpch]$ & Pre & $10.0$ & $9.0$ & $8.5$ & $9.0$\\
        $\sigmas^{\rm fid}\, [\Mpch]$ & Pre & $2.0$ & $2.0$ & $2.0$ & $2.0$\\
        \hline
        $\sigmaper^{\rm fid}\, [\Mpch]$ & Post & $3.0$ & $3.0$ & $3.0$ & $3.0$\\
        $\sigmapar^{\rm fid}\, [\Mpch]$ & Post & $8.0$ & $6.0$ & $6.0$ & $6.0$\\
        $\sigmas^{\rm fid}\, [\Mpch]$ & Post & $2.0$ & $2.0$ & $2.0$ & $2.0$\\
        \hline
    \end{tabular}
    \caption{Mean values of the Gaussian priors (\cref{tab:priors}) for the non-linear BAO damping parameters across and along the line of sight ($\sigmaper$ and $\sigmapar$, respectively) and the Fingers-of-God damping ($\sigmas$), depending on whether they are used in pre- or post-reconstruction BAO fits of each tracer.}
    \label{tab:damping_parameters}
\end{table*}

\subsection{Model fitting}
\label{sec:model_fitting}

We perform the BAO fitting using {\tt desilike},\footnote{\url{https://github.com/cosmodesi/desilike}} which is a {\tt python} package that provides a common framework for writing DESI likelihoods. We sample posterior distributions using the Markov Chain Monte Carlo (MCMC) code {\tt emcee} \citep{emcee:1202.3665}, adopting the priors listed in \cref{tab:priors}. Use flat priors in all cases except for the damping parameters, for which we adopt Gaussian priors centred on the fiducial values listed in \cref{tab:damping_parameters}. These Gaussian priors have been informed by running fits to data vectors averaged over many realizations of the DESI mocks, ensuring that the signal-to-noise ratio (SNR) is large enough to let the damping parameters vary freely during the fit. In later sections, we show that the recovered BAO parameters from fits to DESI data are largely insensitive to this choice of priors. In addition to the spline broadband parameters described in the previous section, we also allow for a linear RSD parameter scaling $d\beta \equiv f/f^{\rm fid}$, where $f^{\rm fid}$ is the fiducial growth rate of structure, and a bias parameter $b_1$ that regulates the clustering amplitude.

Apart from MCMC posterior sampling, we perform posterior profiling using {\tt minuit} \citep{minuit}, which is incorporated in {\tt desilike}. In the following, we focus mainly on the BAO scaling parameters $\alpha_{\text{iso}}$ and $\alpha_{\text{AP}}$ while marginalizing over all other nuisance parameters.

Given the SNR of the clustering measurements of the different DESI samples, only certain tracers and redshift bins allow two-dimensional BAO fits where $\qiso$ and $\qap$ are varied simultaneously. These are the \lrgs, and \elgt. For \bgs, \qso, and \elgo, we fit only the monopole and vary $\qiso$ during the fit. However, in all cases we model the full set of clustering multipoles to account for their potential impact (via convolution with the survey geometry window function) on the clustering monopole.

\section{Results}
\label{sec:results}

\subsection{A qualitative exploration of reconstruction}

\begin{figure*}
    \centering
    \begin{tabular}{c}
      \hspace{-0.3cm}\includegraphics[width=\textwidth]{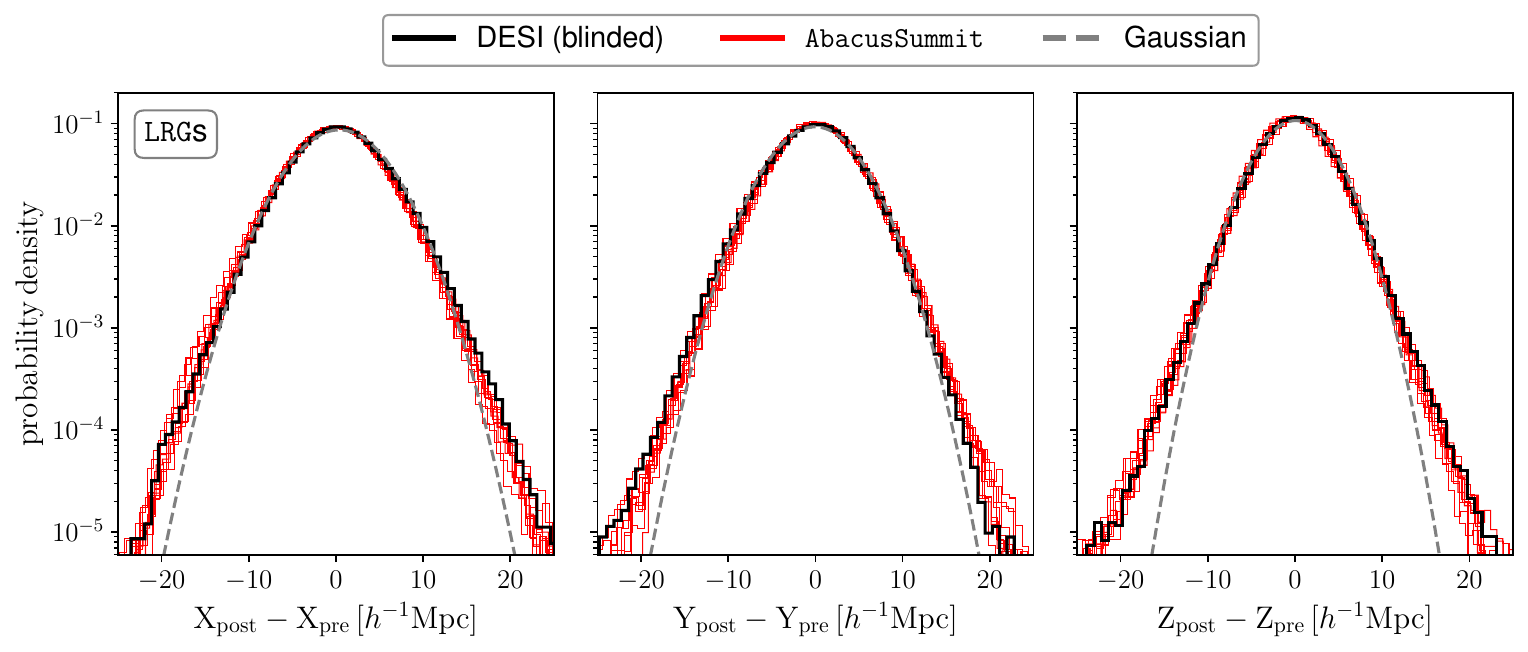}\\
    \end{tabular}
    \caption{Distribution of the displacements vectors from the \lrgs\ reconstruction, projected along the three Cartesian axes of the reconstruction grid. The black line shows the blinded DESI DR1 data, and the thinner blue lines show 10 different realizations of the \abacussummit\ mocks. We compare against a Gaussian with the same mean and dispersion as the DESI data, shown by the grey-dashed line.}
    \label{fig:displacement}
\end{figure*}

We begin by exploring the distribution of displacement vectors resulting from reconstructing the \bgs\ sample from the blinded DESI DR1 data, shown by the thick black line in \cref{fig:displacement}. In this case, we have used a reconstruction smoothing scale of $15\,\Mpch$, and the fiducial values for the linear galaxy bias and the growth rate of structure listed in \cref{subsec:reconstruction}. Since the galaxy overdensity field is estimated on a rectangular grid that encapsulates the survey volume (see \cref{subsubsec:pyrecon}), \cref{fig:displacement} shows the displacement vectors along the three axes of the grid. They are roughly symmetric around zero in each direction and approximately follow a Gaussian distribution. Strong non-linear motions cause deviations from Gaussianity for large displacement values. The \abacussummit\ mocks, of which we show ten independent realizations in the same figure, are largely consistent with the DESI data. More quantitatively, the DESI displacements are centred at $(0.39, -0.08, 0.07) \Mpch$, with a dispersion of $(4.60, 4.29, 3.73) \Mpch$. The mean of 25 mocks yields a mean of $(0.03, 0.05, 0.01) \Mpch$, with a dispersion of $(4.53, 4.19, 3.68) \Mpch$. We have explicitly checked that a similar level of agreement is found for the other tracers, but we do not include the figures here for conciseness. Although for BAO analyses the displacement field is used as a means of recovering the linear density field and sharpening the acoustic feature, the reconstructed velocity field itself is also of great importance for other analyses that focus on the imprints of galaxy velocities on the cosmic microwave background, such as the kinematic Sunyaev-Zel’dovich effect \cite{Hadzhiyska2023:2312.12434, Ried=Guachalla2023:2312.12435}.

\begin{figure*}
    \centering
    \begin{tabular}{cc}
      \includegraphics[width=0.44\columnwidth]{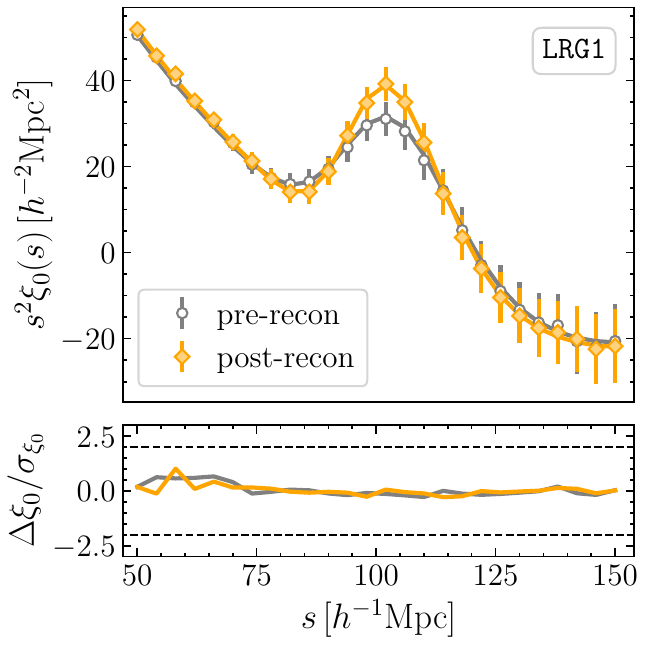} &
      \includegraphics[width=0.48\columnwidth]{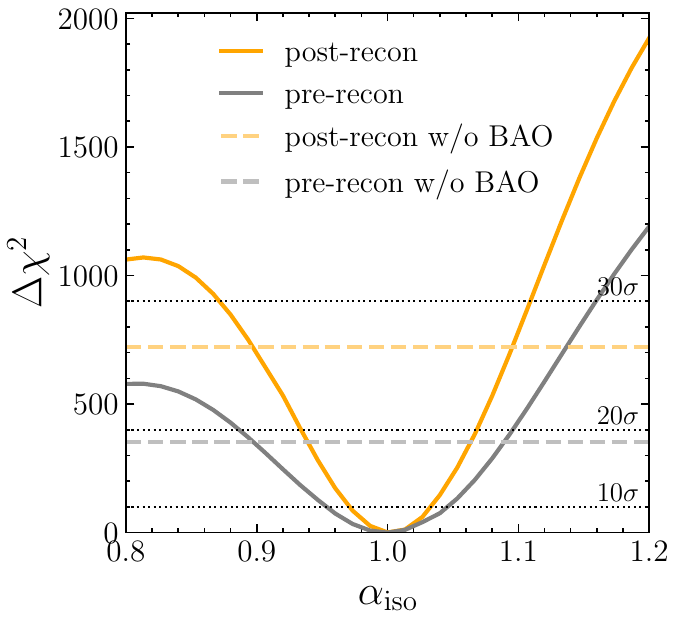}
    \end{tabular}
    \caption{Left: The open and filled markers show the monopole of the galaxy two-point correlation function before and after reconstruction, respectively, averaged over 25 realizations of the \abacussummit\ mocks for LRGs at $0.4 < z < 0.6$. The error bars, which represent the error on the mean, are calculated from a semi-analytic/semi-empirical covariance matrix. The solid lines correspond to the best-fit model. The bottom subpanel shows the relative error between the measurements and the best-fit models. Right: $\Delta \chi^2$ as a function of the BAO scaling parameter $\qiso$. Solid and dashed lines show results from models with and without BAO wiggles, respectively. The horizontal dashed lines show different thresholds of detection.}
    \label{fig:multipoles_and_detection}
\end{figure*}

\begin{figure}
    \centering
    \begin{tabular}{cc}
    \includegraphics[width=0.45\textwidth]{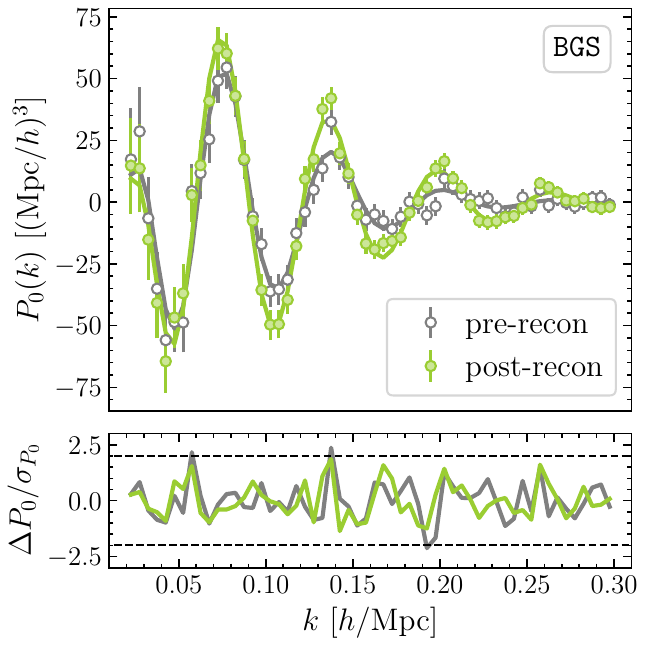} & \hspace{0.5cm}\includegraphics[width=0.4\textwidth]{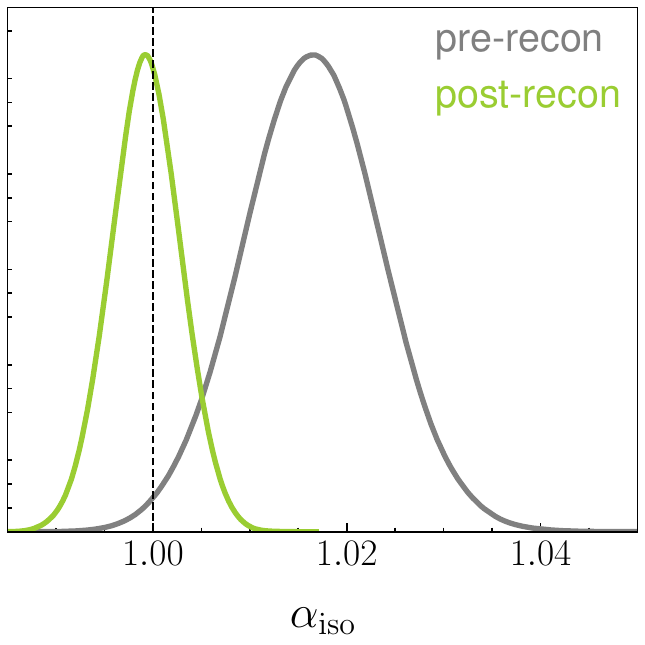}
    \end{tabular}
    \caption{Left: The circles show the broadband-subtracted monopole of the anisotropic power spectrum averaged over 25 realizations of the \abacussummit\ mocks, before and after reconstructing the density field. The solid lines show the best-fit BAO model. Right: Constraints on the isotropic BAO scaling parameter $\qiso$ with and without reconstruction.}
    \label{fig:pk_bgs}
\end{figure}

Having estimated the displacement field, we can use it to estimate the clustering from the reconstructed samples. \cref{fig:multipoles_and_detection} compares the monopole of the anisotropic correlation function for \lrgo\ before and after reconstruction. The pre-reconstruction correlation function shows a prominent BAO peak around $\sim 100 \Mpch$, which is smeared due to non-linear gravitational motions. Reconstruction helps to sharpen the BAO peak by partially restoring the linearity of the density field. The BAO model of \cref{sec:bao_model}, shown by the solid line, is an excellent fit to the data, giving a $\chi^2$ of 33.4 for 37 degrees of freedom after reconstruction. The correlation function is averaged over 25 mock realizations, and the covariance represents the error of the combined volume of all these mocks ($200\,\Mpchcube$). In the panel on the right side, we also display $\Delta \chi^2$ for the pre- and post-reconstruction cases using solid curves. The profiles are well centered around $\qiso = 1$, showing that the model recovers unbiased parameter constraints even for this large volume. The reconstructed case has a sharper distribution due to the better localized BAO bump in the correlation function, which leads to a better determination of the scaling parameter. We can contrast these distributions with the case where we fit a model without BAO wiggles to assess the significance of the BAO detection. The dashed curves show the $\Delta \chi^2$ between the best-fit BAO models and the models without BAO. From this we see that reconstruction increases the statistical significance of the BAO detection from approximately $18 \sigma$ to $27 \sigma$. Such a high detection level is only achieved, of course, since we are fitting the combined set of mocks for which the SNR of the data vector is exquisite.

Reconstruction not only sharpens the BAO feature, but can also help removing systematic shifts in the location of the peak due to non-linear evolution and galaxy biasing. This is exemplified in \cref{fig:pk_bgs}, where we compare the pre- and post-reconstruction power spectrum of the \bgs\ sample from the \abacussummit\ mocks. To better see the BAO wiggles, we have subtracted the broadband component of the power spectrum from both measurements. It can be noticed that the pre-reconstruction case not only exhibits damped wiggles, but also a slight shift in phase towards larger $k$ values. The BAO fit from the pre-reconstruction data, shown in the right-hand panel, is biased high with respect to the expectation of $\qiso = 1$ at a more than 2-$\sigma$ level. In addition to reducing the error on $\qiso$ by 50\%, reconstruction also effectively corrects the bias seen in the mean value.

\subsection{Impact of the smoothing scale}
\label{subsec:smoothing_scale}

\begin{figure}
    \centering
    \begin{tabular}{c}
     \hspace*{-0.3cm}\includegraphics[width=0.9\textwidth]{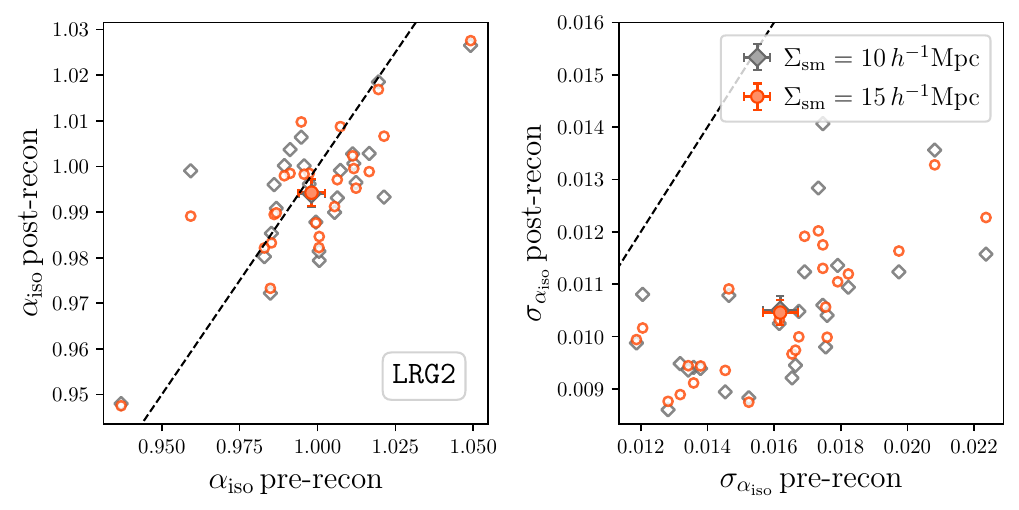}\\
     \hspace*{-0.1cm}\includegraphics[width=0.9\textwidth]{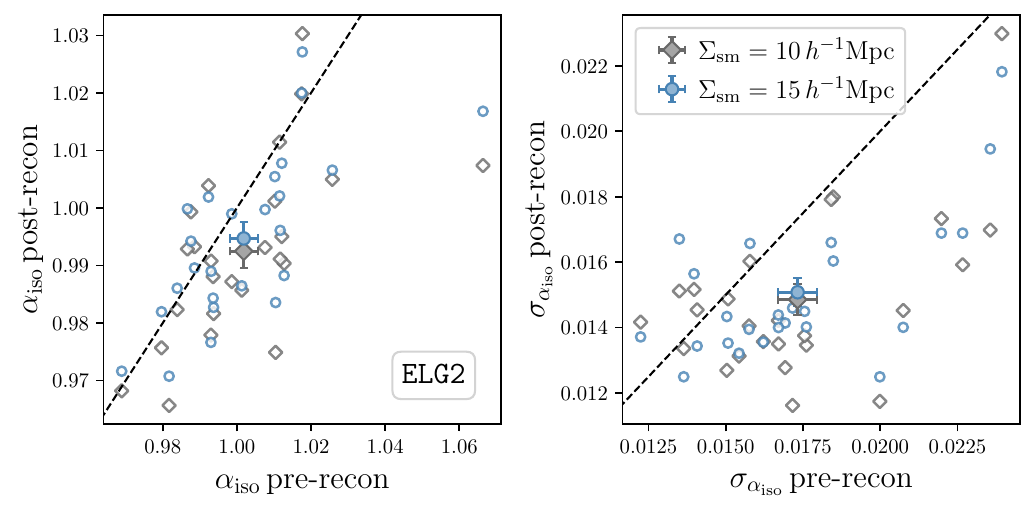}
    \end{tabular}
    \caption{Effect on the constraints on the BAO scaling parameters for \lrgt\ (top) and \elgt\ (bottom), reconstructing the density field using a smoothing scale of $10 \Mpch$ (diamonds) or $15 \Mpch$ (circles). Left: Best-fit values of the isotropic BAO scaling parameter $\qiso$ before and after reconstruction. Right: 1-$\sigma$ errors on $\qiso$ before and after reconstruction. The open markers are individual fits to independent realizations of the \abacussummit\ mocks. The filled markers show the mean of the 25 realizations, and the error bars show the standard deviation.}
    \label{fig:smoothing_scales_qiso}
\end{figure}

\begin{figure}
    \centering
    \begin{tabular}{c}
     \hspace*{-0.2cm}\includegraphics[width=0.9\textwidth]{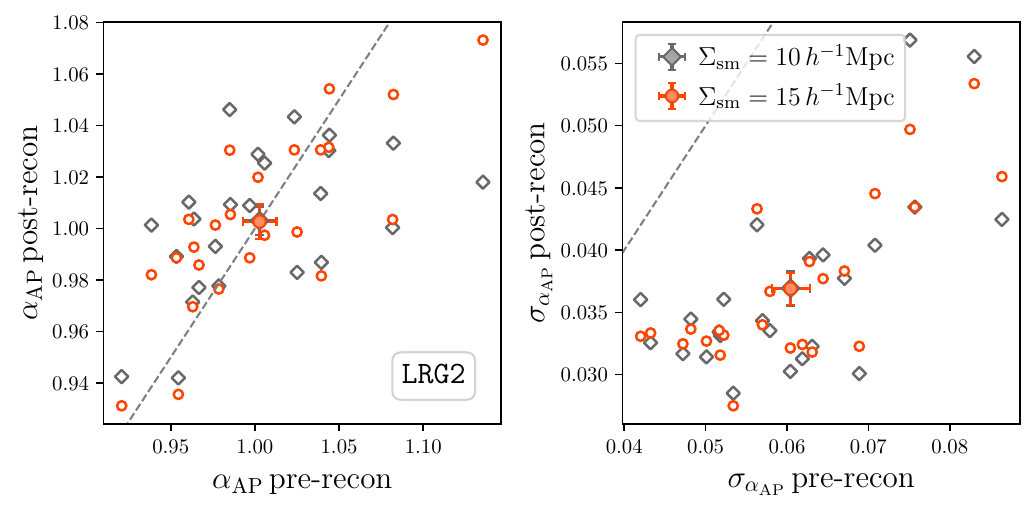}\\
     \hspace*{-0.3cm}\includegraphics[width=0.9\textwidth]{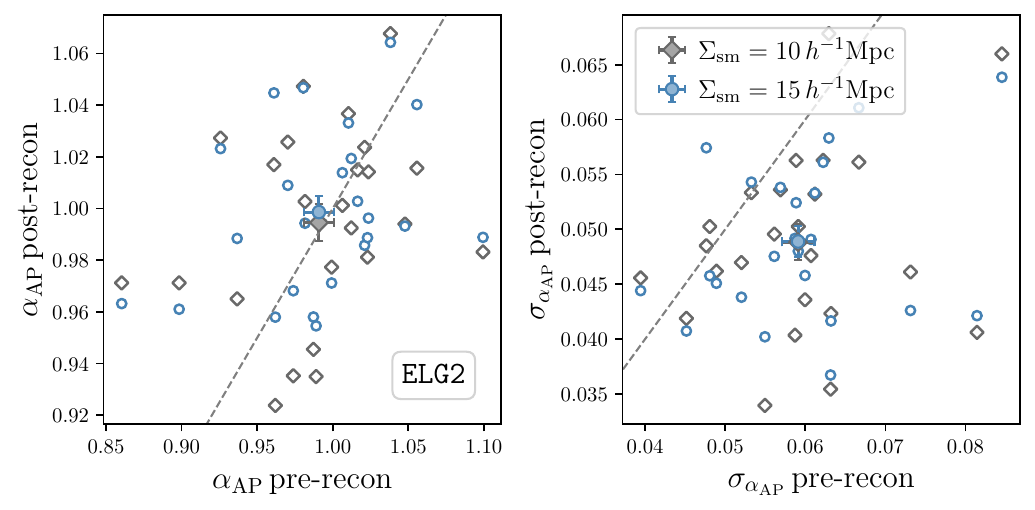}
    \end{tabular}
    \caption{Similar to \cref{fig:smoothing_scales_qiso}, but for the anisotropic BAO scaling parameter $\qap$.}
    \label{fig:smoothing_scales_qap}
\end{figure}

\begin{table}
    \centering
    \small
    \vspace{0.5em}
    \begin{tabular}{|l|c|c|c|c|r|r|}
    \hline
     Tracer       & Recon   & $\Sigma_{\rm sm}$    & $\alpha_{\rm iso}$   & $\alpha_{\rm AP}$   & $\Delta \alpha_{\rm iso}/\sigma$   & $\Delta \alpha_{\rm AP}/\sigma$   \\
    \hline
     ${\tt BGS}$  & Post    & 15 $h^{-1}{\rm Mpc}$ & $0.9943 \pm 0.0039$  & ---                 & $-1.46$                            & ---                               \\
     ${\tt LRG1}$ & Post    & 15 $h^{-1}{\rm Mpc}$ & $0.9980 \pm 0.0026$  & $0.9912 \pm 0.0090$ & $-0.77$                            & $-0.98$                           \\
     ${\tt LRG2}$ & Post    & 15 $h^{-1}{\rm Mpc}$ & $0.9957 \pm 0.0021$  & $1.0040 \pm 0.0074$ & $-2.05$                            & $0.55$                            \\
     ${\tt LRG3}$ & Post    & 15 $h^{-1}{\rm Mpc}$ & $0.9957 \pm 0.0018$  & $0.9928 \pm 0.0065$ & $-2.34$                            & $-1.11$                           \\
     ${\tt ELG1}$ & Post    & 15 $h^{-1}{\rm Mpc}$ & $0.9984 \pm 0.0041$  & ---                 & $-0.39$                            & ---                               \\
     ${\tt ELG2}$ & Post    & 15 $h^{-1}{\rm Mpc}$ & $0.9959 \pm 0.0029$  & $1.0003 \pm 0.0099$ & $-1.39$                            & $0.03$                            \\
     ${\tt QSO}$  & Post    & 30 $h^{-1}{\rm Mpc}$ & $0.9960 \pm 0.0034$  & ---                 & $-1.17$                            & ---                               \\
     \hline
    ${\tt BGS}$  & Post    & 10 $h^{-1}{\rm Mpc}$ & $0.9937 \pm 0.0038$  & ---                 & $-1.64$                            & ---                               \\
     ${\tt LRG1}$ & Post    & 10 $h^{-1}{\rm Mpc}$ & $0.9967 \pm 0.0026$  & $0.9932 \pm 0.0089$ & $-1.31$                            & $-0.77$                           \\
     ${\tt LRG2}$ & Post    & 10 $h^{-1}{\rm Mpc}$ & $0.9960 \pm 0.0021$  & $1.0033 \pm 0.0075$ & $-1.92$                            & $0.44$                            \\
     ${\tt LRG3}$ & Post    & 10 $h^{-1}{\rm Mpc}$ & $0.9957 \pm 0.0019$  & $0.9895 \pm 0.0065$ & $-2.28$                            & $-1.61$                           \\
     ${\tt ELG1}$ & Post    & 10 $h^{-1}{\rm Mpc}$ & $1.0030 \pm 0.0044$  & ---                 & $0.68$                             & ---                               \\
     ${\tt ELG2}$ & Post    & 10 $h^{-1}{\rm Mpc}$ & $0.9955 \pm 0.0030$  & $0.9978 \pm 0.0104$ & $-1.51$                            & $-0.21$                           \\
     ${\tt QSO}$  & Post    & 20 $h^{-1}{\rm Mpc}$ & $0.9930 \pm 0.0039$  & ---                 & $-1.82$                            & ---                               \\
     \hline
     ${\tt BGS}$  & Pre     & ---                  & $1.0077 \pm 0.0070$  & ---                 & $1.10$                             & ---                               \\
     ${\tt LRG1}$ & Pre     & ---                  & $0.9979 \pm 0.0044$  & $0.9726 \pm 0.0167$ & $-0.47$                            & $-1.64$                           \\
     ${\tt LRG2}$ & Pre     & ---                  & $1.0000 \pm 0.0034$  & $1.0042 \pm 0.0130$ & $-0.01$                            & $0.33$                            \\
     ${\tt LRG3}$ & Pre     & ---                  & $0.9996 \pm 0.0026$  & $1.0035 \pm 0.0091$ & $-0.16$                            & $0.38$                            \\
     ${\tt ELG1}$ & Pre     & ---                  & $1.0012 \pm 0.0051$  & ---                 & $0.23$                             & ---                               \\
     ${\tt ELG2}$ & Pre     & ---                  & $1.0015 \pm 0.0033$  & $0.9968 \pm 0.0111$ & $0.45$                             & $-0.29$                           \\
     ${\tt QSO}$  & Pre     & ---                  & $1.0053 \pm 0.0038$  & ---                 & $1.39$                             & ---                               \\
    \hline
\end{tabular}
\caption{Mean values and standard deviations from the marginalized posteriors of the BAO scaling parameters from a fit to the mean of 25 realizations of the \abacussummit\ mocks. We display pre-reconstruction results, as well as post-reconstruction fits obtained with two different choices of smoothing scale $\Sigmasm$. The last two columns show the offset in the mean values with respect to the expected value of $\qiso$ and $\qap$, in units of the corresponding standard deviation, i.e. $\Delta \alpha = (\alpha - 1)/\sigma$. We use a covariance representative of the DESI DR1 volume for each tracer.}
\label{tab:fits_abacussummit}
\end{table}

One of the key hyperparameters that must be tuned to optimally reconstruct the galaxy catalogs is the scale $\Sigmasm$ that is used to smooth the density field before estimating the displacement. Values of $\Sigmasm$ that are too large might degrade the information content by erasing clustering on linear scales, while values that are too small could increase the noise in the reconstruction process and potentially lead to biases due to the inclusion of non-linear modes that break the underlying assumptions in the modeling.

Previous works in the literature have performed careful studies of the impact of the smoothing scale on the reconstruction performance for SDSS galaxy samples \citep{White2010:1004.0250, Burden2014:1408.1348, Anderson2014:1303.4666, Vargas-Magaña:1509.06384}, showing that the optimal choice of smoothing scale has an important dependence on the tracer characteristics: low-density tracers are more affected by shot noise, so larger smoothing scales are usually preferred. The main BOSS and eBOSS clustering analysis adopted a smoothing scale of $15 \Mpch$ for their LRG and ELG samples, obtaining a consistent improvement in the BAO constraints from reconstruction \citep{Alam2017, eboss2020}. Here, we revisit these choices for the DESI DR1 sample and run BAO fits with different $\Sigmasm$ values. Given the various tracers and redshift bins to be explored, we restrict ourselves to only two smoothing scales per subsample, informed by a more extensive exploration of the parameter space presented in \cite{KP4s3-Chen}.

\Cref{fig:smoothing_scales_qiso} compares the constraints on the isotropic BAO scaling parameter, $\qiso$ before and after reconstruction for \lrgth\ and \elgt, setting $\Sigma_{\rm sm}$ to $10 \Mpch$ or $15  \Mpch$. These constraints represent the maximum-posterior values from individual fits to the \abacussummit\ mock realizations (empty circles and diamonds), with the average fit shown by the filled markers with error bars. Looking at the left panel, we see that for both smoothing scales, there is good agreement between the best-fit values before and after reconstruction. This need not necessarily be the case, since non-linear motions will tend to systematically shift the BAO in the correlation function towards smaller distances. This can result in $\qiso > 1$ before reconstruction, even if the fiducial cosmology matches the true one, which can then be corrected by reconstruction (see \cref{fig:pk_bgs} from the previous section). The right panel reveals that for \lrgth, reconstruction significantly improves the errors on $\qiso$ for all mock realizations. This is qualitatively consistent with the results of \cite{Moon2023:2304.08427}, who found that for mocks that match the properties of the DESI EDR LRG sample, reconstruction tends to improve the errors on the isotropic BAO scaling parameter in about 90\% of the cases. However, for \elgt, some of the mocks lie above the reference dashed line, which delineates the region where errors degrade after reconstruction. However, the $\qiso$ errors decrease for most of the mocks, obtaining consistent constraints from both smoothing scales.

\Cref{fig:smoothing_scales_qap} shows the constraints for the anisotropic scaling parameter $\qap$. Both tracers recover $\qap$ around one and show consistent values before and after reconstruction. The choice of smoothing scale has a negligible impact for \lrgt\ and a very small impact on the average best-fit value for \elgt, resulting in a change that is smaller than the error on the mean. For \lrgt, reconstruction improves the precision on $\qap$ for all mocks in a very similar way for the two smoothing scales. For \elgt, there is consistency in the $\qap$ error from both smoothing scales, but a few mocks show increased errors after reconstruction.

In \cref{ap:additional figures}, we show the equivalent figures for the other redshift bins and tracers, and \cref{tab:fits_abacussummit} shows a compilation of the constraints before and after reconstruction with different smoothing scales. To estimate the constraints shown in \cref{tab:fits_abacussummit}, we fit the mean of the 25 \abacussummit\ realizations, using a reduced covariance matrix that matches the combined volume of all realizations. For the \qso, due to the lower tracer number density and shot noise, we test slightly larger smoothing scales of $20 \Mpch$ and $30 \Mpch$, while for \bgs, \lrgo, \lrgt, and \elgt\ we test $\Sigmasm = 10 \Mpch$ and $15 \Mpch$. In terms of the reduction of errors for the scaling parameters, we find a relatively good consistency between the different choices of scale for most tracers. For $\qiso$, the reduction in errors from the first choice of smoothing scale ($30 \Mpch$ for \qso\ and $15 \Mpch$ for the rest) is largest for \bgs, with a 44\%. The smallest improvement is of 11\%, for \qso. For \lrgo, \lrgt, and \lrgth, we get a 31\%, 38\%, and 31\% error reduction, respectively, while for \elgo\ and \elgt, we get a 20\% and 12\% error reduction. The second choice of smoothing scale ($20 \Mpch$ for \qso\ and $10 \Mpch$ for the rest) leads to an error reduction of 46\% and 2.6\% for \bgs\ and \qso, respectively. For \lrgo, \lrgt\ and \lrgth, we get a 41\%, 38\%, and 30\% improvement, while for \elgo\ and \elgt, we get a 14\% and 9\% error improvement. For $\qap$, both smoothing scales lead to consistent reduction in errors for \lrgs, around 46\%, 42\%, and 28\% for \lrgo, \lrgt, and \lrgth, respectively. For \elgt, we see a larger improvement from $\Sigmasm = 15 \Mpch$, which reduces the errors by 11\%, compared to the 6.3\% reduction in error from $\Sigmasm = 10 \Mpch$. Overall, all tracers benefit from reconstruction when looking at the average clustering from all mocks, and the first choice of smoothing scale tends to lead to a better error reduction when comparing the different tracers and scaling parameters.

Focusing on the last two columns of \cref{tab:fits_abacussummit}, we can quantify the bias in the BAO fits with respect to the expectation of $\qiso = 1$ and $\qap = 1$. More formally, the bias is defined as $(\alpha - 1)//\sigma$, where $\sigma$ is the standard deviation of the marginalized posterior quoted in the fourth and fifth columns. We remind the reader that these offsets have been calculated by assuming a covariance matrix associated with the combination of 25 \abacussummit\ realizations, which results in a volume much larger than the DESI DR1 sample. Overall, we do not find evidence that the choice of smoothing scale produces statistically significant differences in the recovered bias from the BAO fits. The largest biases are found for \lrgth, for which $\Sigmasm = 15 \Mpch$ and $\Sigmasm = 10 \Mpch$ produce biases of $-2.34\sigma$ and $-2.28$. Similarly, for \lrgt, these two scales produce shifts of $-2.05\sigma$ and $-1.92\sigma$, respectively. Although none of the tracers show statistically significant offsets with respect to the truth, it is interesting that almost all of them appear to be biased low in $\qiso$, which is not the case for the pre-reconstruction fits. Therefore, in what follows, we further explore possible systematic errors in the modeling by looking at constraints derived from the \ezmocks.

\begin{figure}
    \centering
    \begin{tabular}{cc}
        \hspace{-0.3cm}\includegraphics[width=0.47\textwidth]{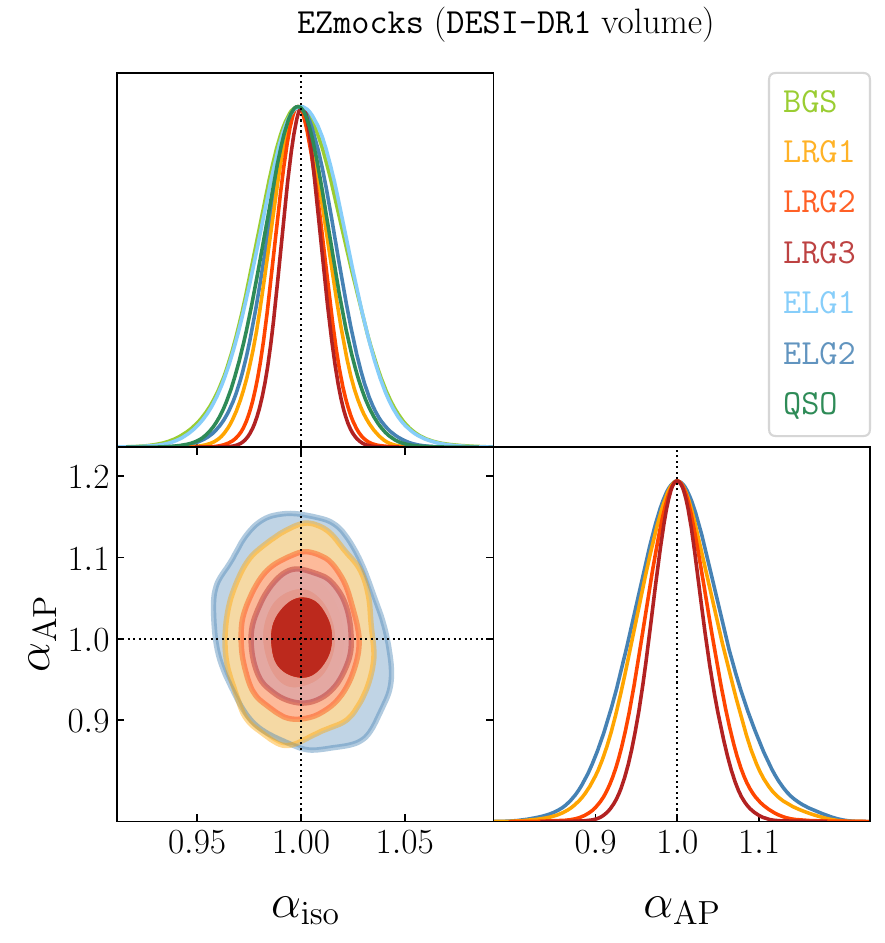} & \hspace{-0.2cm}\includegraphics[width=0.47\textwidth]{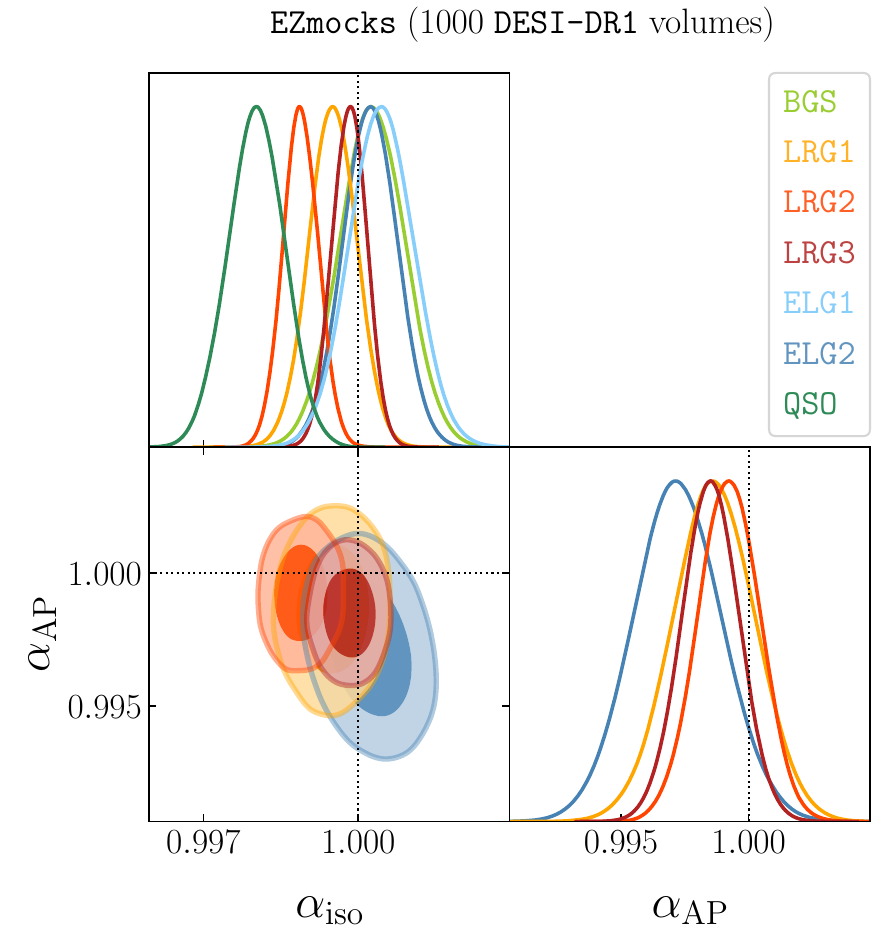}
    \end{tabular}
    \caption{BAO fits to synthetic mock galaxy catalogues generated with the EZmock algorithm. The data vector corresponds to the post-reconstruction correlation function multipoles, averaged over 1000 independent mock realizations. The left  panels show results using a covariance matrix is representative of the DESI DR1 volume, while the right panel displays results using a reduced covariance associated to the suite of 1000 mocks. For tracers for which we only perform isotropic BAO fits (\bgs, \elgo, \qso), we only display the $\qiso$ marginalized posterior in the top subpanels. \href{https://github.com/cosmodesi/desi-y1-kp45/blob/main/y1-papers/optimalrecon/contours_ezmock.py}{\faFileCodeO}}
    \label{fig:fits_ezmock}
\end{figure}

\begin{table}
    \centering
    \small
    \begin{tabular}{|l|c|c|c|c|r|r|}
    \hline
     Tracer       & Recon   & $\Sigma_{\rm sm}$    & $\alpha_{\rm iso}$   & $\alpha_{\rm AP}$   & $\Delta \alpha_{\rm iso}/\sigma$   & $\Delta \alpha_{\rm AP}/\sigma$   \\
    \hline
     ${\tt BGS}$  & Post    & 15 $h^{-1}{\rm Mpc}$ & $1.00027 \pm 0.00063$ & ---                   & $0.43$                             & ---                               \\
     ${\tt LRG1}$ & Post    & 15 $h^{-1}{\rm Mpc}$ & $0.99951 \pm 0.00046$ & $0.99864 \pm 0.00161$ & $-1.06$                            & $-0.84$                           \\
     ${\tt LRG2}$ & Post    & 15 $h^{-1}{\rm Mpc}$ & $0.99889 \pm 0.00034$ & $0.99923 \pm 0.00120$ & $-3.23$                            & $-0.64$                           \\
     ${\tt LRG3}$ & Post    & 15 $h^{-1}{\rm Mpc}$ & $0.99983 \pm 0.00033$ & $0.99850 \pm 0.00113$ & $-0.51$                            & $-1.33$                           \\
     ${\tt ELG1}$ & Post    & 15 $h^{-1}{\rm Mpc}$ & $1.00045 \pm 0.00062$ & ---                   & $0.73$                             & ---                               \\
     ${\tt ELG2}$ & Post    & 15 $h^{-1}{\rm Mpc}$ & $1.00023 \pm 0.00053$ & $0.99723 \pm 0.00173$ & $0.43$                             & $-1.60$                           \\
     ${\tt QSO}$  & Post    & 30 $h^{-1}{\rm Mpc}$ & $0.99802 \pm 0.00054$ & ---                   & $-3.64$                            & ---                               \\
     \hline
     ${\tt BGS}$  & Pre     & ---                  & $1.00081 \pm 0.00108$ & ---                   & $0.75$                             & ---                               \\
     ${\tt LRG1}$ & Pre     & ---                  & $1.00180 \pm 0.00078$ & $0.99807 \pm 0.00298$ & $2.30$                             & $-0.65$                           \\
     ${\tt LRG2}$ & Pre     & ---                  & $1.00265 \pm 0.00057$ & $1.00159 \pm 0.00216$ & $4.63$                             & $0.74$                            \\
     ${\tt LRG3}$ & Pre     & ---                  & $1.00289 \pm 0.00046$ & $1.00012 \pm 0.00173$ & $6.29$                             & $0.07$                            \\
     ${\tt ELG1}$ & Pre     & ---                  & $1.00093 \pm 0.00086$ & ---                   & $1.08$                             & ---                               \\
     ${\tt ELG2}$ & Pre     & ---                  & $1.00055 \pm 0.00057$ & $1.00309 \pm 0.00199$ & $0.97$                             & $1.55$                            \\
     ${\tt QSO}$  & Pre     & ---                  & $1.00143 \pm 0.00064$ & ---                   & $2.25$                             & ---                               \\
    \hline
    \end{tabular}
    \caption{Mean values and standard deviations from the marginalized posteriors of the BAO scaling parameters from a fit to the mean of 1000 realizations of the \ezmocks. We display pre-reconstruction results, as well as post-reconstruction fits obtained with a smoothing scale $\Sigmasm = 15 \Mpch$. The last two columns show the offset in the mean values with respect to the expected value of $\qiso$ and $\qap$, in units of the corresponding standard deviation, i.e. $\Delta \alpha = (\alpha - 1)/\sigma$. We use a reduced covariance matrix representative of the combined volume of all \ezmocks.}
    \label{tab:fits_ezmock}
\end{table}

 In \cref{fig:fits_ezmock} we present fits to the post-reconstruction data vectors averaged over 1000 realizations of the \ezmocks, using a covariance matrix that is representative of the volume of the DESI DR1 samples (left) or a reduced covariance associated to the suite of 1000 mocks. The constraints for the latter case are summarized in \cref{tab:fits_ezmock}, where we have also included the pre-reconstruction results. Given the computational restriction of reconstructing this large number of mocks, we do it only for a single smoothing scale for each tracer, as detailed in the table.

As the left panel of \cref{fig:fits_ezmock} shows, for a DESI DR1 volume, we obtain parameter constraints that are largely unbiased and consistent across the various redshift ranges. The relative constraining power is driven by the effective volume of each sample, with the highest post-reconstruction precision coming from the \lrgth\, and the lowest from the \bgs\ and \elgo. The improvement in errors compared to pre-reconstruction is consistent with what was observed for \abacussummit, with improvements on $\qiso$ ranging from a 7\% for \elgt, to a 41\% for \lrgo. Similarly, for $\qap$, the improvement ranges from 13\% for \elgt\ to 44\% for \lrgt.

Focusing on the constraints from \cref{tab:fits_ezmock}, which are also illustrated in the right panel of \cref{fig:fits_ezmock}, we see that dividing the covariance matrix by a factor of 1000, we find two statistically significant biases in the mean values of the $\qiso$ posteriors: a $-3.23\sigma$ bias for \lrgt, and a $-3.64\sigma$ bias for \qso. Interestingly, when looking at the collection of all tracers, the post-reconstruction biases are not as coherent as seen in \abacussummit, as they can be biased high (\bgs, \elgo, \elgt) or low (\lrgs\ and \qso). In terms of percentage, these offsets are much milder than in \abacussummit. For example, the $-3.24\sigma$ bias in the \ezmocks\ \lrgt\ corresponds to $0.11$\%, while in \abacussummit, the $-2.59\sigma$ bias is of $0.52$\%. Furthermore, for the same tracers and redshift bins, the offsets can point in different directions, indicating that some fraction of these biases could be attributed to noise fluctuations given the small number of \abacussummit\ realizations to perform this test. However, we cannot fully discard the possibility that there are systematic effects in the mocks that are currently unaccounted for. In the next section, we assess the level at which fiber assignment incompleteness affects the BAO constraints, showing that no statistically significant detections of systematics are found due to this effect.

Overall, given the small sensitivity of the BAO constraints to the smoothing scales we have tested here as seen in \abacussummit, and considering the robustness of the BAO constraints when validated against the \ezmocks, we opt for $\Sigma_{\rm sm} = 15 \Mpch$ for BGS, LRG, and ELG, and $\Sigma_{\rm sm} = 30 \Mpch$ for QSO. In later sections, we show that the blinded DESI data are also insensitive to this choice. Furthermore, these tests not only inform the decision about the smoothing scale, but also indirectly and approximately reveal the size of the systematic error budget in the BAO modeling that can be inferred from these mocks. A detailed description of the different components of the systematic error budget and how this is added to the statistical error of the unblinded galaxy BAO analysis is presented in \cite{DESI2024.III.KP4}.

\subsection{Impact of the fiber assignment}

\begin{figure}
    \centering
    \begin{tabular}{ccc}
       \includegraphics[width=0.3\textwidth]{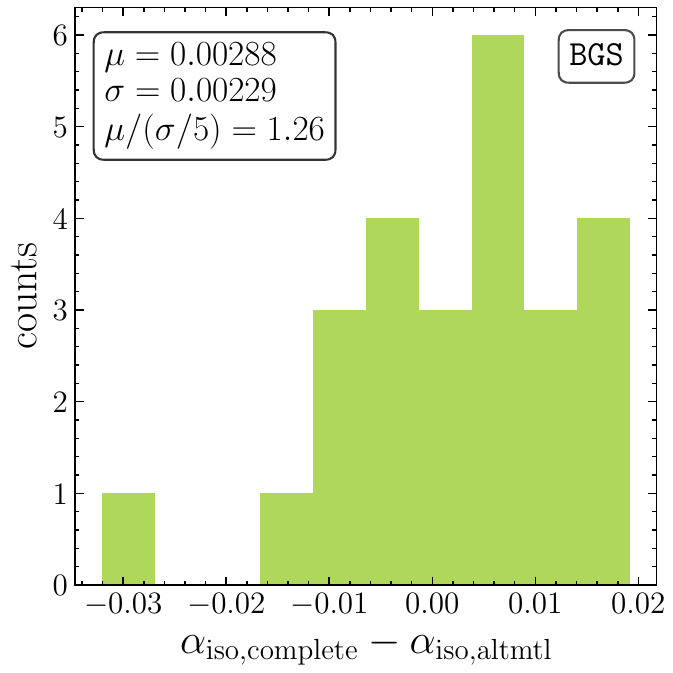}  & \includegraphics[width=0.3\textwidth]{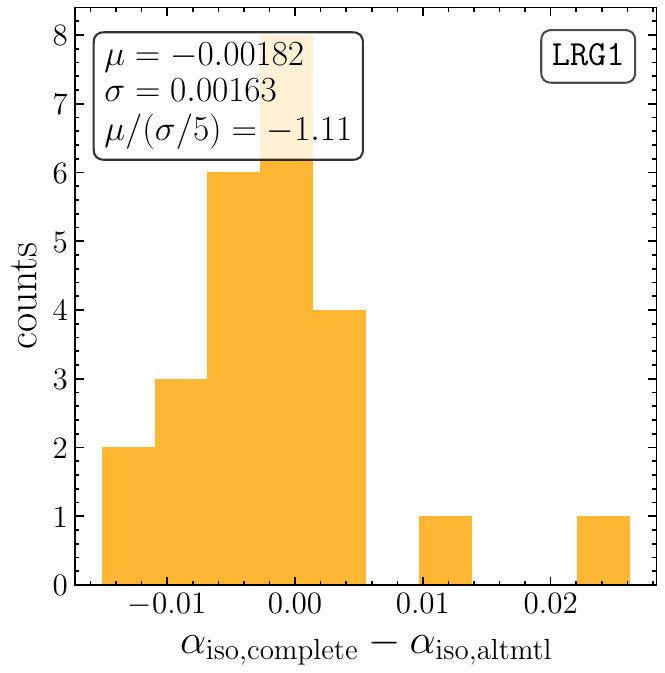} & \includegraphics[width=0.31\textwidth]{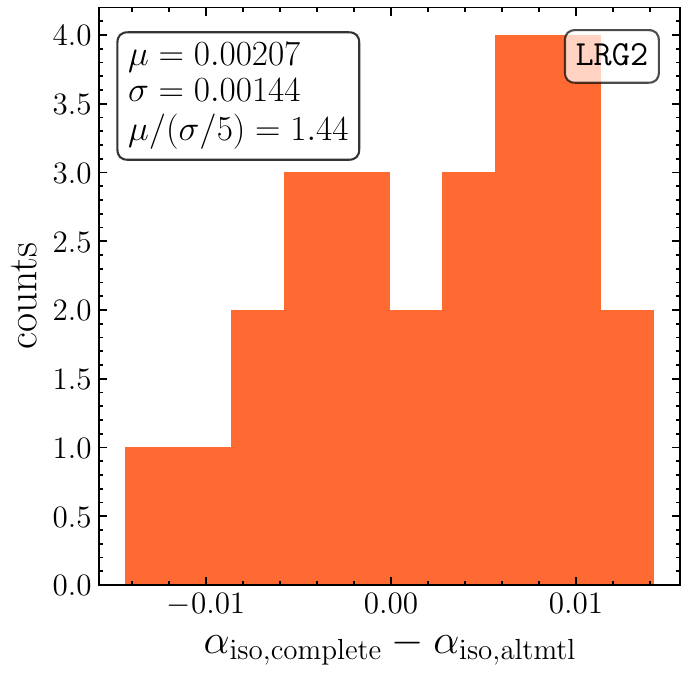} \\
       \includegraphics[width=0.31\textwidth]{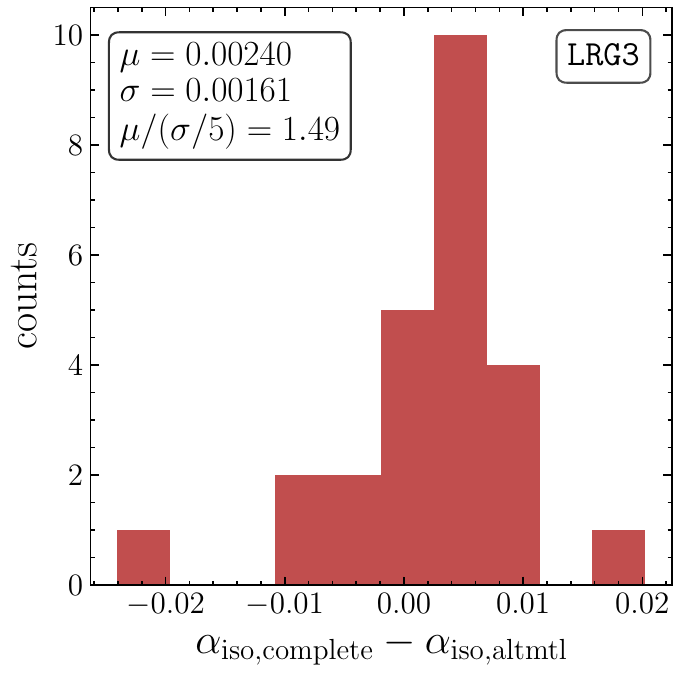} & \includegraphics[width=0.3\textwidth]{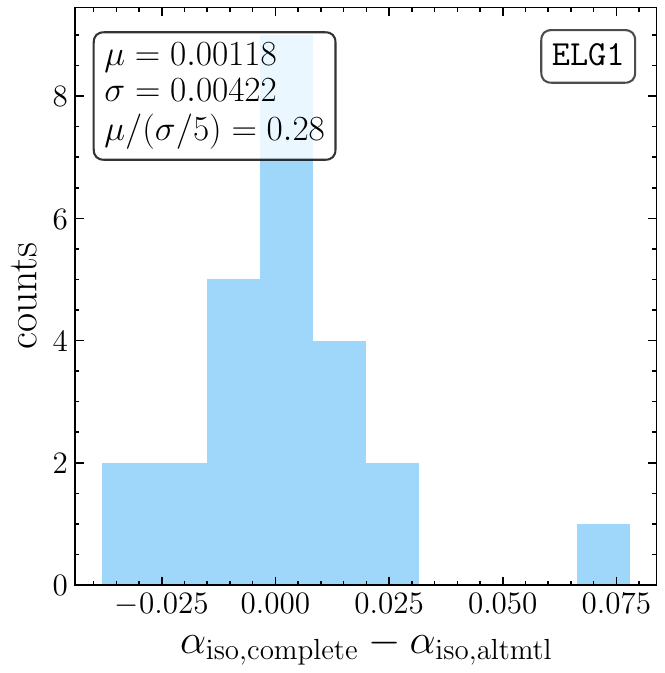} & \includegraphics[width=0.31\textwidth]{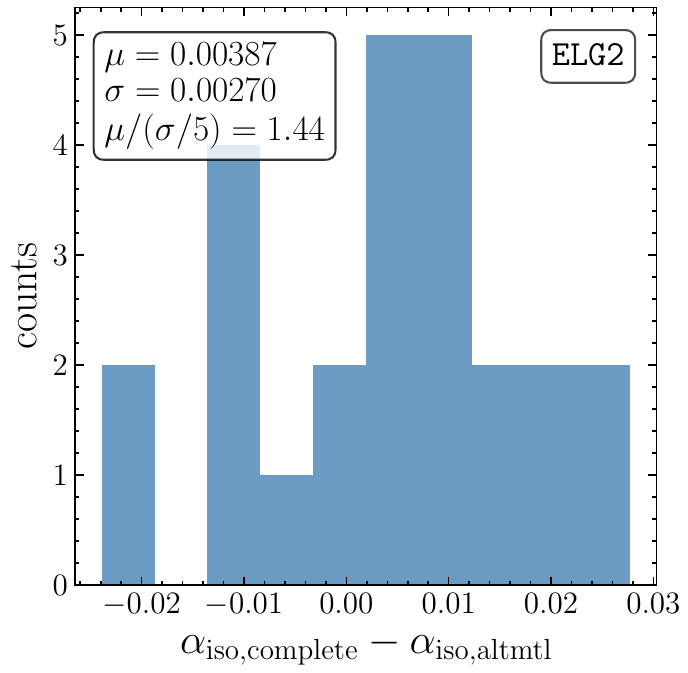} \\
       & \includegraphics[width=0.305\textwidth]{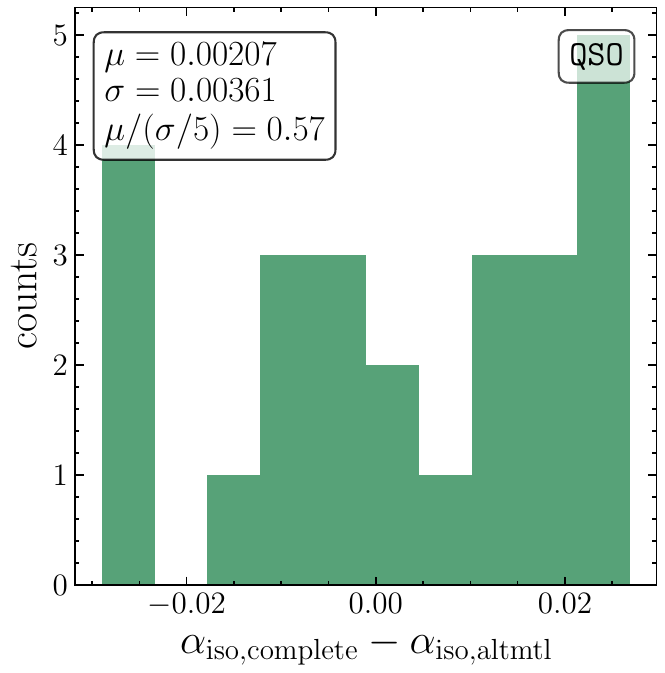} &
    \end{tabular}
    \caption{Difference in the mean value of the marginalized posterior of the isotropic BAO scaling parameter $\qiso$ between the complete (no fiber assignment incompleteness) and altmtl (realistic fiber assignment incompleteness) mocks, derived from fits to inidividual realizations of the \abacussummit\ mocks. The legend shows the mean and dispersion of this difference estimated from all realizations, as well as the standard error on the mean.}
    \label{fig:fits_abacussummit_fiberassign_qiso}
\end{figure}

\begin{figure}
    \centering
    \begin{tabular}{ccc}
       \includegraphics[width=0.3\textwidth]{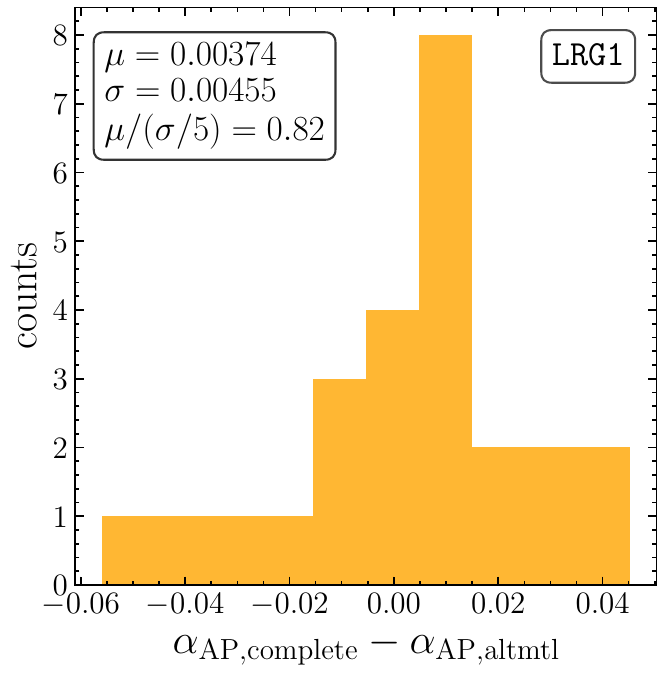} & \includegraphics[width=0.315\textwidth]{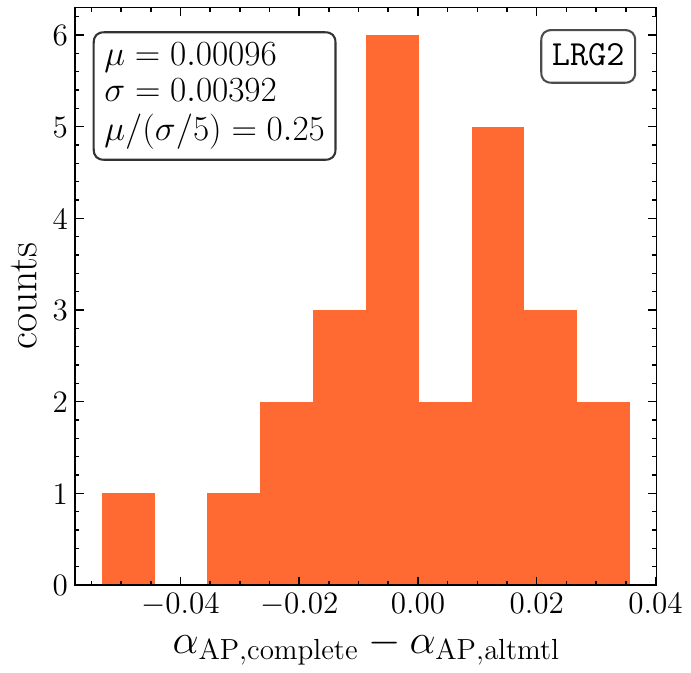} & \includegraphics[width=0.3\textwidth]{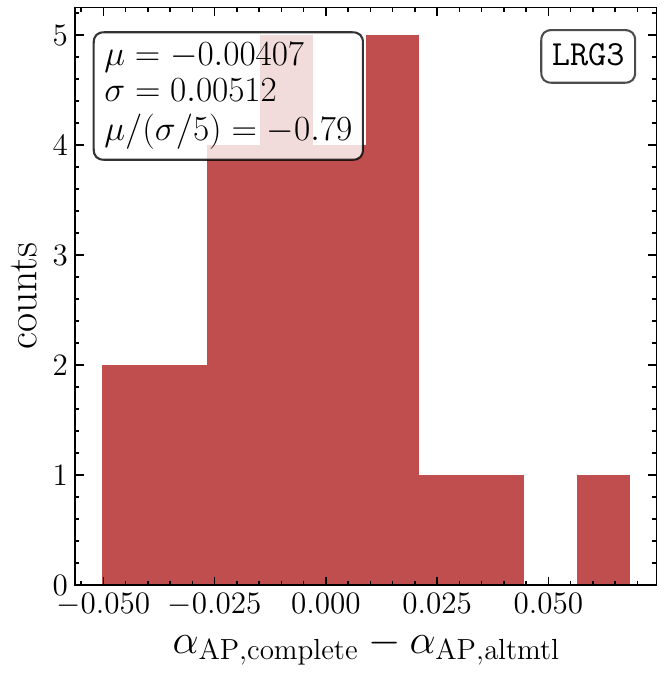}\\
       & \includegraphics[width=0.315\textwidth]{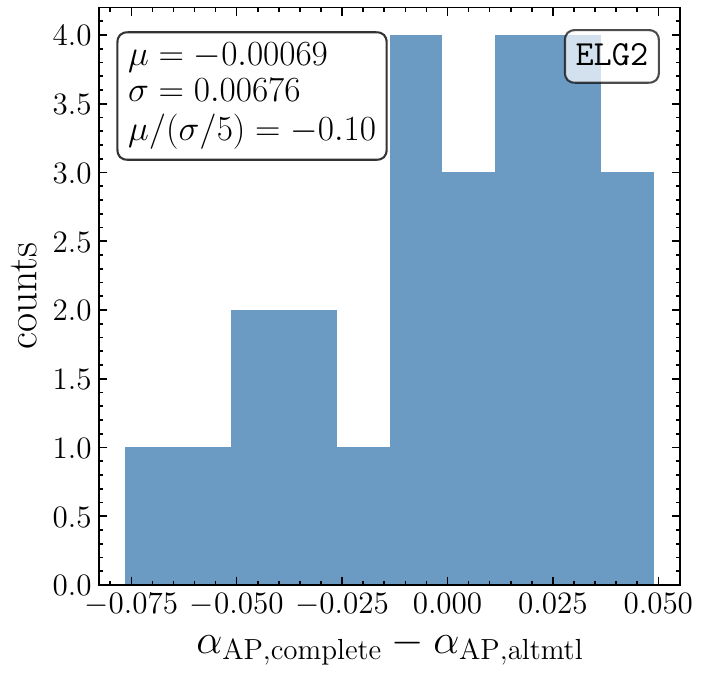} & 
    \end{tabular}
    \caption{Similar to \cref{fig:fits_abacussummit_fiberassign_qiso}, but showing results for the anisotropic BAO scaling parameter, $\qap$.}
    \label{fig:fits_abacussummit_fiberassign_qap}
\end{figure}

\begin{table}[]
    \centering
    \small
    \begin{tabular}{|l|c|c|c|r|r|}
    \hline
     Tracer       & Recon   & $\langle \alpha^{\rm complete}_{\rm iso} \rangle$   & $\langle \alpha^{\rm complete}_{\rm AP}\rangle$   & $\langle \alpha^{\rm complete}_{\rm iso} - \alpha^{\rm altmtl}_{\rm iso}\rangle$   & $\langle\alpha^{\rm complete}_{\rm AP} - \alpha^{\rm altmtl}_{\rm AP}\rangle$ \\
    \hline
     ${\tt BGS}$  & Post    & $0.9998 \pm 0.0038$                 & ---                                & $0.00288 \pm 0.00229$                                                 & ---                                                                   \\
     ${\tt LRG1}$ & Post    & $0.9996 \pm 0.0025$                 & $0.9934 \pm 0.0091$                & $-0.00182 \pm 0.00163$                                                & $0.00374 \pm 0.00455$                                               \\
     ${\tt LRG2}$ & Post    & $0.9967 \pm 0.0022$                 & $1.0091 \pm 0.0075$                & $0.00207 \pm 0.00144$                                                 & $0.00096 \pm 0.00392$                                              \\
     ${\tt LRG3}$ & Post    & $0.9958 \pm 0.0019$                 & $0.9854 \pm 0.0063$                & $0.00240 \pm 0.00161$                                                 & $-0.00407 \pm 0.00512$                                              \\
     ${\tt ELG1}$ & Post    & $1.0015 \pm 0.0035$                 & ---                                & $0.00118 \pm 0.00422$                                                 & ---                                                               \\
     ${\tt ELG2}$ & Post    & $0.9988 \pm 0.0030$                 & $1.0025 \pm 0.0100$                & $0.00387 \pm 0.00270$                                                 & $0.00387 \pm 0.00676$                                              \\
     ${\tt QSO}$  & Post    & $0.9968 \pm 0.0033$                 & ---                                & $0.00207 \pm 0.00361$                                                 & ---                                                               \\
    \hline
    \end{tabular}
    \caption{Constraints on the BAO scaling parameters averaged over individual fits to post-reconstruction correlation functions from 25 realizations of the complete \abacussummit\ mocks, which do not include any fiber assignment incompleteness. We also show the mean of the differences in the $\alpha$ values between the complete mocks and the altmtl mocks, which mimic the fiber assignment incompleteness of the DESI target samples.}
    \label{tab:complete_vs_altmtl}
\end{table}

The assignment of fibers that are used to measure the redshifts of DESI tracers plays an important role in their clustering properties. To capture these properties in the mock catalogs, the DESI {\tt FiberAssign} code mimics the effect of assignment incompleteness by trying to reproduce the DESI fiber assignment from real targets. Here, we distinguish between two flavors of mocks to study the impact of this effect on the BAO constraints after reconstruction: `complete' mocks which do not include any assignment incompleteness, and thus can be treated as a baseline for the assessment of the incompleteness, and `altmtl' mocks, which represent our most realistic simulations of the DR1 fiber assignment scheme. They apply the algorithm described in \cite{KP3s7-Lasker}, using the {\tt FiberAssign} code perfectly reproduce the assignment of DESI fibers on real targets, without approximations. Our results from \abacussummit\ in previous section were derived from the altmtl mocks.

In \cref{tab:complete_vs_altmtl}, we show the constraints on $\qiso$ and $\qap$ derived from the complete mocks. These constraints have been obtained by averaging individual fits to the post-correlation functions from all 25 mock realizations. We note that even in the complete case, there are deviations from $\qiso = 1$ for all tracers, which, although not statistically significant, point in the same direction as the altmtl mocks that were used to derive the constraints from the previous section. This suggests that the fiber assignment process is not the driver of the biases seen in the \abacussummit\ fits, and reinforces the idea that they could be partially attributed to statistical fluctuations given the small number of mock realizations. 

The fifth and sixth columns show the mean of the differences in $\qiso$ and $\qap$, respectively, between the complete and altmtl mocks, as averaged over all mock realizations. The distribution of individual fits are also shown in \cref{fig:fits_abacussummit_fiberassign_qiso,fig:fits_abacussummit_fiberassign_qap}. Overall, we do not observe any statistically significant shift in the constraints. For $\qiso$, the largest offset is of $1.49\sigma$ for \lrgth. For $\qap$, it is for \lrgo\ at $0.82\sigma$.

\begin{figure}
    \centering
    \begin{tabular}{c}
    \includegraphics[width=\textwidth]{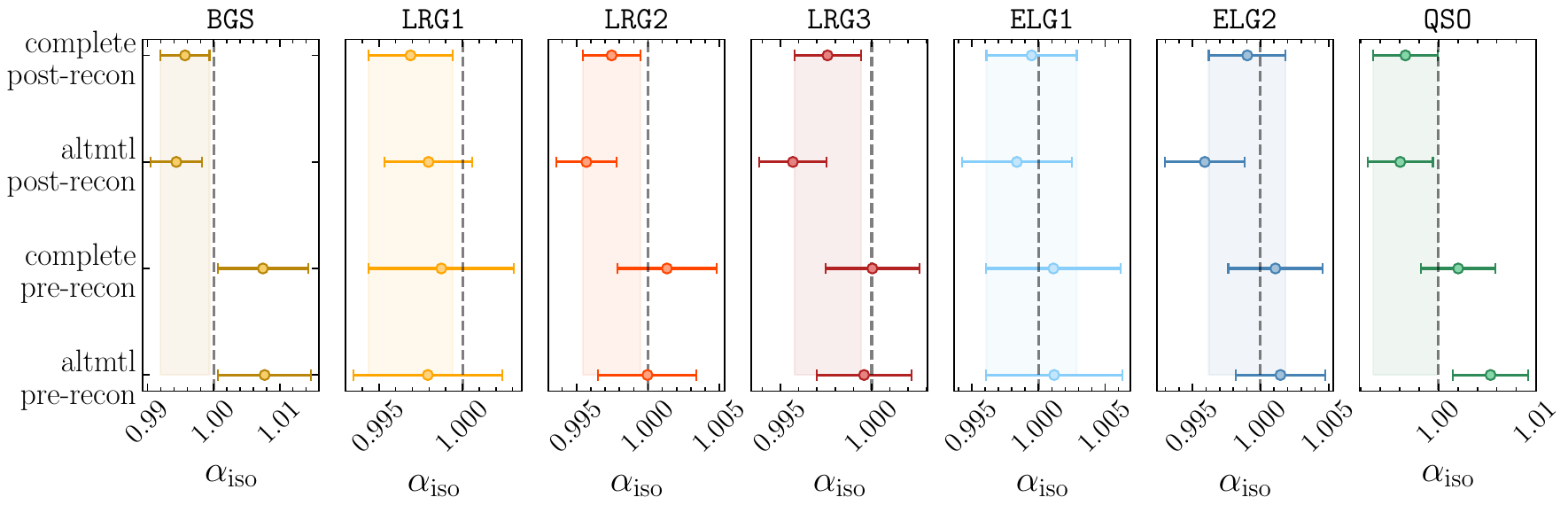}       \\
    \includegraphics[width=0.55\textwidth]{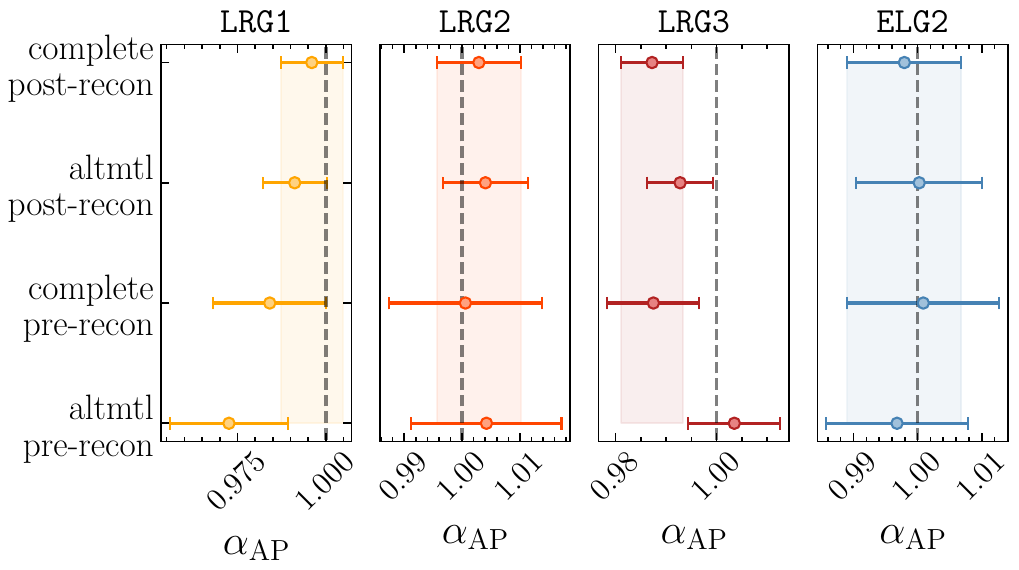}
    \end{tabular}
    \caption{A comparison of the constraints on the isotropic BAO scaling parameter $\qiso$ from fits to the mean of 25 \abacussummit\ mock realizations, after having passed the mocks through a code that realistically mimics the DESI DR1 fiber assignment (altmtl) or using mocks that do not include any fiber assignment incompleteness (complete). We have used a reduced covariance matrix associated to the combined volume of all 25 mock realizations.}
    \label{fig:whisker_fiberassign}
\end{figure}

In \cref{fig:whisker_fiberassign}, we show a fit to the mean of 25 mocks for complete and altmtl mocks, where we have also included pre-reconstruction results. For all these fits, we use a reduced covariance matrix that is representative of the combined volume of all mock realizations. For pre-reconstruction, we find that the inclusion of fiber assignment incompleteness makes little effect on the constraints for most tracers. Overall, we do not find statistically significant features that could suggest that the fiber assignment scheme affects pre- and post-reconstruction measurements differently. In general, our findings corroborate that the effect of fiber assignment incompleteness on BAO measurements is relatively mild, since we restrict the analysis to fairly large scales. Moreover, we are isolating the BAO feature in the modeling, and potential large-scale systematics are partially absorbed by the broadband nuisance parameters. However, we need to keep in mind that the small number of mock realizations available restricts the precision at which we can perform this test. Although the offsets we observed in this and previous sections are much smaller than the statistical error from the DESI-DR1 samples, subsequent analyses with more data will demand stringent tests based on a larger collection of mock catalogs to match the increase level of precision in the BAO constraints.

\subsection{Consistency between the DESI data and mocks}

\begin{figure}
    \centering
    \begin{tabular}{ccc}
        \includegraphics[width=0.3\textwidth]{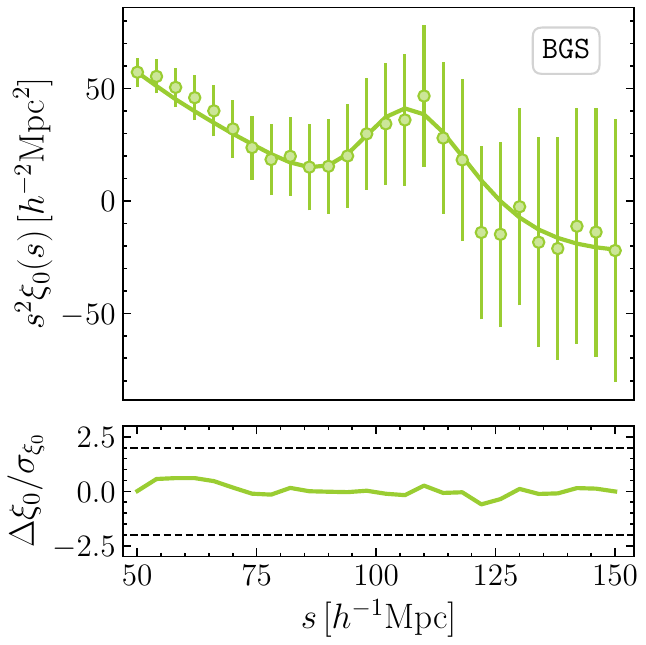} &
        \includegraphics[width=0.3\textwidth]{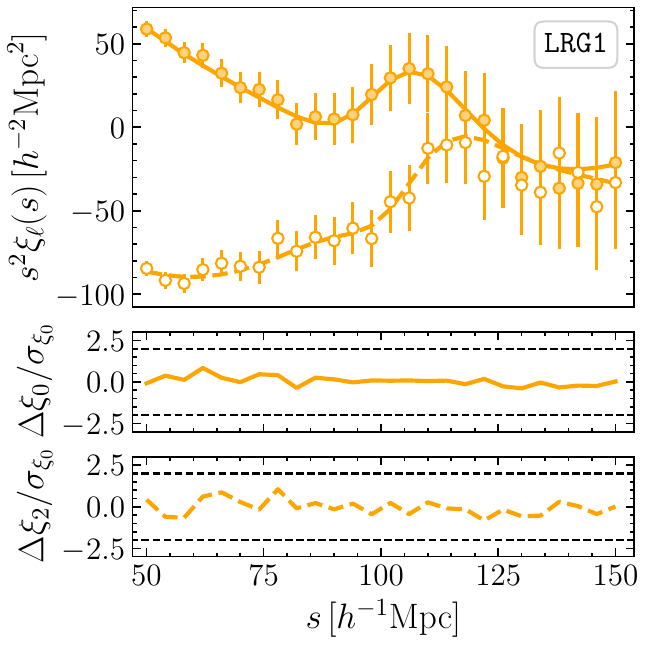} &
        \includegraphics[width=0.3\textwidth]{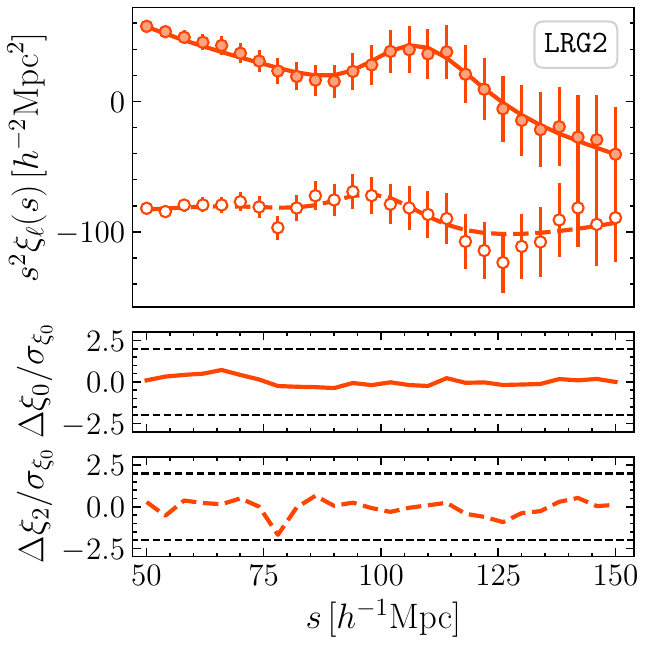}\\
        \includegraphics[width=0.3\textwidth]{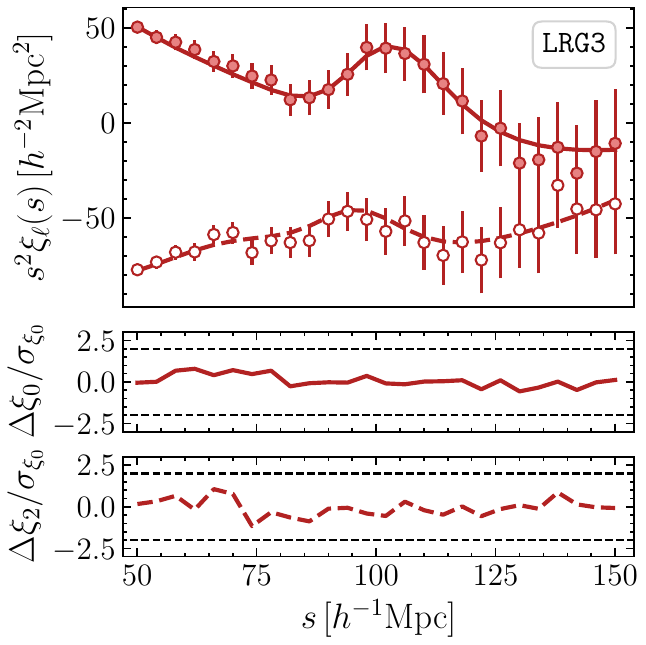} &
        \includegraphics[width=0.3\textwidth]{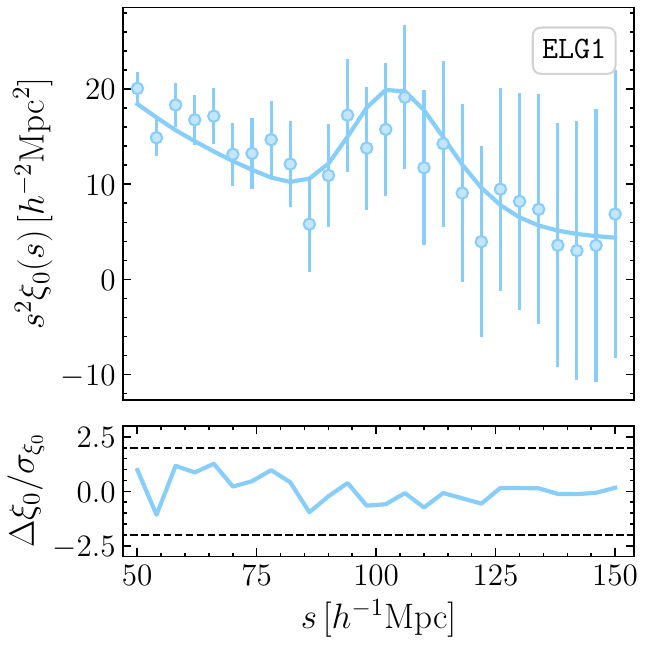} &
        \includegraphics[width=0.3\textwidth]{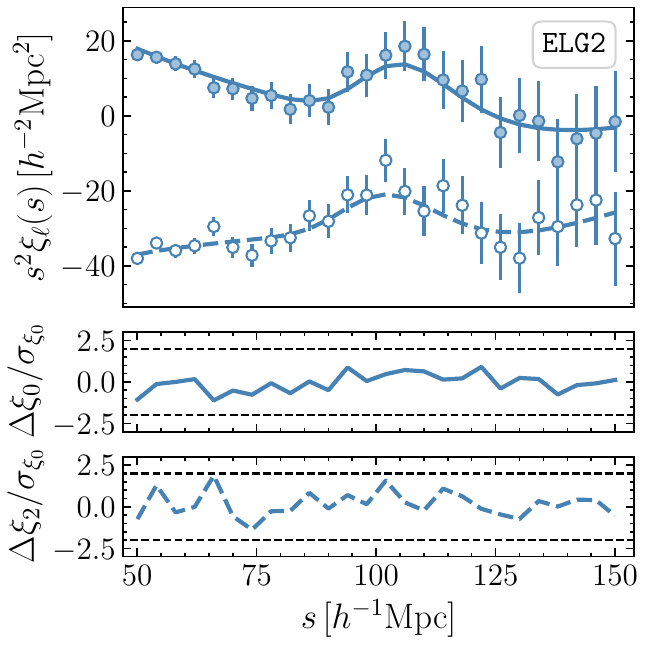}\\
         & \includegraphics[width=0.3\textwidth]{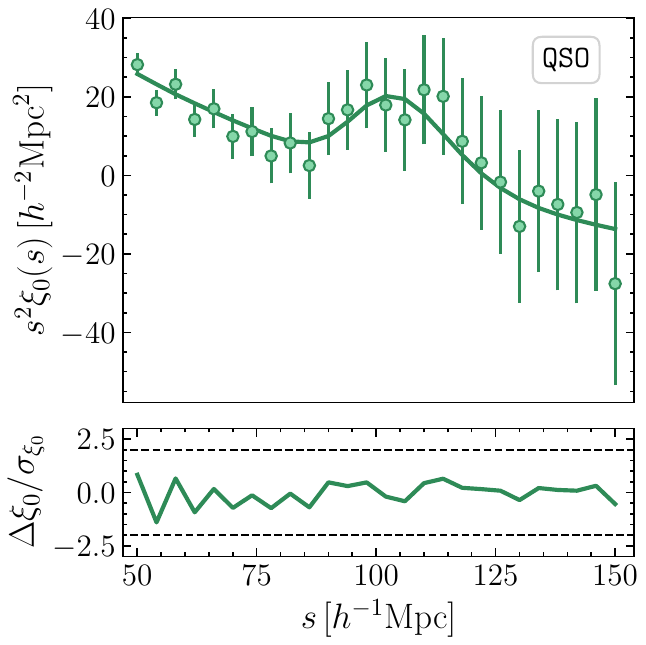} & 
    \end{tabular}
    \caption{Measured multipoles from the post-reconstruction correlation functions of the blinded DESI DR1 data. Filled and open circles show the monopole and quadrupole moments, respectively, with their corresponding best-fit models shown by the solid and dashed lines, respectively. For \bgs, \elgo\, and \qso, we only fit the monopole, due to the lower SNR of the data. The lower sub-panels show the residuals between the best-fit models and the data in each case.}
    \label{fig:multipoles_dr1}
\end{figure}

Having explored the BAO constraints from the mocks, the main question we wish to answer in this section is \textit{are the BAO constraints obtained from the blinded DESI data consistent with the results obtained from the distribution of mocks}? The tests presented here constitute some of the criteria that the DESI Collaboration used to decide whether we were ready to fix the end-to-end analysis pipeline, unblind the LSS catalogs and go ahead with the main BAO cosmological analysis.

In \cref{fig:multipoles_dr1}, we display the post-reconstruction correlation function multipoles of the blinded DESI DR1 data, where we have adopted the fiducial smoothing scales determined in the previous section. For the cases where we only perform 1D BAO fits (\bgs, \elgo\, and \qso), only the monopole moment is shown. We find that the BAO model accurately fits the data for all scales considered in the fit, with residuals between the model and the data that are within $2\sigma$. In addition to potential systematic errors due to non-linear evolution/biasing and large-scale systematics, the observed position of the BAO feature can be affected by Alcock-Paczynski distortions arising due to a potential mismatch between our fiducial cosmology and the true cosmology of the Universe, as well as due to the blinding scheme that can change the position of the BAO feature and the RSD signal. The error bars come from the semi-analytic semi-empirical {\tt RascalC} covariances that were tuned to match the clustering of the DESI DR1 samples. These constraints are summarized in \cref{tab:fits_desi_blinded}, along with the pre-reconstruction constraints from the same data.

\begin{table}
    \centering
    \label{tab:fits_desi_blinded}
    \vspace{0.5em}
    \begin{tabular}{|l|c|c|c|c|c|r|}
    \hline
     Tracer       & Redshift   & Recon   & $\alpha_{\rm iso}$    & $\alpha_{\rm AP}$     & $r$         & $\chi^2 / {\rm dof}$   \\
    \hline
     ${\tt BGS}$  & 0.1--0.4    & Post    & $0.98416 \pm 0.01601$ & ---                   & ---             & 24.5/19                \\
     ${\tt LRG1}$ & 0.4--0.6    & Post    & $0.98416 \pm 0.00988$ & $0.99679 \pm 0.03201$ & -0.12           & 40.1/39                \\
     ${\tt LRG2}$ & 0.6--0.8    & Post    & $0.98416 \pm 0.01169$ & $0.99679 \pm 0.04577$ & -0.09           & 40.2/39                \\
     ${\tt LRG3}$ & 0.8--1.1    & Post    & $0.98416 \pm 0.00891$ & $0.99679 \pm 0.03279$ & -0.03           & 35.0/39                \\
     ${\tt ELG1}$ & 0.8--1.1    & Post    & $0.98416 \pm 0.02450$ & ---                   & ---             & 23.4/19                \\
     ${\tt ELG2}$ & 1.1--1.6    & Post    & $0.98416 \pm 0.01376$ & $0.99679 \pm 0.04727$ & -0.31           & 45.7/39                \\
     ${\tt QSO}$  & 0.8--2.1    & Post    & $0.98416 \pm 0.02509$ & ---                   & ---             & 16.9/19                \\
     \hline
     ${\tt BGS}$  & 0.1--0.4    & Pre     & $0.98416 \pm 0.02650$ & ---                   & ---             & 13.9/19                \\
     ${\tt LRG1}$ & 0.4--0.6    & Pre     & $0.98416 \pm 0.01877$ & $0.99679 \pm 0.06980$ & 0.44            & 32.6/39                \\
     ${\tt LRG2}$ & 0.6--0.8    & Pre     & $0.98416 \pm 0.01869$ & $0.99679 \pm 0.07317$ & 0.46            & 51.0/39                \\
     ${\tt LRG3}$ & 0.8--1.1    & Pre     & $0.98416 \pm 0.01162$ & $0.99679 \pm 0.04469$ & 0.15            & 42.9/39                \\
     ${\tt ELG1}$ & 0.8--1.1    & Pre     & $0.98416 \pm 0.05708$ & ---                   & ---             & 23.6/19                \\
     ${\tt ELG2}$ & 1.1--1.6    & Pre     & $0.98416 \pm 0.01813$ & $0.99679 \pm 0.06523$ & -0.06           & 34.0/39                \\
     ${\tt QSO}$  & 0.8--2.1    & Pre     & $0.98416 \pm 0.02304$ & ---                   & ---             & 9.7/19                 \\
    \hline
    \end{tabular}
    \caption{Mean values and standard deviations of the marginalized posteriors on the BAO scaling parameters, from fits to the blinded DESI DR1 correlation functions. We display pre- and post-reconstruction constraints, where the reconstruction adopts the fiducial settings determined in \cref{subsec:smoothing_scale}. We additionally show the cross-correlation coefficient between $\qiso$ and $\qap$, labeled as $r$, and the $\chi^2$ per degree of freedom at the best fit.}
\end{table}

\begin{figure}
    \centering
    \includegraphics[width=0.7\textwidth]{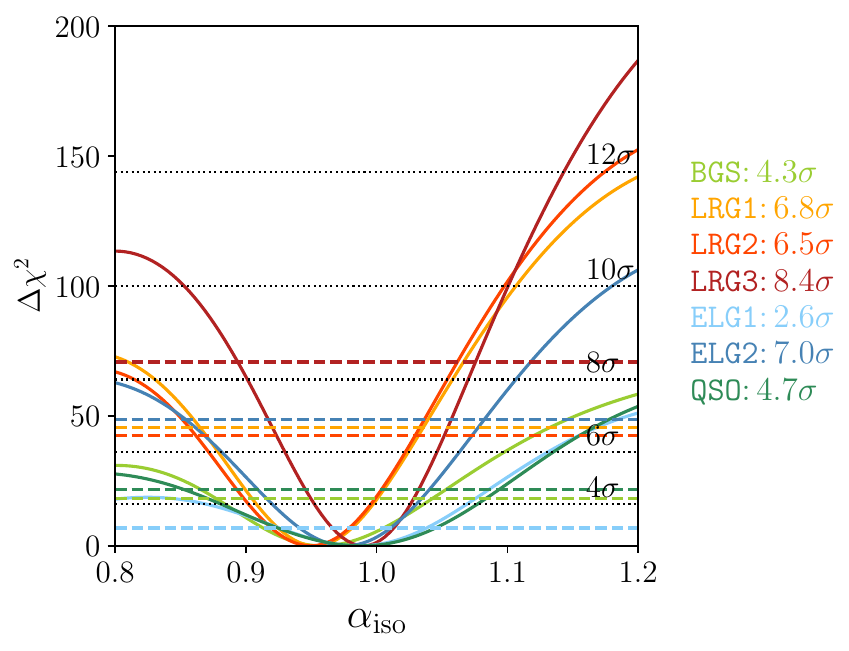}
    \caption{BAO detection level signicance from fits to the correlation functions from the blinded DESI DR1 data. The solid and dashed lines in different colors show the $\Delta \chi^2$ for models with and without BAO wiggles, respectively. We display horizontal dotted lines at various levels of detection significance for reference.}
    \label{fig:bao_detection_dr1blinded}
\end{figure}

\begin{figure}
    \centering
    \begin{tabular}{ccc}
        \includegraphics[width=0.3\textwidth]{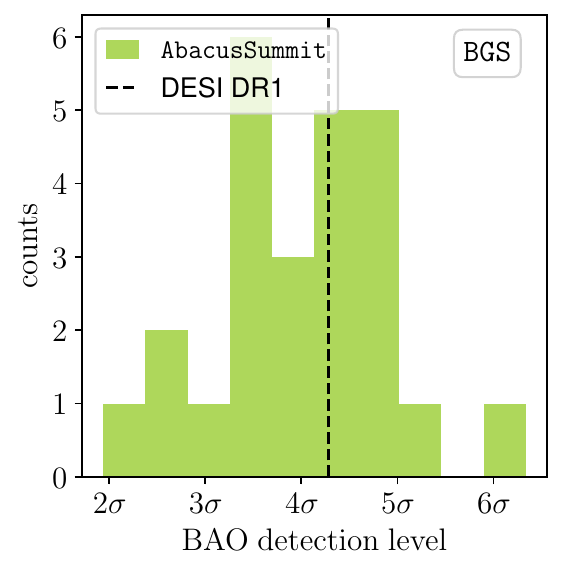} &
        \includegraphics[width=0.3\textwidth]{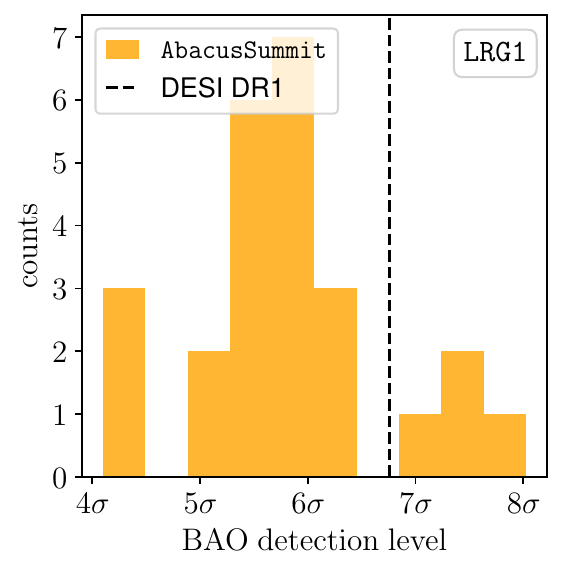} &
        \includegraphics[width=0.3\textwidth]{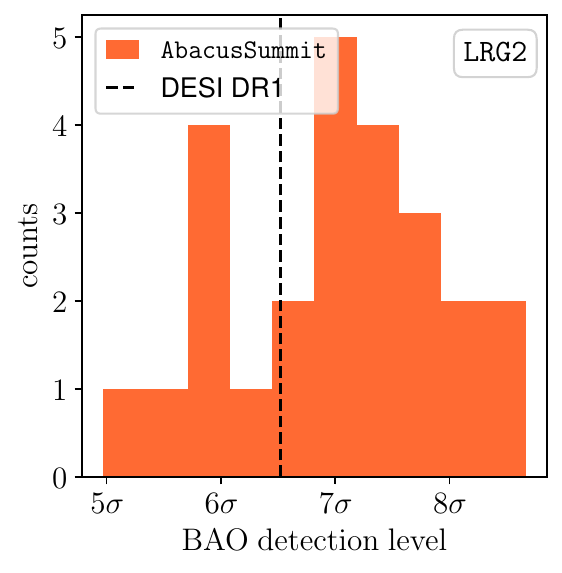}\\
        \includegraphics[width=0.3\textwidth]{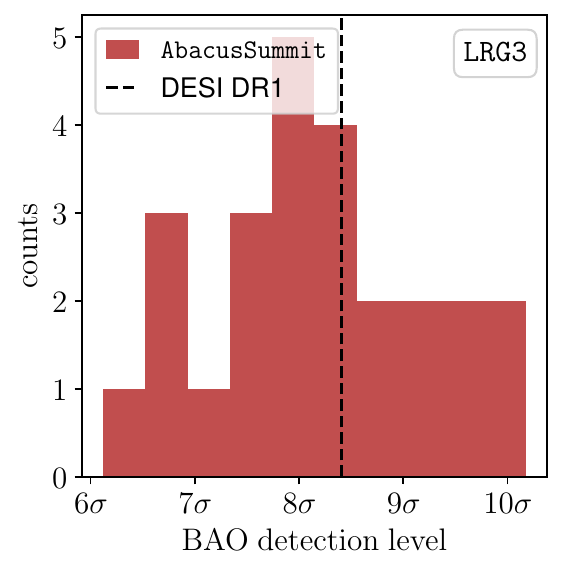} &
        \includegraphics[width=0.3\textwidth]{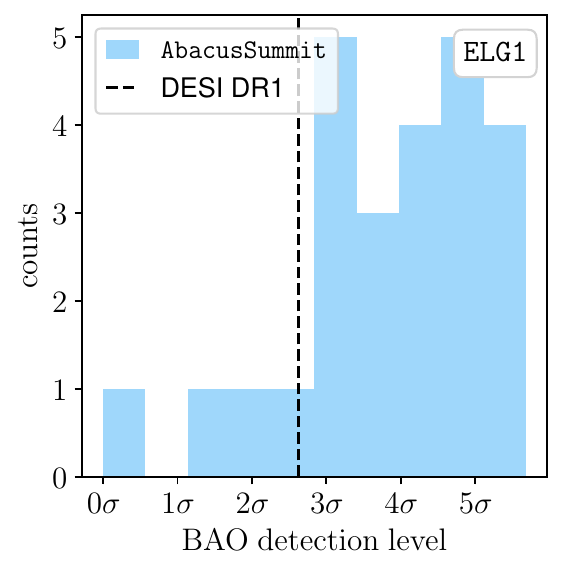} &
        \includegraphics[width=0.3\textwidth]{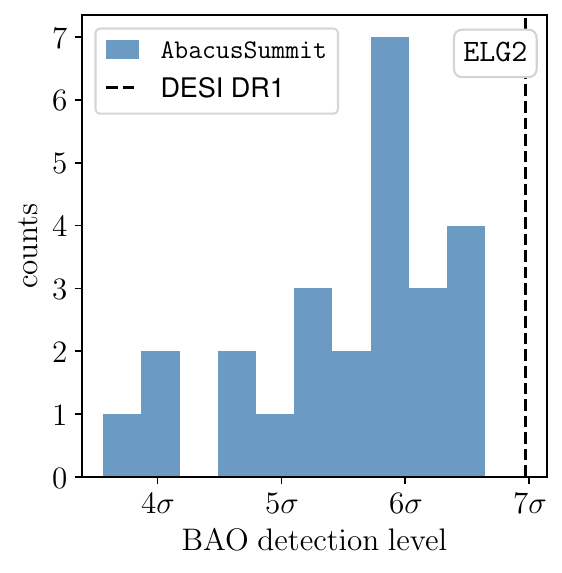}\\
         & \includegraphics[width=0.3\textwidth]{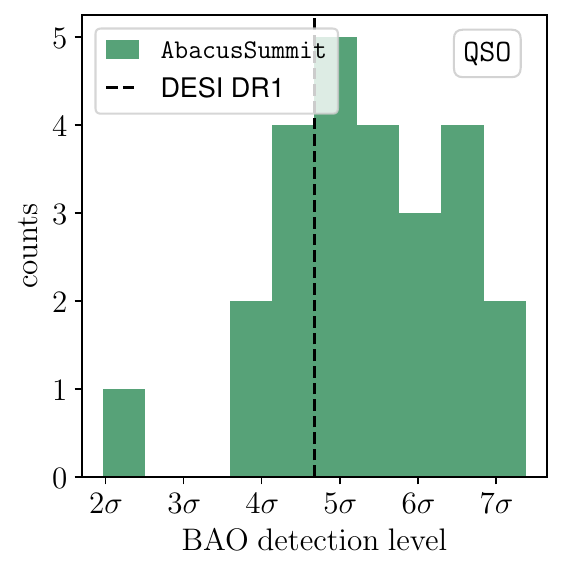} &
    \end{tabular}
    \caption{A comparison of the BAO detection level from the blinded DESI DR1 data (vertical dashed line) and individual realizations of the \abacussummit\ simulations (colored histograms).}
    \label{fig:bao_detection_abacussummit}
\end{figure}

To assess the statistical significance of the BAO detection from the blinded DR1 data, we run $\qiso$ fits and compare the difference in $\chi^2$ at the best fit using models with and without BAO wiggles, where the square root of the $\Delta \chi^2$ is quoted as the detection significance. We display the results in \cref{fig:bao_detection_dr1blinded} for all the different tracers. We obtain the strongest detection level from \lrgth\ at $8.4\sigma$, while the weakest detection occurs for \elgo\ at $2.1\sigma$. To verify that these detection levels are consistent with the mocks, in \cref{fig:bao_detection_abacussummit} show the distribution of detection significance values from fits to individual realizations of the \abacussummit mocks. Overall, we observe great consistency between the mocks and the data for all tracers except \elgt, for which the blinded DR1 data gives a higher detection level than all mock realizations.

\begin{figure}
    \centering
    \includegraphics[width=0.9\textwidth]{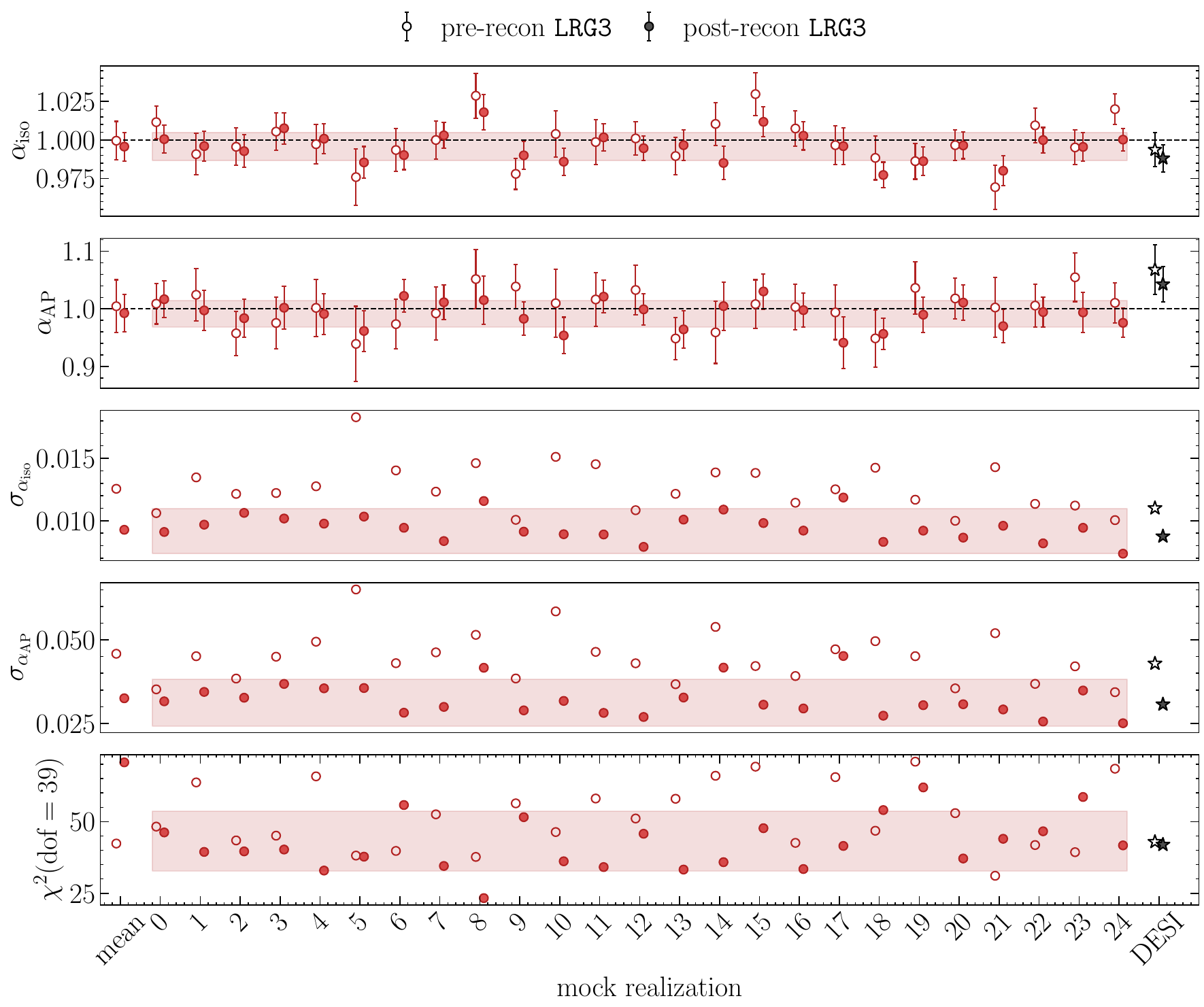}
    \caption{BAO fits of the anisotropic correlation function of the AbacusSummit mocks and the blinded DESI DR1 for \lrgth. Open and filled markers show pre- and post-reconstruction fits, respectively. The first and second panels show constraints on $\qiso$ and $\qap$, respectively, fitting the mean of the 25 mocks (first column, where we have scaled the errors to match that of a single DESI DR1 volume), individual mock realizations (consecutive 25 columns), and the blinded DESI DR1 data (last column). The third and fourth panels show the errors of these fits (equivalent to the size of the error bars from the first two panels). The bottom panel shows the $\chi^2$ value for each fit, where the number of degrees of freedom is specified in the label as a reference. The shaded red region in each panel shows the scatter from the 25 mock fits post-reconstruction.}
    \label{fig:fits_abacussummit_lrg_0.8_1.1}
\end{figure}

\begin{figure*}
    \centering
    \includegraphics[width=0.9\textwidth]{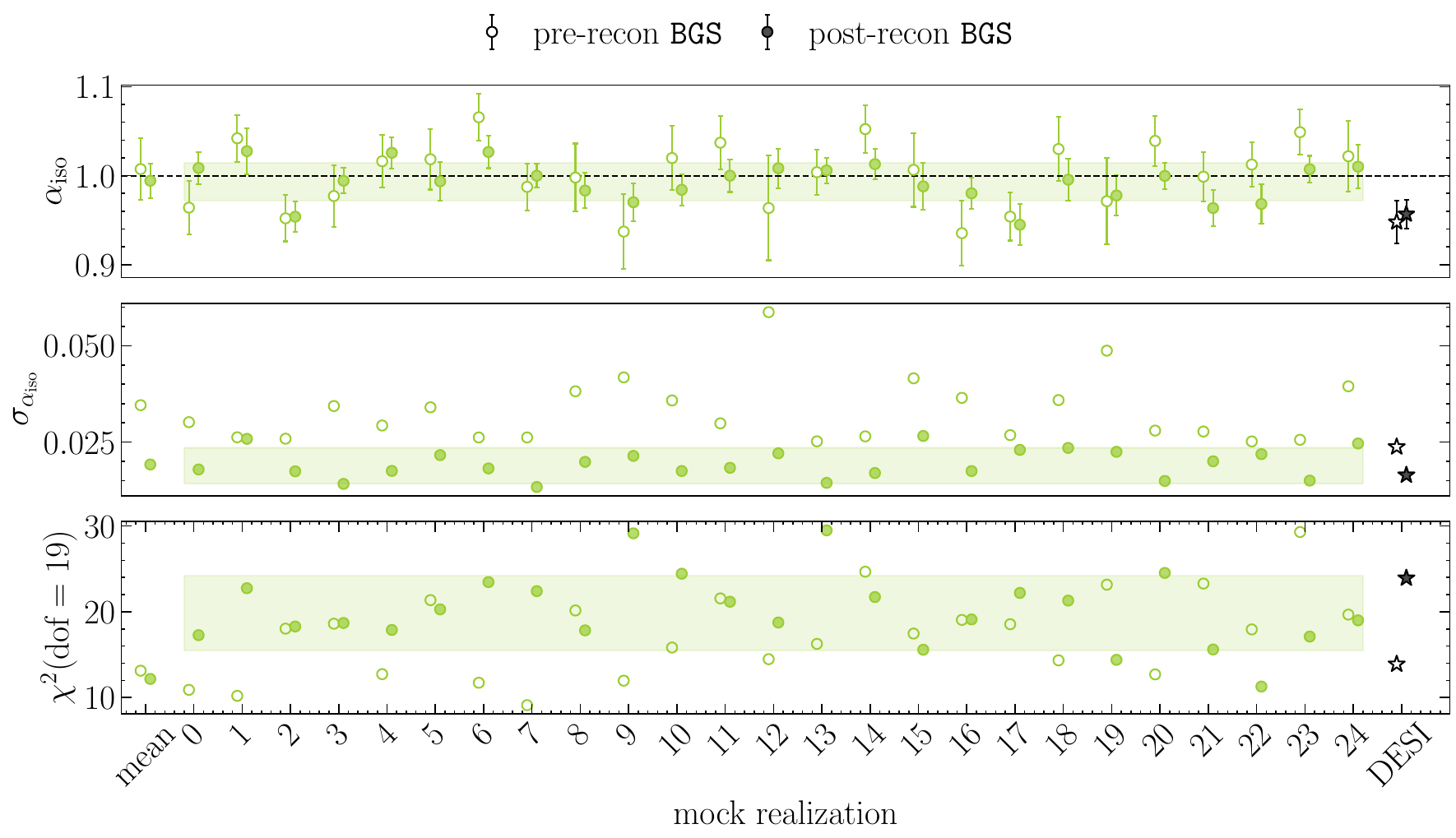}
    \caption{Similar to \cref{fig:fits_abacussummit_lrg_0.6_0.8}, but showing fits to \bgs. Similar to the \qso\ and \elgo, we only perform an isotropic BAO fit for this sample, where $\qiso$ is the only scaling parameter that is varied.}
    \label{fig:fits_abacussummit_bgs_0.1_0.4}
\end{figure*}

In \cref{fig:fits_abacussummit_lrg_0.8_1.1}, we compare the constraints from the blinded DESI \lrgth\ (black stars) with the \abacussummit simulations (colored circles), where we distinguish between pre- and post-reconstruction constraints by the empty and solid markers, respectively. We find excellent agreement between the average post-reconstruction error on $\qiso$ and $\qap$ (red bands) from \abacussummit, and the blinded DESI data. The average $\chi^2$ value per degree of freedom from the mocks also closely follows that of the data, from which we get $\chi^2 = 35$ for 39 degrees of freedom, corresponding to a p-value of 34.7\%. Results for \bgs\ are displayed in \cref{fig:fits_abacussummit_bgs_0.1_0.4}, which shows that the $\qiso$ errros and the $\chi^2$ at the best fit from the blinded data are consistent with the mocks at the $1-\sigma$ level. The $\chi^2$ is 24.5 for 19 degrees of freedom, corresponding to a p-value of 17.8\%. We show equivalent figures for the remaining tracers, where we find similar results in most cases. A noteworthy case are the \qso, for which the blinded DESI DR1 error on $\qiso$ is worse than all mocks after reconstruction, although they are still consistent to within $2\sigma$. For \lrgo, \lrgth, and \elgt, the constraints $\qap$ are not displayed for a few or the \abacussummit\ mock realizations for which noise fluctuations shifted the quadrupole BAO features to scales that are not supported by the data, resulting in fits that hit the prior limit on $\qap$. In these cases, we have instead resorted to 1D BAO fits, thus only providing constraints for $\qiso$.

\subsection{Robustness of the blinded DESI constraints}

\begin{figure*}
    \centering
    \includegraphics[width=\textwidth]{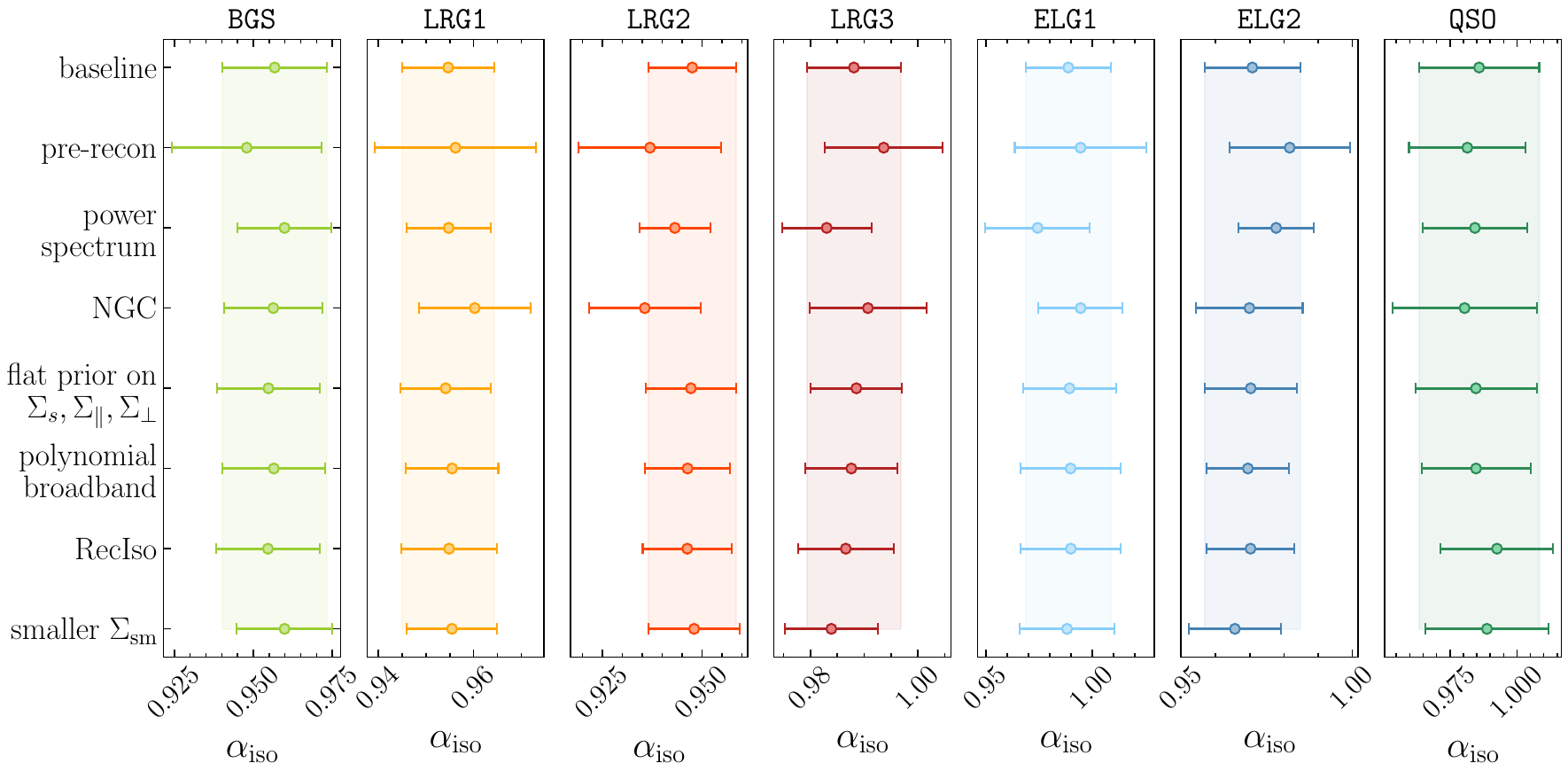}
    \caption{Response of the constraints on the isotropic BAO scaling parameter, $\qiso$, to changes in the data or theory vector, priors, or parametrization. All constraints come from fits to the blinded DESI-DR1 data. The baseline configuration that is used for the main BAO analysis is shown at the top, while the other rows single-variations around the baseline (see main text for a description of each variation). The circles show the best-fit values and the error bars correspond to 68\% confidence levels.}
    \label{fig:whisker_alpha_iso}
\end{figure*}

\begin{figure*}
    \centering
    \includegraphics[width=0.6\textwidth]{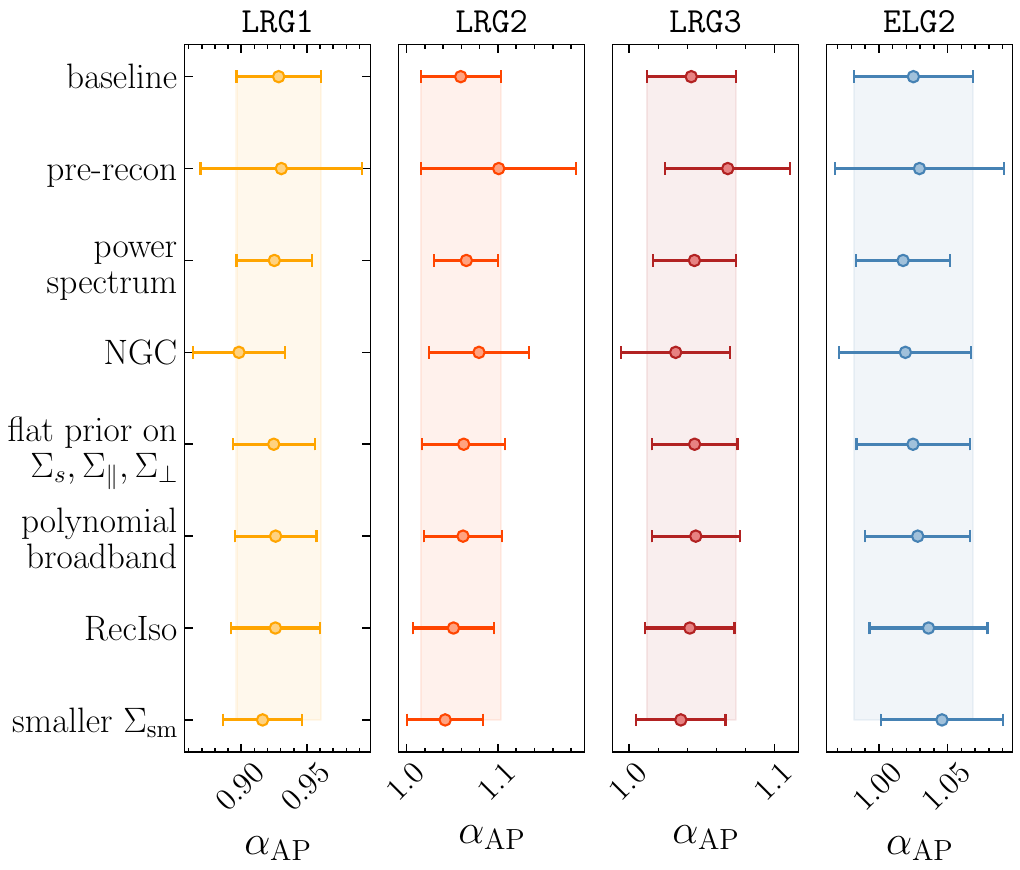}
    \caption{Same as \cref{fig:whisker_alpha_iso}, but showing constraints for the anisotropic BAO scaling parameter, $\qap$.}
    \label{fig:whisker_alpha_ap}
\end{figure*}

\cref{fig:whisker_alpha_iso,fig:whisker_alpha_ap} show how the constraints on $\qiso$ and $\qap$ from the blinded DESI 2024 data respond to different variations in the analysis configuration. To speed-up calculations for this test, we have used posterior maximization with {\tt Minuit} rather than MCMC chains, but we expect the two methods to give comparable results under the assumption of a Gaussian likelihood. For the baseline case, we fit the post-reconstruction correlation function multipoles. For \lrgs\ and \elgt, we simultaneously vary $\qiso$ and $\qap$ and fit the monopole and quadrupole moments. For the remaining samples, we only vary $\qiso$ and exclusively fit the monopole. The BAO damping parameters have Gaussian priors with means and widths as detailed in \cref{tab:priors}. The broadband of the correlation function is parameterized with a piecewise cubic spline basis (\cref{sec:bao_model}).

The second row shows the pre-reconstruction constraints. Reconstruction improves the precision on $\qiso$ by 47\%, 37\% and 23\% for \lrgo, \lrgt, and \lrgth, respectively, by 24\% for \elgt, and by 40\% for \bgs. For \elgo, the pre-reconstruction errors appear smaller than the chain values reported in \cref{tab:fits_desi_blinded}. We have verified that this is due to a noise fluctuation that causes the pre-reconstruction marginalized posterior on $\qiso$ to appear highly non-Gaussian, making the standard deviation from the chains an unsuitable measure of the error. Measuring the region at which $\Delta \chi^2 = 1$, we obtain $\qiso = 0.987 \pm 0.031$, while reconstruction improves the error by 29\%. For \qso, reconstruction ends up slightly degrading the precision on $\qiso$, by 8.9\%. The tests on the mocks from \cref{fig:smoothing_scales_qiso_3} in \cref{ap:additional figures} show that even though reconstruction tends to help improve the errors on the scaling parameters for \qso-like samples on average, there are a few mock realizations for which the errors get worse, which is consistent with what is found for the DESI blinded data. This is also in line with previous analyses by eBOSS, which found that reconstruction did not improve the BAO constraints from their QSO sample due to the high shot noise contamination \citep{HouJiaminThesis}. For $\qap$, the errors after reconstruction improve by 54\%, 37\%, 27\%, and 28\% for \lrgo, \lrgt, \lrgth, and \elgt, respectively.

The Fourier-space constraints from the power spectrum multipoles, which are displayed on the third row, show a good agreement with the configuration-space results, with comparable error bars and offsets that are never larger than $1\sigma$. In the fourth row, we display the constraints obtained by only fitting data from the northern galactic cap (NGC), in contrast with the baseline scenario where the northern and southern galactic caps are combined at the clustering level. In most cases, we only see a small degradation in the parameter constraints. We note that for dark-time tracers, the DR1 NGC has a factor of $\sim 1.74$ larger sky area than the SGC, while for bright-time tracers, the ratio is $\sim 2.38$ \citep{DESI2024.II.KP3}.

The fifth row shows the resulting parameter constraints when we adopt uniform priors $\unif(0, 10)$ for the BAO damping parameters. In general, we see that the results are almost equivalent to the baseline case, where we adopt Gaussian priors. Tests on mock galaxy catalogs by \cite{KP4s2-Chen} have shown that using non-informative priors on the damping parameters for data with a low SNR can potentially lead to weaker constraints and even potential biases. Indeed, as we have previously discussed, some tracers and redshift bins, such as the \elgo, do not have a sufficiently large SNR to reliably perform an anisotropic BAO fit, and we have explicitly checked that when doing so, the priors on the damping parameters have a strong impact on the recovered best-fit values and precision of the BAO scaling parameters. However, these low-SNR cases are not included in the main DESI BAO analysis, and the constraints from the remaining target samples are robust against this choice.

Previous BAO analyses from BOSS \cite{Alam2017} and eBOSS \cite{eboss2020} parameterized the broadband component of the multipole correlation function / power spectrum using polynomials of slightly varying degrees and functional forms, with coefficients that vary freely during the fit. In this work, we have adopted the new spline-basis parameterization for the broadband proposed by \cite{KP4s2-Chen}, which gives constraints that are in excellent agreement with those obtained with the (e)BOSS polynomial parameterization, shown in the fifth row of \cref{fig:whisker_alpha_iso,fig:whisker_alpha_ap}. 

One of the main differences with respect to previous applications of reconstruction in SDSS in the use of the \recsym\ convention. As discussed in \cref{subsec:reconstruction}, \recsym\ shifts galaxies and randoms in the same way, using the full displacement vector that includes the $(1 + f)$ factor along the line of sight, which preserves large-scale RSD in the post-reconstruction clustering. The convention used in (e)BOSS was \reciso, where the $(1 + f)$ factor is not applied when displacing the randoms, which has the effect of erasing RSD post-reconstruction. This is exemplified in \cref{fig:recsym_vs_reciso}, where we see how the quadrupole moment of the \lrgt\ is strongly suppressed in \reciso. The right-hand side panel shows how the distribution of displacements along the line of sight is larger for \recsym\ due to the additional $(1 + f)$ factor. Despite this important difference, it can be seen that the BAO features in the correlation function are largely preserved. The last row in \cref{fig:whisker_alpha_iso,fig:whisker_alpha_ap} show the constraints derived using the \reciso\ convention. Since the spline-basis broadband parametrization has only been validated for \recsym, we run this test using the BOSS polynomial parameterization instead, which was shown to give consistent results with the spline method for \recsym. The BAO scaling parameters are largely unaffected by the reconstruction mode, with the largest shift being less than $0.09 \sigma$ for the QSO. 

\begin{figure}
    \centering
    \begin{tabular}{cc}
    \includegraphics[width=0.46\textwidth]{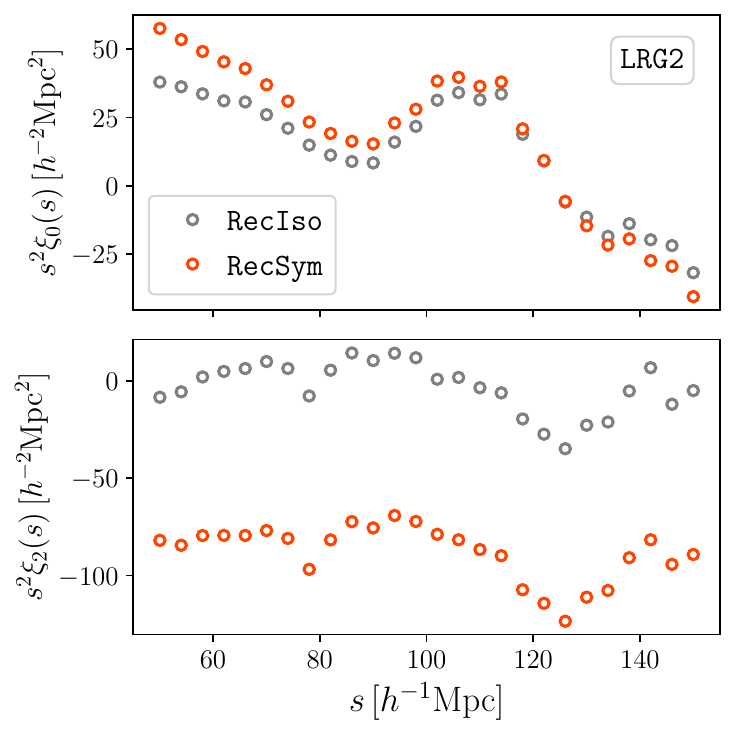}     & \includegraphics[width=0.48\textwidth]{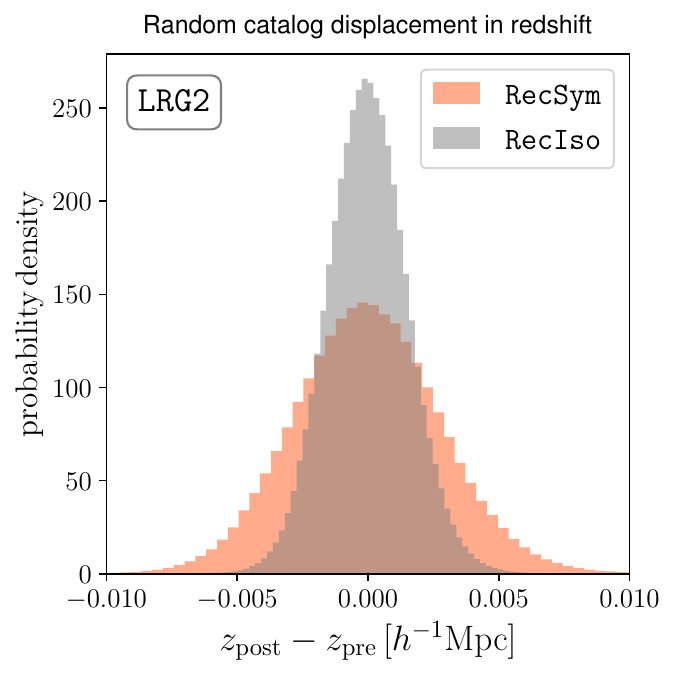}
    \end{tabular}
    \caption{Density-field reconstruction applied to the blinded DESI \lrgt\ data using the \recsym\ or \reciso\ conventions. In the left, we show the monopole and quadrupole moments of the correlation function, and in the right, we show the distribution of displacement vectors for the random catalogue along the radial direction, given by the difference in the redshift of the random particles before and after reconstruction.}
    \label{fig:recsym_vs_reciso}
\end{figure}

Finally, in the last rows we include the alternative choice of smoothing scale that was tested in \cref{subsec:smoothing_scale}. Similar to what was found with the mocks, the resulting BAO constraints are only mildly affected by this choice, with changes in the recovered best-fit values on $\qiso$ and $\qap$ that are smaller than $0.2\sigma$. We note, however, than this is due to the fact that in this paper we have narrowed down our options of smoothing scales to only two choices per tracer with values that are relatively close to each to other, based on more extensive tests that are presented in \cite{KP4s3-Chen}. More extreme values of the smoothing scale are expected to have stronger impacts on the BAO constraints, as has been also shown in previous works \citep{White2010:1004.0250, Burden2014:1408.1348, Anderson2014:1303.4666, Vargas-Magaña:1509.06384}

\subsection{Gaussianity of the BAO posterior}

\begin{figure}
    \centering
    \begin{tabular}{cc}
    \includegraphics[width=0.45\textwidth]{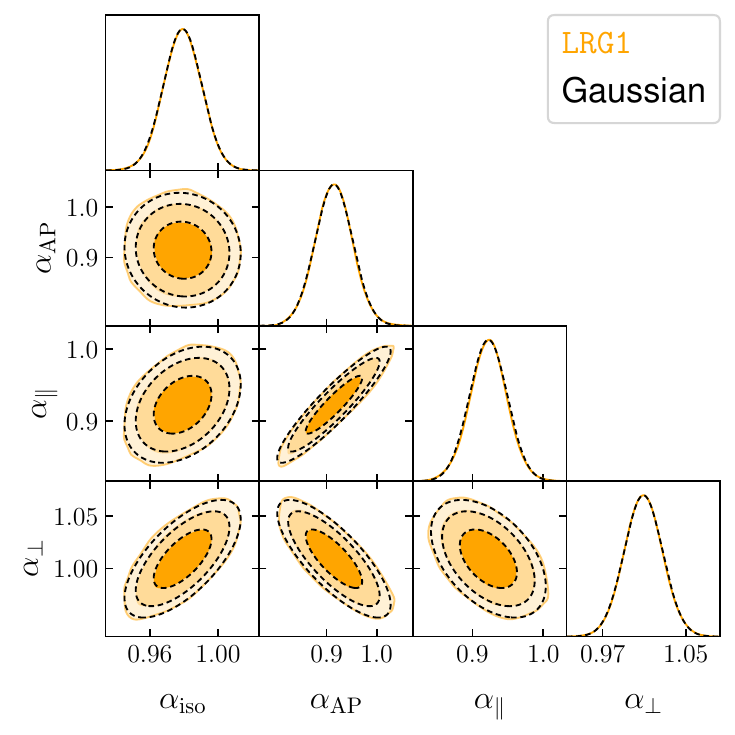} & \includegraphics[width=0.45\textwidth]{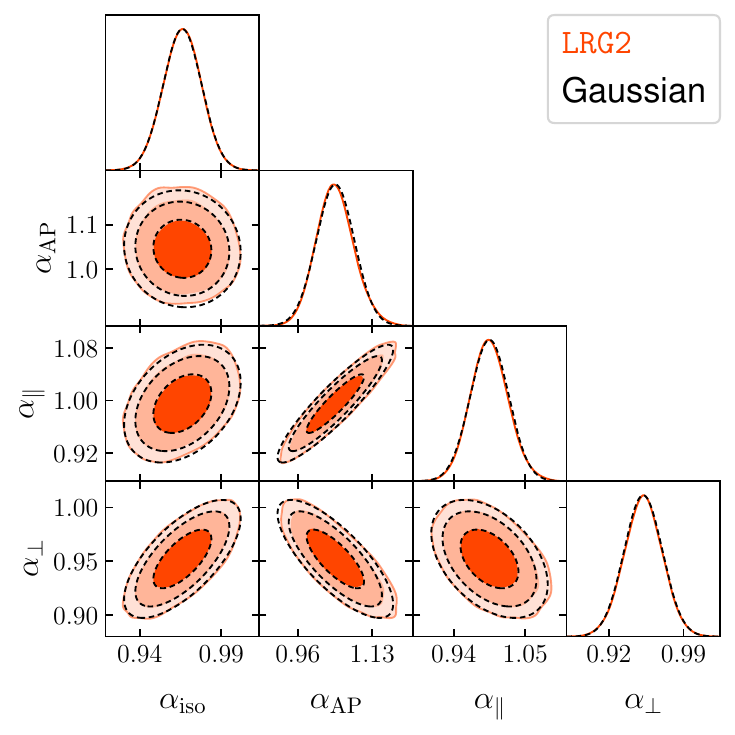}\\
    \includegraphics[width=0.45\textwidth]{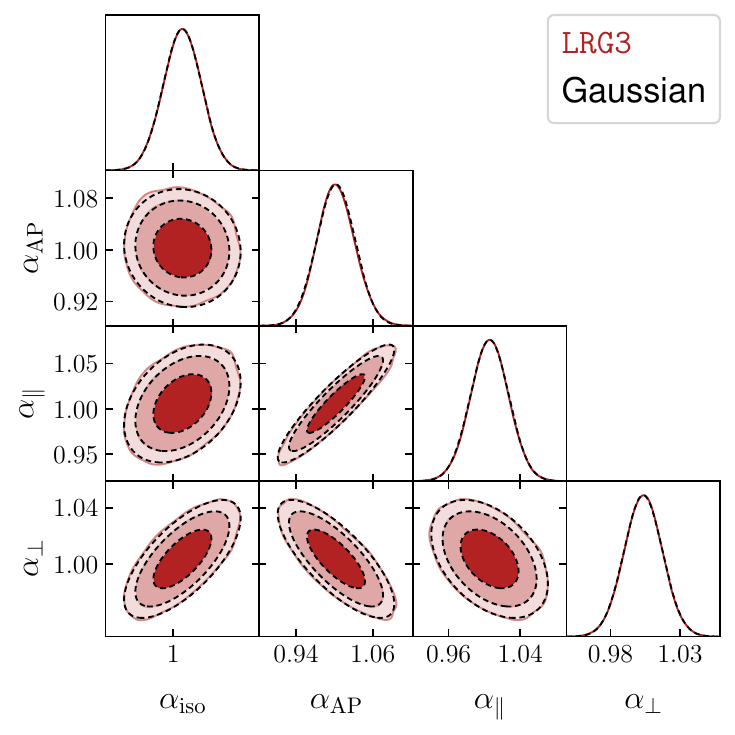} & \includegraphics[width=0.45\textwidth]{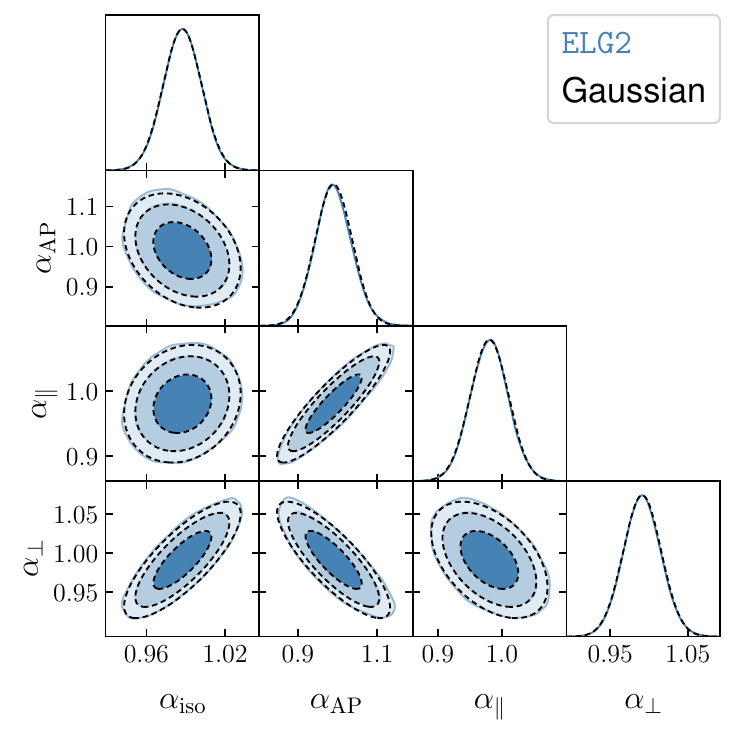}
    \end{tabular}
    \caption{A comparison of the BAO posterior form the blinded \lrgs\ and \elgt\ (filled contours in color), and a multivariate Gaussian distribution with the same mean and covariance (black contours in dashed line style). We sample the parameter space in the $\qiso$-$\qap$ basis, and report $\qpar$ and $\qper$ as derived parameters.}
    \label{fig:gaussianity_2d}
\end{figure}

\begin{figure}
    \centering
    \begin{tabular}{ccc}
    \includegraphics[width=0.3\textwidth]{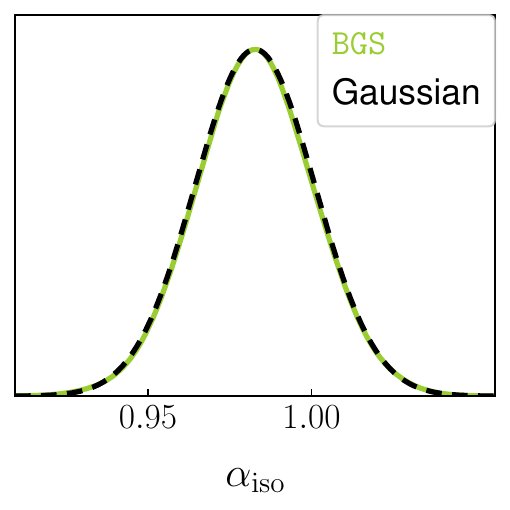} & \includegraphics[width=0.3\textwidth]{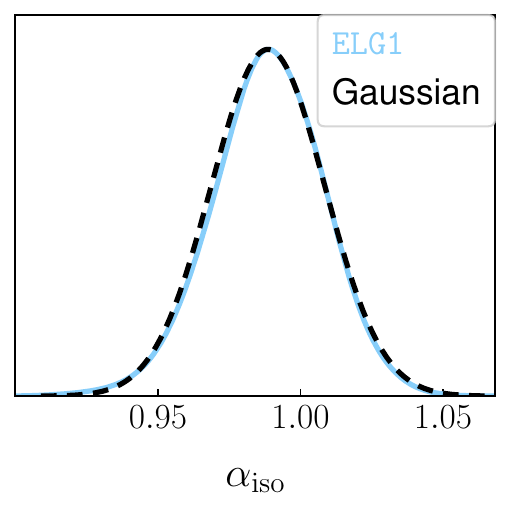} &
    \includegraphics[width=0.3\textwidth]{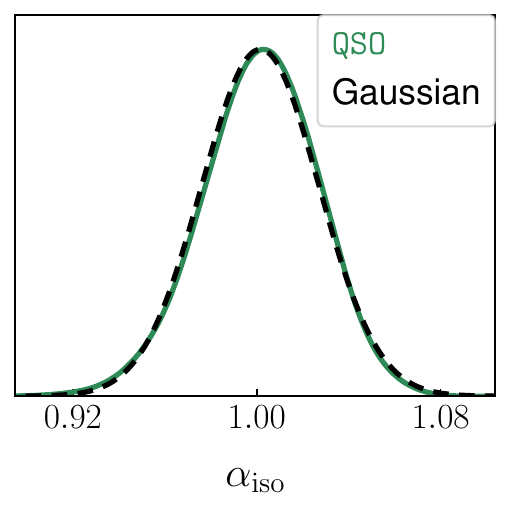} 
    \end{tabular}
    \caption{Similar to \cref{fig:gaussianity_2d}, but showing results for the 1D BAO fits on \bgs, \elgo, and \qso.}
    \label{fig:gaussianity_1d}
\end{figure}

As a final sanity check, we assess whether the posterior distribution of BAO parameters from the blinded DESI data follows a Gaussian distribution, which would guarantee that the mean and covariance of the posterior capture all the information from the fit.

In \cref{fig:gaussianity_2d}, we show the posterior of the BAO scaling parameters from fits to the post-reconstruction correlation functions of \lrgs\ and \elgt, comparing them with a multivariate Gaussian distribution with the same mean and covariance. The contours, which correspond to 1-$\sigma$, 2-$\sigma$, and 3-$\sigma$ levels, show an excellent consistency, indicating that the BAO posterior closely follows the Gaussian approximation. For \lrgs, we notice that there is almost no correlation between $\qiso$ and $\qap$, which is also indicated by the small cross-correlation coefficients seen in \cref{tab:fits_desi_blinded}, which are -0.12, -0.09, and -0.03 for \lrgo, \lrgt, and \lrgth, respectively. For \elgt, we observe a higher correlation coefficient of -0.31. We have also derived constraints for $\qpar$ and $\qper$, following \cref{eq:qiso,eq:qap}. We find that the constraints are much more correlated on this basis, with a correlation coefficient between $\qpar$ and $\qper$ of -0.45 for all tracers. However, the posterior appears equally Gaussian under both parameterizations.

For completeness, we also show the equivalent figures for the 1D BAO fits performed on \bgs, \elgo\, and \qso\ in \cref{fig:gaussianity_1d}, which shows that the $\qiso$ posteriors are also consistent with a Gaussian distribution in all cases.

Although we have performed some standardized tests of multivariate normality, such as the Shapiro-Wilk test \citep{Shapiro-Wilk1965}, we found the results to be very sensitive to the number of chain points in the fits. In fact, in the limit of infinite chain points, any test of normality will reject the null hypothesis, except for a problem that is truly Gaussian by construction. Although the assumption of a Gaussian likelihood has been the standard for the cosmological interpretation of the BAO constraints \cite{Alam2017, eboss2020}, the ever-increasing level of precision from spectroscopic galaxy surveys will require more quantitative assessments of the impact of this assumption at the cosmological parameter level. We plan to explore this further in future work.

\section{Summary and Conclusions}

We have studied the impact of the density-field reconstruction technique on the DESI DR1 galaxy BAO analysis. Reconstruction is a method that aims at improving the signal-to-noise ratio of the BAO signal from galaxy clustering data, improving its detection significance, and therefore increasing the precision of cosmological constraints derived from BAO. In addition, reconstruction helps to remove systematic shifts in the observed location of the BAO feature that are caused by non-linear structure growth and biasing.  

To assess the effectiveness of reconstruction in a controlled setting, we have used two suites of mock galaxy catalogs that match the clustering and selection properties of the DESI DR1 target samples: the \abacussummit\ mocks, which correspond to high-fidelity HOD catalogs generated from the suite of N-body simulations presented in \cite{Maksimova2021}, and a collection of 1000 approximate mocks generated with the {\tt EZmock} algorithm \citep{Chuang2015:1409.1124}. We have also performed a series of tests on the DESI DR1 blinded data, assessing the robustness of the BAO constraints against different assumptions in our analysis configuration. In this context, the blinding of the data refers to the deliberate modification of the large-scale structure catalogs that are analyzed while the BAO pipeline is being tested and calibrated, in order to avoid confirmation bias in the cosmological analysis.  A detailed description of the blinding scheme in DESI can be found in \cite{KP3s9-Andrade}.

Our main findings can be summarized as follows.

\begin{itemize}
    \item[-] Except for the \qso\ sample, reconstruction ubiquitously helps to increase the signal-to-noise ratio of the BAO feature for the blinded DESI data, increasing the precision of the recovered on the isotropic BAO scaling parameter $\qiso$ from 23\% to 47\%, depending on the tracer and redshift range. This increase in precision is consistent with results obtained from the mock galaxy catalogs.

    \item[-] For \qso, while the lack of improvement from the reconstruction of the blinded DESI data is not unusual due to the high shot noise contamination of this sample, we find that in the majority of the cases, reconstruction increases the recovered precision on $\qiso$ from the \qso\ mocks. This motivates the decision to adopt reconstruction as the default for all samples of the DESI galaxy BAO analysis, including \qso.
    
    \item[-] We calibrated the optimal smoothing scale at which we smooth the density field for reconstruction, in order to null out small-scale modes that are highly non-linear. Among the options we tested, we find that the reconstruction of the DESI DR1 samples is not particularly sensitive to the choice of smoothing scale, and values close to those adopted by previous SDSS BAO analyses are suitable in most cases. The \qso\ sample is an exception due to the higher shot noise contribution, and larger smoothing scales are required to optimally reconstruct the density field. We adopt $\Sigmasm = 15 \Mpch$ for \bgs, \lrgs, and \elgs, and $\Sigmasm = 30 \Mpch$ for the \qso, as our fiducial smoothing scales for the main galaxy BAO analysis.
    
    \item[-] Using mocks that have been passed through a pipeline that mimics the fiber assignment scheme of DESI DR1, we have tested the impact of the fiber assignment incompleteness on the BAO constraints before and after reconstruction. Overall, we do not find statistically significant signatures from this effect on the recovered constraints on the BAO scaling parameters, regardless of whether the galaxy catalogs are reconstructed or not.

    \item[-] We have performed a series of unblinding tests that the DESI Collaboration used to assess whether we were ready to fix the galaxy BAO pipeline, unblind the LSS catalogs, and perform the official DR1 BAO analysis. These tests analyzed the consistency between the BAO statistics measured from the blinded DESI data and the \abacussummit\ mocks, as well as the robustness of the constraints from the blinded data to different configurations in our pipeline.
    
    \item[-] The BAO constraints from the DESI blinded data are robust against a wide variety of measurement and model choices, including the parameterization of the broadband component of the power spectrum, the priors on the parameters that capture the BAO non-linear damping, and different reconstruction conventions. We also find excellent consistency between the constraints derived from configuration space (the correlation function) and Fourier space (the power spectrum).

    \item[-] After reconstruction, the posteriors of the BAO scaling parameters closely follow a multivariate Gaussian distribution, regardless of whether the scaling parameters are expressed in the basis that isolates the isotropic and anisotropic scaling components ($\qiso$ and $\qap$), or the perpendicular and parallel to the line of sight components ($\qpar$ and $\qper$).
\end{itemize}

In this work, we have focused on a single reconstruction algorithm that uses an iterative Fast Fourier Transforms scheme to recover the real-space displacement field, as introduced in \cite{Burden2014:1408.1348, Burden2015:1504.02591}. In \cite{KP4s3-Chen}, we present a thorough comparison of different reconstruction algorithms in the context of BAO analyses. Our companion papers also present detailed studies of various potential sources of uncertainty in our BAO constraints, including: the choice of fiducial cosmology \cite{KP4s9-Perez-Fernandez}, the galaxy-halo connection modeling \citep{KP4s11-Garcia-Quintero, KP4s10-Mena-Fernandez}, the calibration of the covariance matrix \cite{KP4s8-Alves, KP4s7-Rashkovetskyi}, and the theory model \cite{KP4s2-Chen}. The unblinded galaxy BAO measurements from DESI are presented in \cite{DESI2024.III.KP4}, while the Lyman-$\alpha$ forest BAO measurements are described in \citep{DESI2024.IV.KP6}. Both of these sets are combined to infer cosmology in \citep{DESI2024.VI.KP7A}.

The tests presented in this work and in our companion papers are a further confirmation that BAO remain as one of the most robust probes of large-scale structure cosmology to date, even under the unprecedented statistical precision of the DESI DR1 sample. However, future data releases from DESI, as well as  other on-going and upcoming galaxy surveys, such as Euclid \citep{euclid} and the Nancy Grace Roman Space Telescope \cite{roman}, will require ever more stringent tests and validation, which are likely to push the limits of our understanding of cosmology and large-scale structure science.

\section{Data Availability}
The data will be publicly available once the paper is accepted for publication.

\acknowledgments
ZD acknowledges support from the National Key R\&D Program of China (2023YFA1607800, 2023YFA1607802), the National Science Foundation of China (grant numbers 12273020 and 11621303), and the science research grant from the China Manned Space Project with NO. CMS-CSST-2021-A03. H-JS acknowledges support from the U.S. Department of Energy, Office of Science, Office of High Energy Physics under grant No. DE-SC0023241. H-JS also acknowledges support from Lawrence Berkeley National Laboratory and the Director, Office of Science, Office of High Energy Physics of the U.S. Department of Energy under Contract No. DE-AC02-05CH1123 during the sabbatical visit.
SN acknowledges support from an STFC Ernest Rutherford Fellowship, grant reference ST/T005009/2.

This material is based upon work supported by the U.S. Department of Energy (DOE), Office of Science, Office of High-Energy Physics, under Contract No. DE–AC02–05CH11231, and by the National Energy Research Scientific Computing Center, a DOE Office of Science User Facility under the same contract. Additional support for DESI was provided by the U.S. National Science Foundation (NSF), Division of Astronomical Sciences under Contract No. AST-0950945 to the NSF’s National Optical-Infrared Astronomy Research Laboratory; the Science and Technology Facilities Council of the United Kingdom; the Gordon and Betty Moore Foundation; the Heising-Simons Foundation; the French Alternative Energies and Atomic Energy Commission (CEA); the National Council of Humanities, Science and Technology of Mexico (CONAHCYT); the Ministry of Science and Innovation of Spain (MICINN), and by the DESI Member Institutions: \url{https://www.desi.lbl.gov/collaborating-institutions}. Any opinions, findings, and conclusions or recommendations expressed in this material are those of the author(s) and do not necessarily reflect the views of the U. S. National Science Foundation, the U. S. Department of Energy, or any of the listed funding agencies.

The authors are honored to be permitted to conduct scientific research on Iolkam Du’ag (Kitt Peak), a mountain with particular significance to the Tohono O’odham Nation.

\bibliographystyle{JHEP}
\bibliography{references, DESI2024}

\appendix

\section{Additional Figures} \label{ap:additional figures}

In this section, we show additional figures for some of the tests carried out in \cref{sec:results}, including the results for those tracers that we did not add to the main body of the manuscript. These additional results are still discussed in the main text, but we reserve the placement of the figures to this appendix for formatting convenience. This includes the \abacussummit\ $\qiso$ constraints for different reconstruction smoothing scales for \lrgo\ and \lrgth\ (\cref{fig:smoothing_scales_qiso_2}), and for \bgs, \elgo, and \qso\ (\cref{fig:smoothing_scales_qiso_3}). \cref{fig:smoothing_scales_qap_2} shows the \abacussummit\ constraints on $\qap$ for different smoothing scales from fits to the \lrgo\ and \lrgth\ samples. \cref{fig:fits_abacussummit_lrg_0.4_0.6,fig:fits_abacussummit_lrg_0.6_0.8,fig:fits_abacussummit_ELG_0.8_1.1,fig:fits_abacussummit_ELG_1.1_1.6,fig:fits_abacussummit_QSO_0.8_2.1} show the summary of pre- and post-reconstruction constraints from the \abacussummit\ mocks and the DESI DR1 blinded data, for the \lrgo, \lrgt, \elgs, and \qso\ samples.

\begin{figure}
    \centering
    \begin{tabular}{c}
     \hspace*{-0.5cm}\includegraphics[width=0.9\textwidth]{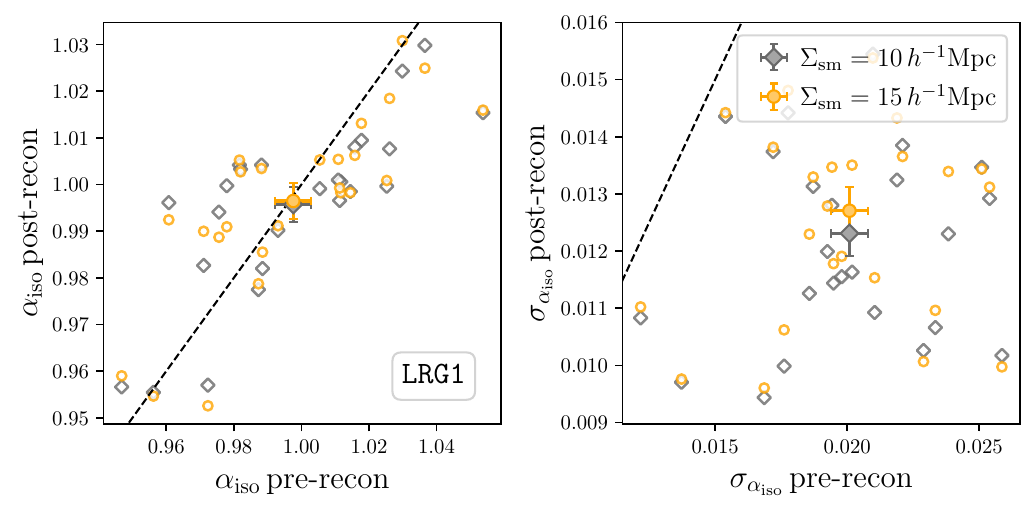}\\
     \hspace*{-0.5cm}\includegraphics[width=0.9\textwidth]{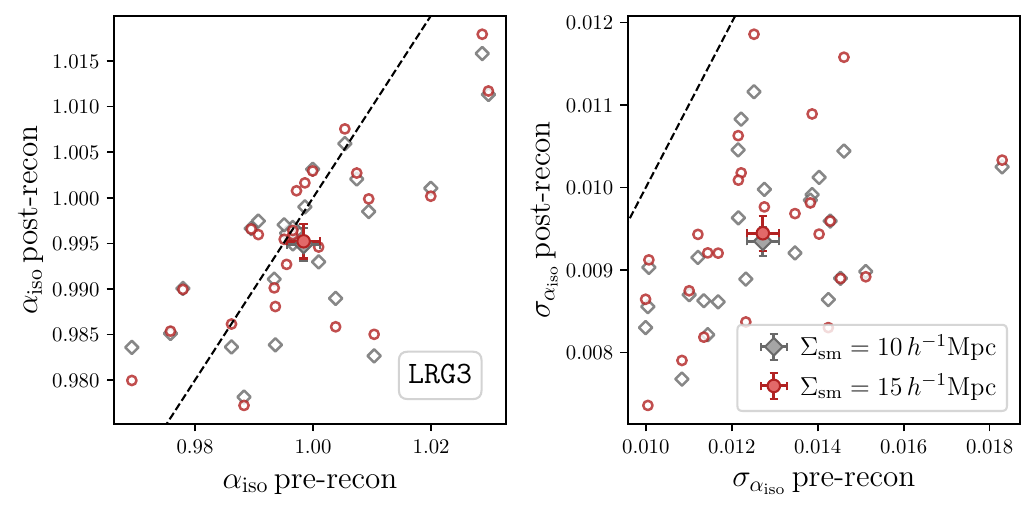}
    \end{tabular}
    \caption{Similar to \cref{fig:smoothing_scales_qiso}, but showing results for \lrgo\ and \lrgth.}
    \label{fig:smoothing_scales_qiso_2}
\end{figure}

\begin{figure}
    \centering
    \begin{tabular}{c}
     \hspace*{-0.5cm}\includegraphics[width=0.9\textwidth]{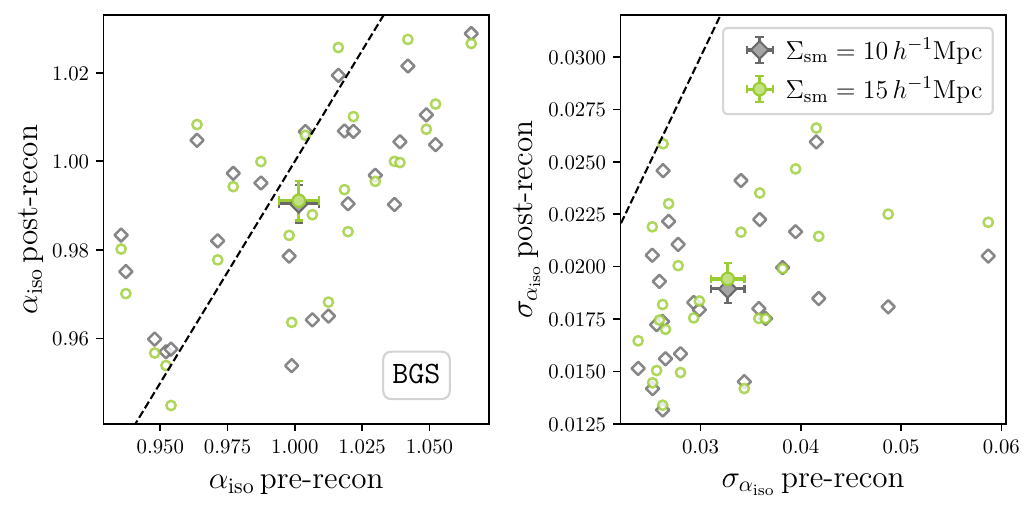}\\
     \hspace*{-0.5cm}\includegraphics[width=0.9\textwidth]{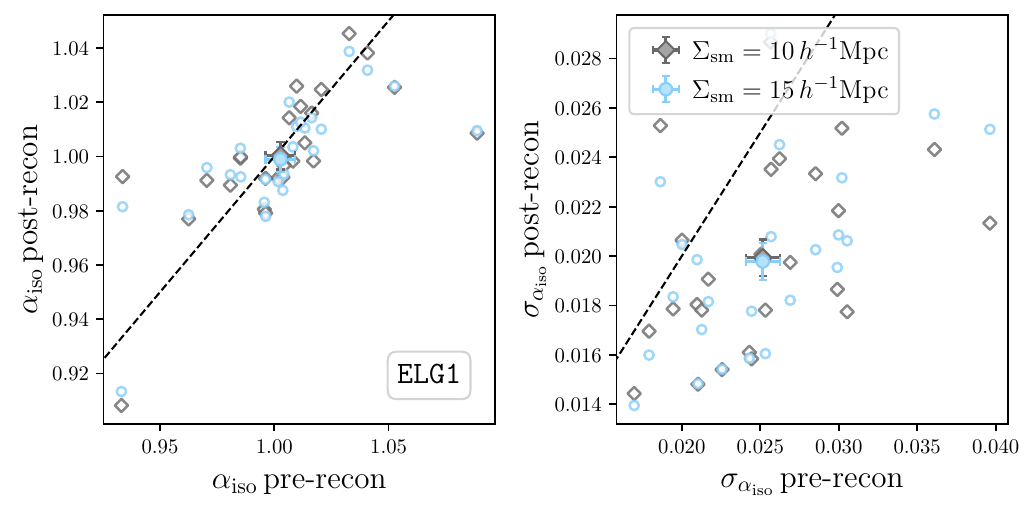}\\
     \hspace*{-0.5cm}\includegraphics[width=0.9\textwidth]{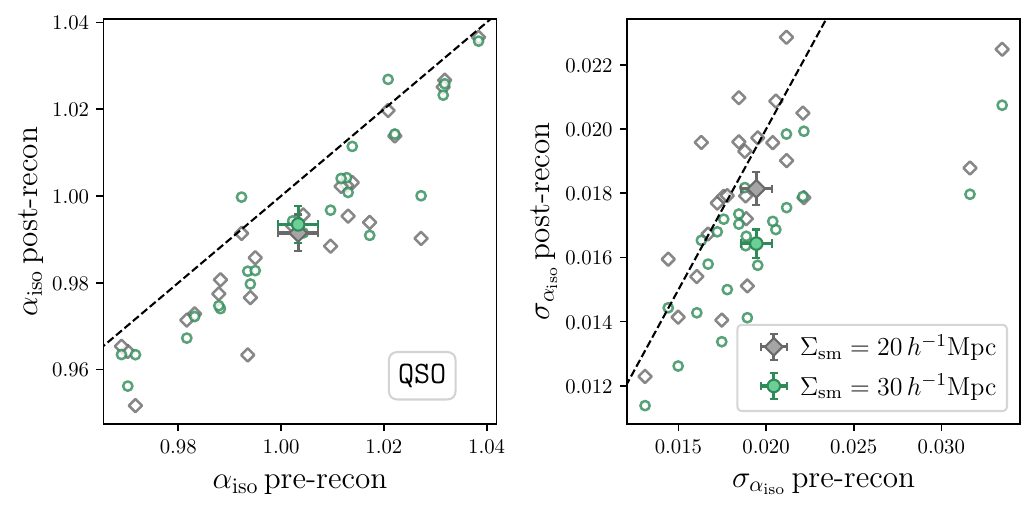}
    \end{tabular}
    \caption{Similar to \cref{fig:smoothing_scales_qiso}, but showing results for \bgs, \qso, and \elgo.}
    \label{fig:smoothing_scales_qiso_3}
\end{figure}

\begin{figure}
    \centering
    \begin{tabular}{c}
     \hspace*{-0.5cm}\includegraphics[width=0.9\textwidth]{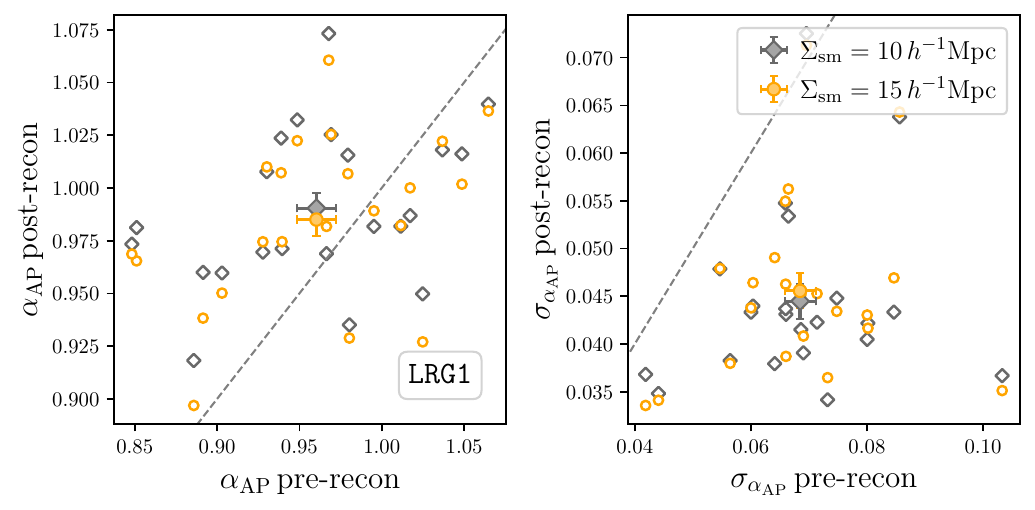}\\
     \hspace*{-0.5cm}\includegraphics[width=0.9\textwidth]{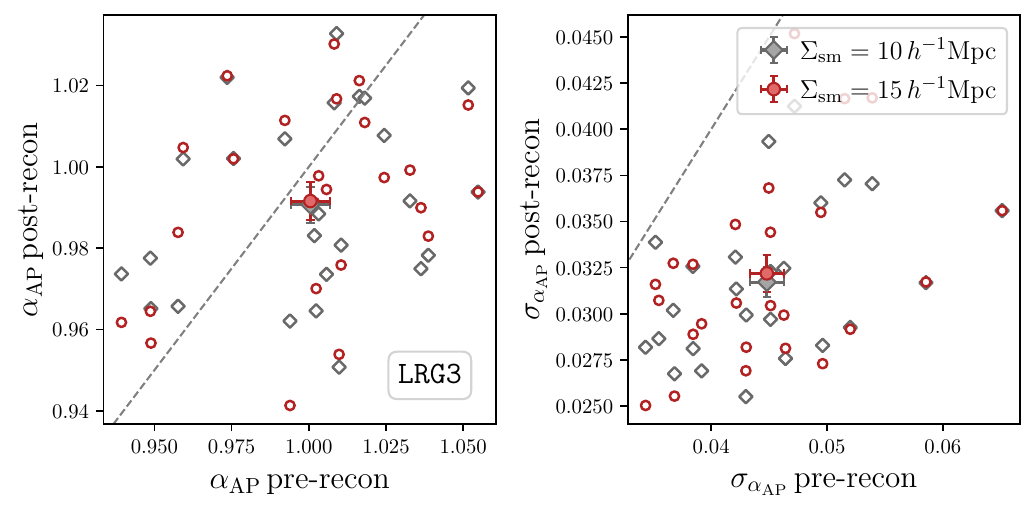}
    \end{tabular}
    \caption{Similar to \cref{fig:smoothing_scales_qap}, but showing results for \lrgo\ and \lrgth.}
    \label{fig:smoothing_scales_qap_2}
\end{figure}

\begin{figure}
    \centering
    \includegraphics[width=0.9\textwidth]{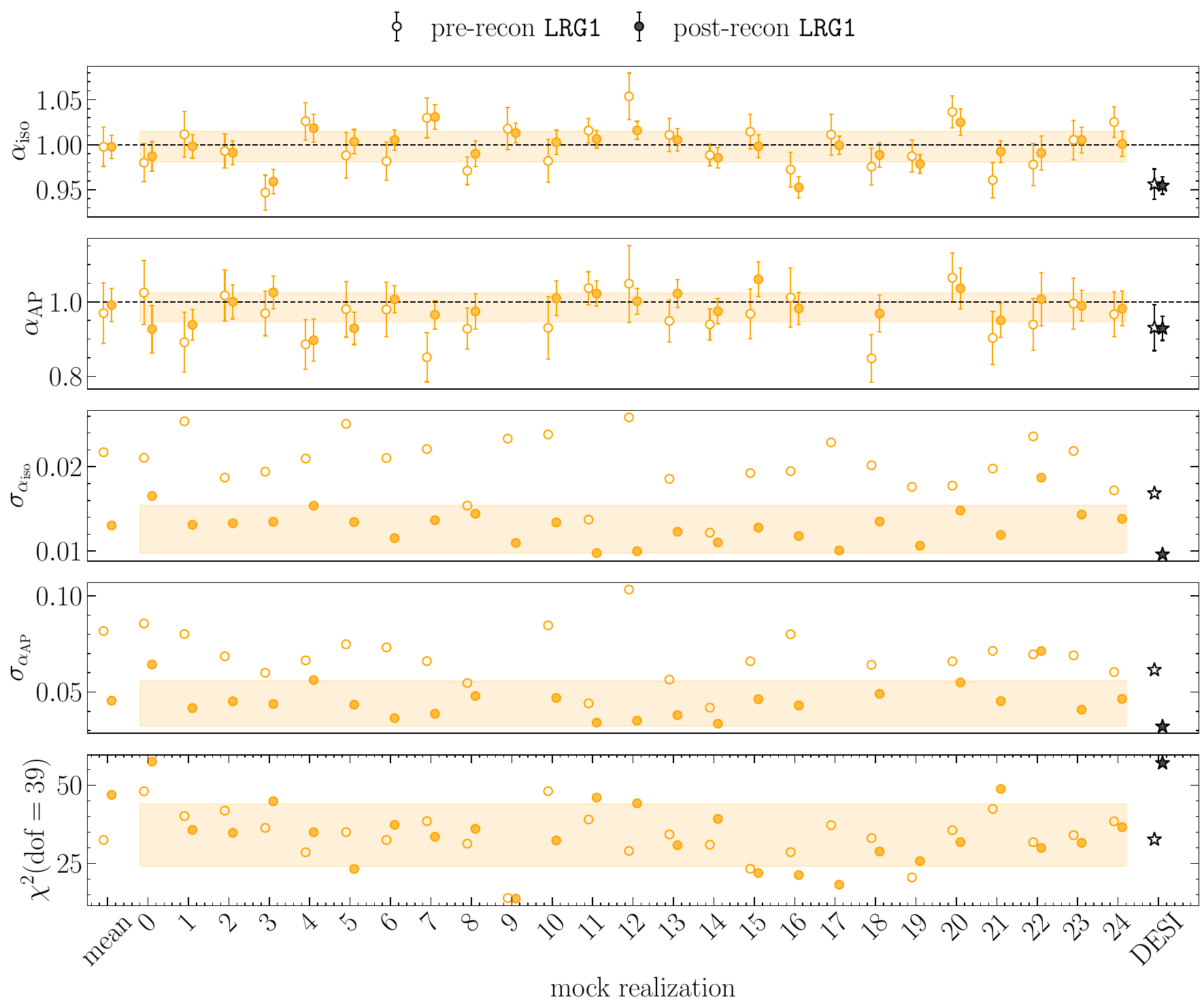}
    \caption{Similar to \cref{fig:fits_abacussummit_lrg_0.8_1.1}, but for \lrgo.}
    \label{fig:fits_abacussummit_lrg_0.4_0.6}
\end{figure}

\begin{figure}
    \centering
    \includegraphics[width=0.9\textwidth]{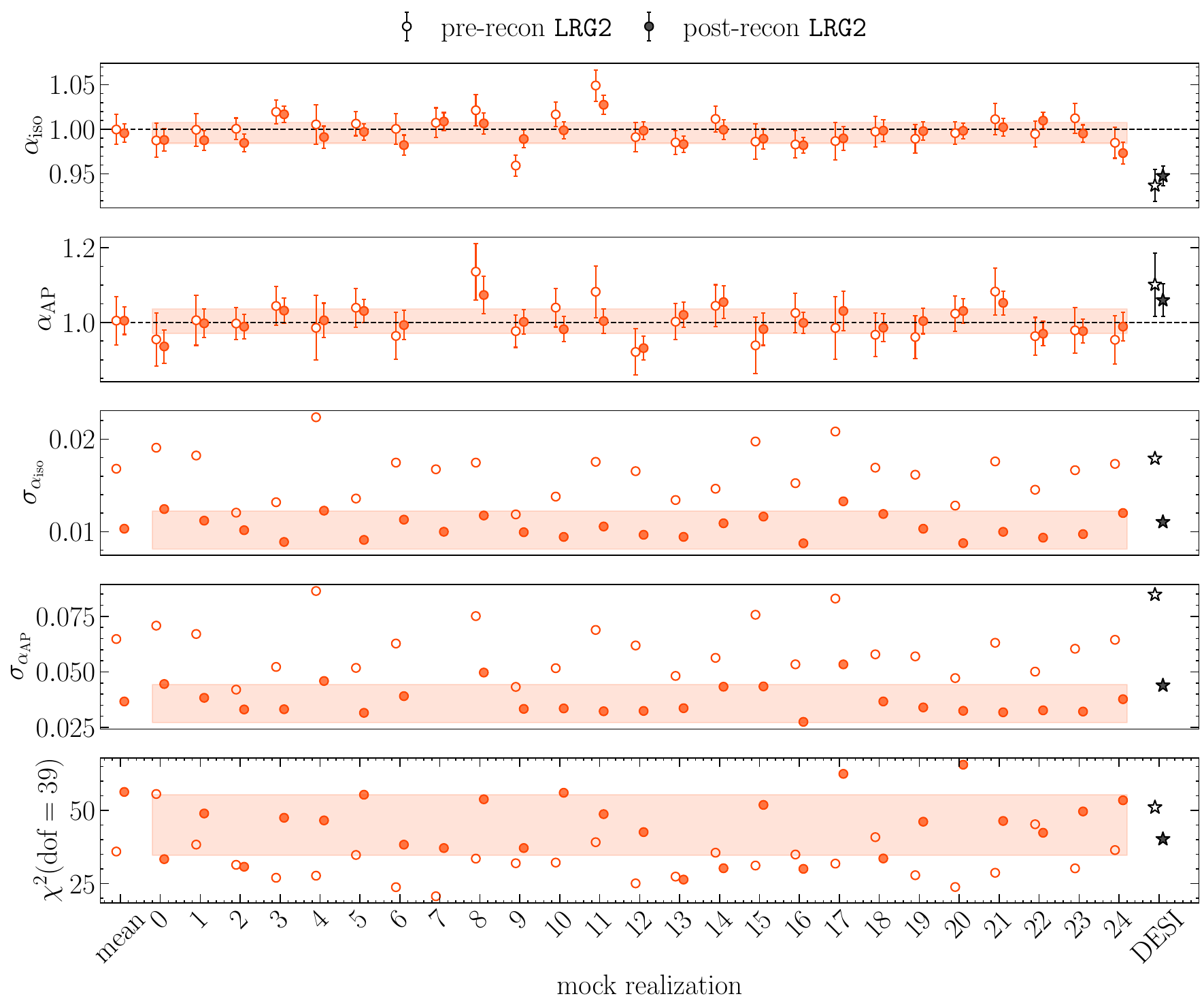}
    \caption{Same as \cref{fig:fits_abacussummit_lrg_0.8_1.1}, but for \lrgt.}
    \label{fig:fits_abacussummit_lrg_0.6_0.8}
\end{figure}

\begin{figure}
    \centering
    \includegraphics[width=0.9\textwidth]{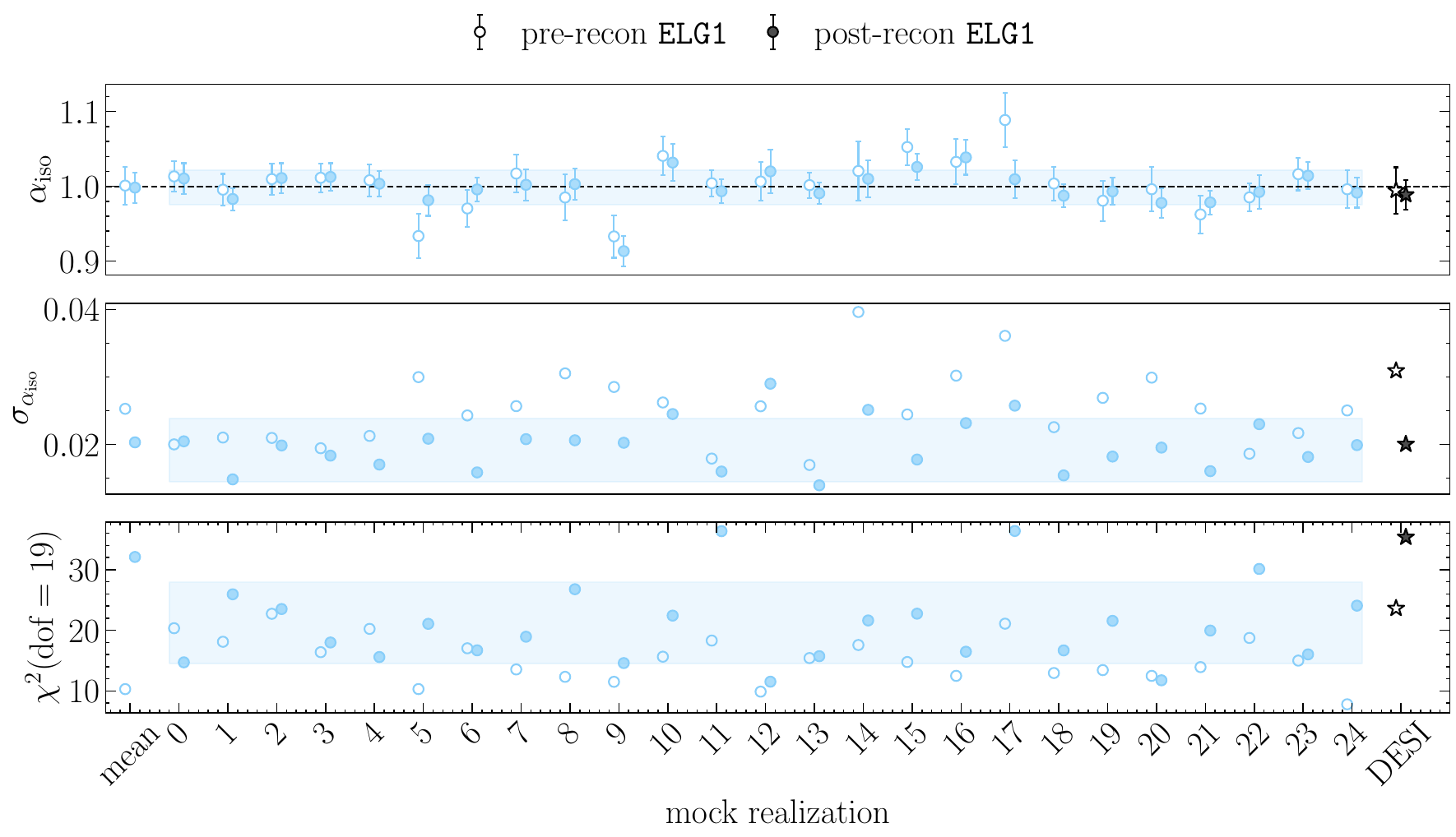}
    \caption{Similar to \cref{fig:fits_abacussummit_bgs_0.1_0.4}, but for \elgo.}
    \label{fig:fits_abacussummit_ELG_0.8_1.1}
\end{figure}

\begin{figure}
    \centering
    \includegraphics[width=0.9\textwidth]{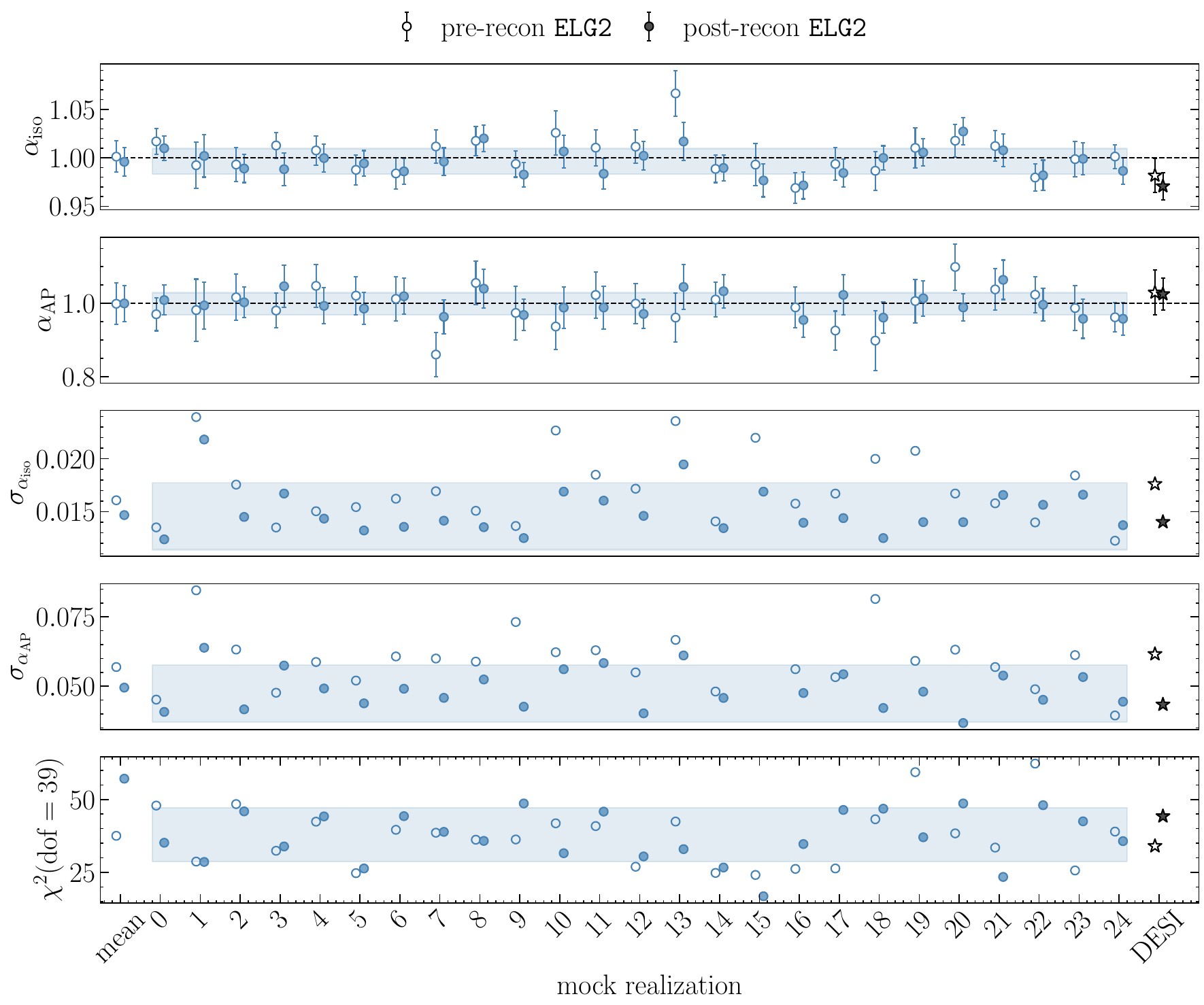}
    \caption{Similar to \cref{fig:fits_abacussummit_lrg_0.8_1.1}, but for \elgt.}
    \label{fig:fits_abacussummit_ELG_1.1_1.6}
\end{figure}

\begin{figure}
    \centering
    \includegraphics[width=0.9\textwidth]{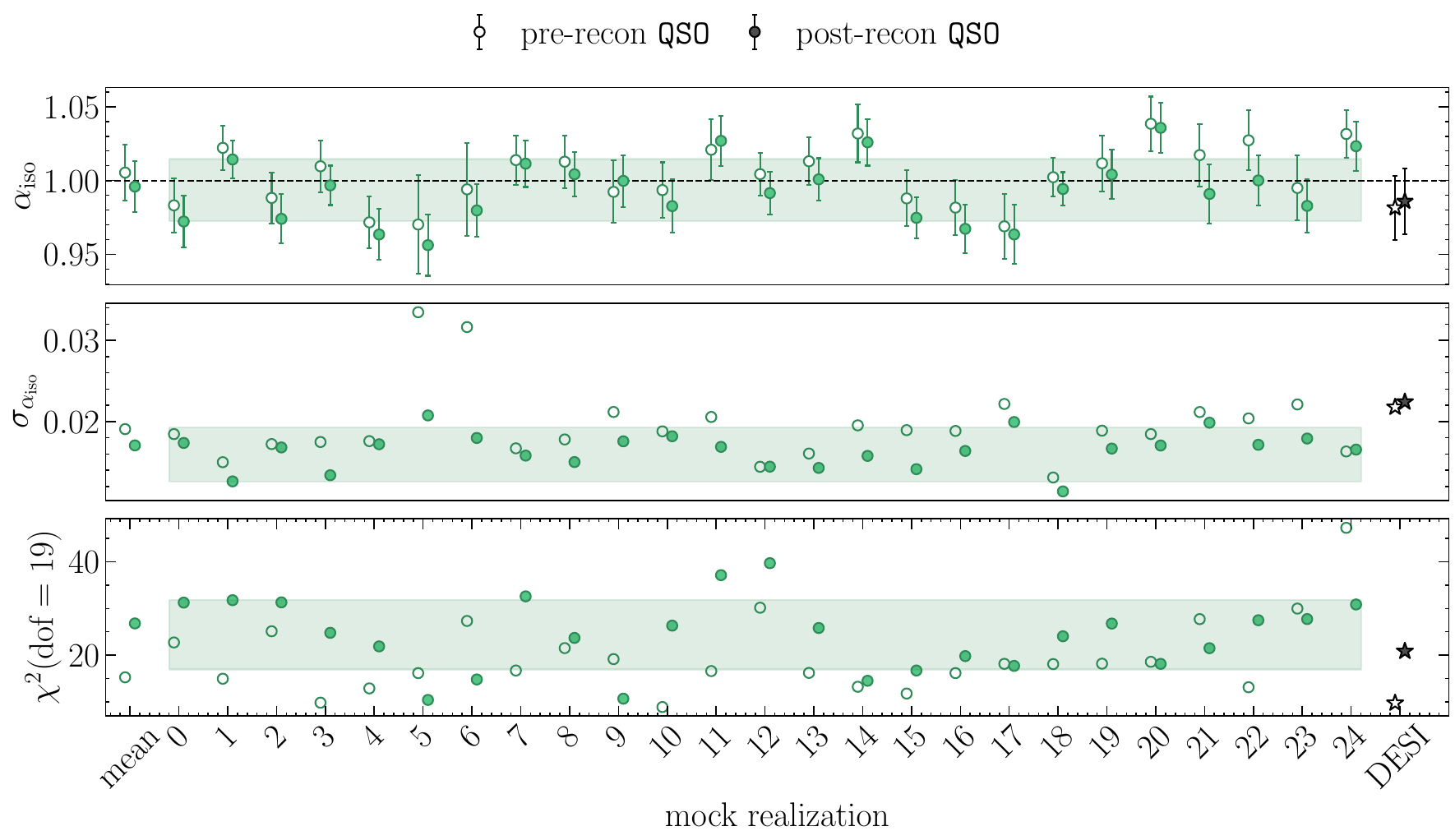}
    \caption{Similar to \cref{fig:fits_abacussummit_bgs_0.1_0.4}, but for \qso.}
    \label{fig:fits_abacussummit_QSO_0.8_2.1}
\end{figure}

\end{document}

%% file: authors.tex
\emailAdd{enrique.paillas@uwaterloo.ca}
\emailAdd{zhejied@sjtu.edu.cn}
\emailAdd{xinyi.chen@yale.edu}

\author[1,2]{{E.~Paillas}\orcidlink{0000-0002-4637-2868},}
\author[3]{{Z.~Ding}\orcidlink{0000-0002-3369-3718},}
\author[4]{{X.~Chen},}
\author[5]{{H.~Seo}\orcidlink{0000-0002-6588-3508},}
\author[4]{{N.~Padmanabhan},}
\author[6]{{A.~de~Mattia},}
\author[7,8,9]{{A.~J.~Ross}\orcidlink{0000-0002-7522-9083},}
\author[10]{{S.~Nadathur}\orcidlink{0000-0001-9070-3102},}
\author[11]{{C.~Howlett}\orcidlink{0000-0002-1081-9410},}
\author[12]{{J.~Aguilar},}
\author[13]{{S.~Ahlen}\orcidlink{0000-0001-6098-7247},}
\author[14]{{O.~Alves},}
\author[15,14]{{U.~Andrade}\orcidlink{0000-0002-4118-8236},}
\author[16]{{D.~Brooks},}
\author[17,18]{{E.~Buckley-Geer},}
\author[6]{{E.~Burtin},}
\author[19]{{S.~Chen}\orcidlink{0000-0002-5762-6405},}
\author[12]{{T.~Claybaugh},}
\author[20]{{S.~Cole}\orcidlink{0000-0002-5954-7903},}
\author[21]{{K.~Dawson},}
\author[22]{{A.~de la Macorra}\orcidlink{0000-0002-1769-1640},}
\author[23]{{Arjun~Dey}\orcidlink{0000-0002-4928-4003},}
\author[16]{{P.~Doel},}
\author[24,25]{{K.~Fanning}\orcidlink{0000-0003-2371-3356},}
\author[12,26]{{S.~Ferraro}\orcidlink{0000-0003-4992-7854},}
\author[27,28]{{J.~E.~Forero-Romero}\orcidlink{0000-0002-2890-3725},}
\author[29]{{C.~Garcia-Quintero}\orcidlink{0000-0003-1481-4294},}
\author[30,10,31]{{E.~Gaztañaga},}
\author[32,30,33]{{H.~Gil-Mar\'in}\orcidlink{0000-0003-0265-6217},}
\author[12]{{S.~Gontcho A Gontcho}\orcidlink{0000-0003-3142-233X},}
\author[18]{{G.~Gutierrez},}
\author[34]{{C.~Hahn}\orcidlink{0000-0003-1197-0902},}
\author[14]{{M.~M.~S~Hanif}\orcidlink{0009-0006-2583-5006},}
\author[7,35,9]{{K.~Honscheid},}
\author[29]{{M.~Ishak}\orcidlink{0000-0002-6024-466X},}
\author[36]{{R.~Kehoe},}
\author[12]{{A.~Kremin}\orcidlink{0000-0001-6356-7424},}
\author[12]{{M.~Landriau}\orcidlink{0000-0003-1838-8528},}
\author[37]{{L.~Le~Guillou}\orcidlink{0000-0001-7178-8868},}
\author[12]{{M.~E.~Levi}\orcidlink{0000-0003-1887-1018},}
\author[38,39]{{M.~Manera}\orcidlink{0000-0003-4962-8934},}
\author[7,8,9]{{P.~Martini}\orcidlink{0000-0002-4279-4182},}
\author[29]{{L.~Medina-Varela},}
\author[23]{{A.~Meisner}\orcidlink{0000-0002-1125-7384},}
\author[40]{{J.~Mena-Fern\'andez}\orcidlink{0000-0001-9497-7266},}
\author[41,39]{{R.~Miquel},}
\author[42]{{J.~Moustakas}\orcidlink{0000-0002-2733-4559},}
\author[43]{{E.~Mueller},}
\author[22]{{A.~Muñoz-Gutiérrez},}
\author[44]{{A.~D.~Myers},}
\author[45]{{J.~ A.~Newman}\orcidlink{0000-0001-8684-2222},}
\author[46]{{J.~Nie}\orcidlink{0000-0001-6590-8122},}
\author[47,48]{{G.~Niz}\orcidlink{0000-0002-1544-8946},}
\author[6,12]{{N.~Palanque-Delabrouille}\orcidlink{0000-0003-3188-784X},}
\author[1,49,2]{{W.~J.~Percival}\orcidlink{0000-0002-0644-5727},}
\author[12,50,26]{{C.~Poppett},}
\author[51]{{F.~Prada}\orcidlink{0000-0001-7145-8674},}
\author[22,52]{{A.~P\'{e}rez-Fern\'{a}ndez}\orcidlink{0009-0006-1331-4035},}
\author[53]{{M.~Rashkovetskyi}\orcidlink{0000-0001-7144-2349},}
\author[54]{{M.~Rezaie}\orcidlink{0000-0001-5589-7116},}
\author[5]{{A.~Rosado-Marin},}
\author[55]{{G.~Rossi},}
\author[56,11]{{R.~Ruggeri}\orcidlink{0000-0002-0394-0896},}
\author[57]{{E.~Sanchez}\orcidlink{0000-0002-9646-8198},}
\author[52]{{C.~Saulder}\orcidlink{0000-0002-0408-5633},}
\author[58]{{E.~F.~Schlafly}\orcidlink{0000-0002-3569-7421},}
\author[12]{{D.~Schlegel},}
\author[59,14]{{M.~Schubnell},}
\author[23]{{D.~Sprayberry},}
\author[14]{{G.~Tarl\'{e}}\orcidlink{0000-0003-1704-0781},}
\author[5]{{D.~Valcin}\orcidlink{0000-0003-0129-0620},}
\author[22]{{M.~Vargas-Maga\~na}\orcidlink{0000-0003-3841-1836},}
\author[60]{{J.~Yu},}
\author[25]{{S.~Yuan}\orcidlink{0000-0002-5992-7586},}
\author[12]{{R.~Zhou}\orcidlink{0000-0001-5381-4372},}
\author[46]{and {H.~Zou}\orcidlink{0000-0002-6684-3997}}

\affiliation[1]{Department of Physics and Astronomy, University of Waterloo, 200 University Ave W, Waterloo, ON N2L 3G1, Canada}
\affiliation[2]{Waterloo Centre for Astrophysics, University of Waterloo, 200 University Ave W, Waterloo, ON N2L 3G1, Canada}
\affiliation[3]{Department of Astronomy, School of Physics and Astronomy, Shanghai Jiao Tong University, Shanghai 200240, China}
\affiliation[4]{Physics Department, Yale University, P.O. Box 208120, New Haven, CT 06511, USA}
\affiliation[5]{Department of Physics \& Astronomy, Ohio University, Athens, OH 45701, USA}
\affiliation[6]{IRFU, CEA, Universit\'{e} Paris-Saclay, F-91191 Gif-sur-Yvette, France}
\affiliation[7]{Center for Cosmology and AstroParticle Physics, The Ohio State University, 191 West Woodruff Avenue, Columbus, OH 43210, USA}
\affiliation[8]{Department of Astronomy, The Ohio State University, 4055 McPherson Laboratory, 140 W 18th Avenue, Columbus, OH 43210, USA}
\affiliation[9]{The Ohio State University, Columbus, 43210 OH, USA}
\affiliation[10]{Institute of Cosmology and Gravitation, University of Portsmouth, Dennis Sciama Building, Portsmouth, PO1 3FX, UK}
\affiliation[11]{School of Mathematics and Physics, University of Queensland, 4072, Australia}
\affiliation[12]{Lawrence Berkeley National Laboratory, 1 Cyclotron Road, Berkeley, CA 94720, USA}
\affiliation[13]{Physics Dept., Boston University, 590 Commonwealth Avenue, Boston, MA 02215, USA}
\affiliation[14]{University of Michigan, Ann Arbor, MI 48109, USA}
\affiliation[15]{Leinweber Center for Theoretical Physics, University of Michigan, 450 Church Street, Ann Arbor, Michigan 48109-1040, USA}
\affiliation[16]{Department of Physics \& Astronomy, University College London, Gower Street, London, WC1E 6BT, UK}
\affiliation[17]{Department of Astronomy and Astrophysics, University of Chicago, 5640 South Ellis Avenue, Chicago, IL 60637, USA}
\affiliation[18]{Fermi National Accelerator Laboratory, PO Box 500, Batavia, IL 60510, USA}
\affiliation[19]{Institute for Advanced Study, 1 Einstein Drive, Princeton, NJ 08540, USA}
\affiliation[20]{Institute for Computational Cosmology, Department of Physics, Durham University, South Road, Durham DH1 3LE, UK}
\affiliation[21]{Department of Physics and Astronomy, The University of Utah, 115 South 1400 East, Salt Lake City, UT 84112, USA}
\affiliation[22]{Instituto de F\'{\i}sica, Universidad Nacional Aut\'{o}noma de M\'{e}xico,  Cd. de M\'{e}xico  C.P. 04510,  M\'{e}xico}
\affiliation[23]{NSF NOIRLab, 950 N. Cherry Ave., Tucson, AZ 85719, USA}
\affiliation[24]{Kavli Institute for Particle Astrophysics and Cosmology, Stanford University, Menlo Park, CA 94305, USA}
\affiliation[25]{SLAC National Accelerator Laboratory, Menlo Park, CA 94305, USA}
\affiliation[26]{University of California, Berkeley, 110 Sproul Hall \#5800 Berkeley, CA 94720, USA}
\affiliation[27]{Departamento de F\'isica, Universidad de los Andes, Cra. 1 No. 18A-10, Edificio Ip, CP 111711, Bogot\'a, Colombia}
\affiliation[28]{Observatorio Astron\'omico, Universidad de los Andes, Cra. 1 No. 18A-10, Edificio H, CP 111711 Bogot\'a, Colombia}
\affiliation[29]{Department of Physics, The University of Texas at Dallas, Richardson, TX 75080, USA}
\affiliation[30]{Institut d'Estudis Espacials de Catalunya (IEEC), 08034 Barcelona, Spain}
\affiliation[31]{Institute of Space Sciences, ICE-CSIC, Campus UAB, Carrer de Can Magrans s/n, 08913 Bellaterra, Barcelona, Spain}
\affiliation[32]{Departament de F\'{\i}sica Qu\`{a}ntica i Astrof\'{\i}sica, Universitat de Barcelona, Mart\'{\i} i Franqu\`{e}s 1, E08028 Barcelona, Spain}
\affiliation[33]{Institut de Ci\`encies del Cosmos (ICCUB), Universitat de Barcelona (UB), c. Mart\'i i Franqu\`es, 1, 08028 Barcelona, Spain.}
\affiliation[34]{Department of Astrophysical Sciences, Princeton University, Princeton NJ 08544, USA}
\affiliation[35]{Department of Physics, The Ohio State University, 191 West Woodruff Avenue, Columbus, OH 43210, USA}
\affiliation[36]{Department of Physics, Southern Methodist University, 3215 Daniel Avenue, Dallas, TX 75275, USA}
\affiliation[37]{Sorbonne Universit\'{e}, CNRS/IN2P3, Laboratoire de Physique Nucl\'{e}aire et de Hautes Energies (LPNHE), FR-75005 Paris, France}
\affiliation[38]{Departament de F\'{i}sica, Serra H\'{u}nter, Universitat Aut\`{o}noma de Barcelona, 08193 Bellaterra (Barcelona), Spain}
\affiliation[39]{Institut de F\'{i}sica d’Altes Energies (IFAE), The Barcelona Institute of Science and Technology, Campus UAB, 08193 Bellaterra Barcelona, Spain}
\affiliation[40]{Laboratoire de Physique Subatomique et de Cosmologie, 53 Avenue des Martyrs, 38000 Grenoble, France}
\affiliation[41]{Instituci\'{o} Catalana de Recerca i Estudis Avan\c{c}ats, Passeig de Llu\'{\i}s Companys, 23, 08010 Barcelona, Spain}
\affiliation[42]{Department of Physics and Astronomy, Siena College, 515 Loudon Road, Loudonville, NY 12211, USA}
\affiliation[43]{Department of Physics and Astronomy, University of Sussex, Brighton BN1 9QH, U.K}
\affiliation[44]{Department of Physics \& Astronomy, University  of Wyoming, 1000 E. University, Dept.~3905, Laramie, WY 82071, USA}
\affiliation[45]{Department of Physics \& Astronomy and Pittsburgh Particle Physics, Astrophysics, and Cosmology Center (PITT PACC), University of Pittsburgh, 3941 O'Hara Street, Pittsburgh, PA 15260, USA}
\affiliation[46]{National Astronomical Observatories, Chinese Academy of Sciences, A20 Datun Rd., Chaoyang District, Beijing, 100012, P.R. China}
\affiliation[47]{Departamento de F\'{i}sica, Universidad de Guanajuato - DCI, C.P. 37150, Leon, Guanajuato, M\'{e}xico}
\affiliation[48]{Instituto Avanzado de Cosmolog\'{\i}a A.~C., San Marcos 11 - Atenas 202. Magdalena Contreras, 10720. Ciudad de M\'{e}xico, M\'{e}xico}
\affiliation[49]{Perimeter Institute for Theoretical Physics, 31 Caroline St. North, Waterloo, ON N2L 2Y5, Canada}
\affiliation[50]{Space Sciences Laboratory, University of California, Berkeley, 7 Gauss Way, Berkeley, CA  94720, USA}
\affiliation[51]{Instituto de Astrof\'{i}sica de Andaluc\'{i}a (CSIC), Glorieta de la Astronom\'{i}a, s/n, E-18008 Granada, Spain}
\affiliation[52]{Max Planck Institute for Extraterrestrial Physics, Gie\ss enbachstra\ss e 1, 85748 Garching, Germany}
\affiliation[53]{Center for Astrophysics $|$ Harvard \& Smithsonian, 60 Garden Street, Cambridge, MA 02138, USA}
\affiliation[54]{Department of Physics, Kansas State University, 116 Cardwell Hall, Manhattan, KS 66506, USA}
\affiliation[55]{Department of Physics and Astronomy, Sejong University, Seoul, 143-747, Korea}
\affiliation[56]{Centre for Astrophysics \& Supercomputing, Swinburne University of Technology, P.O. Box 218, Hawthorn, VIC 3122, Australia}
\affiliation[57]{CIEMAT, Avenida Complutense 40, E-28040 Madrid, Spain}
\affiliation[58]{Space Telescope Science Institute, 3700 San Martin Drive, Baltimore, MD 21218, USA}
\affiliation[59]{Department of Physics, University of Michigan, Ann Arbor, MI 48109, USA}
\affiliation[60]{Ecole Polytechnique F\'{e}d\'{e}rale de Lausanne, CH-1015 Lausanne, Switzerland}